\documentclass[11pt]{article}
\usepackage{graphicx,amscd,amsmath,amssymb,amsfonts,verbatim}
\usepackage[a4paper, total={6.15 in, 10 in}]{geometry}
\usepackage{mathrsfs}
\usepackage{bbm}
\usepackage[T1]{fontenc}
\usepackage{lmodern}
\usepackage[english]{babel}
\usepackage{amsthm}
\usepackage[all]{xy}
\usepackage{mathtools}
\usepackage{algorithm}
\usepackage{authblk}
\usepackage{caption}
\usepackage{subfigure}
\usepackage{chngcntr}
\usepackage{setspace}
\usepackage{booktabs}
\usepackage{pgfplots}
\usepackage{adjustbox}

\usepackage{natbib}
\usepackage{authblk}
\usepackage{multirow}
\usepackage{enumitem}
\usepackage{fancyhdr}
\DeclareMathOperator*{\plim}{plim}
\usepackage{hyperref}
\hypersetup{
	colorlinks=true,
	citecolor = blue
}
\usepackage{lastpage}
\usepackage{pdflscape}
\graphicspath{}

\newtheorem{theorem}{Theorem}
\newtheorem{lemma}{Lemma}

\newtheorem{example}{Example}
\newtheorem{corollary}{Corollary}
\newtheorem{definition}{Definition}
\newtheorem{assumption}{Assumption}

\DeclareMathAlphabet{\mathpzc}{OT1}{pzc}{m}{it}
\author{Raphaël Langevin\footnote{Department of Economics, McGill Unversity, 855 Sherbrooke W, Montréal, QC, Canada, H3A 0C4. E-mail: raphael.langevin@mail.mcgill.ca.}}

\begin{document}

	\title{Bias-Reduced Estimation of Finite Mixtures~: An Application to Latent Group Structures in Panel Data\thanks{This is a revised version of the earlier draft entitled ``Consistent Estimation of Finite Mixture: An Application to Latent Group Panel Structures''. I am grateful to my supervisors, Erin Strumpf and Saraswata Chaudhuri, and to Philippe Goulet Coulombe, William McCausland, Victoria Zinde-Walsh, John W Galbraith, Byoung Park, JoonHwan Cho, Marine Carrasco, Lynda Khalaf, Umut Oguzoglu, Enrique Pinzon, David Stephens, Abbas Khalili, Partha Deb, Fabrice Etilé, Brantly Callaway, Peng Shao, Jad Beyhum, and Aristide Houndetoungan for helpful comments on early drafts of this paper. Special thanks to Pierre-Carl Michaud, David Boisclair, and François Laliberté-Auger for providing access to computational hardware and technical advice. I also thank the participants of NY Camp Econometrics XVII, the 18$^{th}$ CIREQ PhD Student Conference, the $57^{th}$ Annual Meeting of the Canadian Economics Association, the 2023 Stata Conference in Stanford, the 2024 RCEA International Conference at Brunel University London, the 63$^{th}$ SCSE Annual Congress at HEC Montréal, the 2024 NASMES at Vanderbilt University, the 2024 IAAE Annual Conference in Thessaloniki, and the 2024 AHEW in Washington D.C. for advice and comments. This work was supported by the grant no.767-2020-2809 of the Social Sciences and Humanities Research Council of Canada (SSHRC).}}

	\maketitle
	\thispagestyle{empty} 
	\begin{abstract}
		Finite mixture models are widely used in econometric analyses to capture unobserved heterogeneity. This paper shows that maximum likelihood estimation of finite mixtures of parametric densities can suffer from substantial finite-sample bias in all parameters under mild regularity conditions. The bias arises from the influence of outliers in component densities with unbounded or large support and increases with the degree of overlap among mixture components. I show that maximizing the classification-mixture likelihood function, equipped with a consistent classifier, yields parameter estimates that are less biased than those obtained by standard maximum likelihood estimation (MLE). I then derive the asymptotic distribution of the resulting estimator and provide conditions under which oracle efficiency is achieved. Monte Carlo simulations show that conventional mixture MLE exhibits pronounced finite-sample bias, which diminishes as the sample size or the statistical distance between component densities tends to infinity. The simulations further show that the proposed estimation strategy generally outperforms standard MLE in finite samples in terms of both bias and mean squared errors under relatively weak assumptions. An empirical application to latent group panel structures using health administrative data shows that the proposed approach reduces out-of-sample prediction error by approximately 17.6\% relative to the best results obtained from standard MLE procedures.
	\end{abstract}
	
\newpage
\renewcommand{\thefootnote}{\arabic{footnote}}
\section{Introduction}\label{sec1:1}
\setstretch{1.3}
\pagenumbering{arabic}
Finite mixtures are extensively used in statistics, computer science, and machine learning for pattern recognition and unsupervised classification to account for various types of unobserved heterogeneity \citep{bishop_pattern_2006, fruhwirth-schnatter_finite_2006}. Several applications of finite mixtures can also be found in labor and health economics \citep{heckman_method_1984,deb_demand_1997, keane_career_1997, jones_healthcare_2015}. For instance, \cite{deb_demand_1997} use finite mixtures to distinguish two unobserved, latent types (i.e. the ``healthy'' and the ``ill'') regarding the demand for medical care and find substantial differences in fitted distributions for each type. Methods to account for unobserved heterogeneity can reduce bias \citep{hsiao_analysis_2014}, improve inference and forecast \citep{boot_optimal_2018}, and also allow for the estimation of heterogeneous treatment effects \citep{ahn_dierence_2024}.
\par
A finite mixture distribution is a convex combination of a small number of distinct parametric densities where the combination weights, known as \textit{mixing weights}, correspond to the proportion of observations that originate from each density in the population. The resulting density, known as the \textit{mixture density}, fully describes the distribution of the observed data. Finite mixtures are related to unsupervised clustering methods since each \textit{component density} within the mixture fully describes its corresponding group of observations. Finite mixtures are also known to be highly flexible since they can ``approximate any distribution to a sufficient level of accuracy'' and easily accommodate the presence of covariates and nonlinear components \citep{nguyen_approximations_2019}. Estimation of the parameters contained in the mixture density is usually performed using maximum likelihood estimation (MLE) procedures, including nonparametric MLE \citep{compiani_using_2016, kitamura_nonparametric_2018}.
\par
Although it has been shown that maximizing the (constrained) likelihood of a finite mixture of parametric densities yields consistent estimates, little is known about the finite-sample behavior of the MLE of finite mixtures \citep{balakrishnan_statistical_2017,kwon_minimax_2021}. In this paper, I show that maximizing the likelihood of a mixture density using standard parametric MLE methods can lead to finite-sample biases that are much larger than those that are typically expected to be obtained from standard MLE approaches under generic identifiability and correct specification of the mixture density. The biases come from the presence of outliers that are generated by the tails of component densities with unbounded or large support. When components are not sufficiently separated, the global maximum of the likelihood shifts away from the true parameters such that at least one component density becomes strongly influenced by these ``abnormal observations'' \citep{chen_statistical_2023}. The presence of large finite-sample biases in the mixture parameters is confirmed via Monte Carlo simulations under mixtures of normal, Poisson, and exponential distributions. The size of the bias depends on the components' size and on the degree of separation between each component density.
\par
The contributions of the paper are twofold. First, I show why standard MLE of finite mixtures can lead to severely biased estimates of all parameters in the mixture under weak regularity conditions and generic identifiability \citep{fruhwirth-schnatter_finite_2006}.\footnote{By ``all parameters'' in the mixture, I refer to the mixing weights and the set of parameters governing each density in the mixture. Note also that ``components'' and ``groups'' are used interchangeably throughout the paper.} Given the analytical complexities associated with the general behavior of mixture densities in finite samples, I only present the general intuition behind the existence of the finite-sample bias using the two-component mixture of normal distributions as a practical example. Extension of the same intuition to mixtures of other distributions in the exponential family is straightforward. By standard MLE of finite mixtures, I refer to procedures that attempt to globally maximize the mixture likelihood function, as defined in Section \ref{subsec1:31}. The well-known expectation-maximization (EM) algorithm of \cite{dempster_maximum_1977} and all Newton-type algorithms, when applied to the mixture likelihood, both fall into this category.
\par
The second contribution of the paper is to show under which conditions the maximization of a different objective function, the classification-mixture likelihood (C-ML) \citep{mclachlan_iterative_1975,mclachlan_finite_2019}, will lead to less biased, consistent, and asymptotically efficient estimates of all parameters in any generically identifiable mixture models. Similar to \cite{dzemski_convergence_2021}, I show that consistency in estimation and asymptotic efficiency can be obtained by maximizing the C-ML function if it is combined with a consistent classifier, meaning that the proportion of misclassified observations goes to zero in the limit when evaluated at the true parameter values.
\par
Although MLE is asymptotically unbiased in general, it is not unbiased in finite samples. Therefore, obtaining unbiased estimates (up to a term of order $N^{-1}$) of all parameters in the mixture in finite samples is impossible without bias-reduction techniques \citep{kosmidis_bias_2014}. In the case of finite mixtures, I show that it is possible to reduce the size of the bias from the standard MLE by classifying all observations in their true component density. This latter requirement is much stronger than the one used for consistency, and it cannot be guaranteed for all data-generating processes (DGPs) with a given sample size. Nonetheless, conditions under which maximizing the C-ML objective will lead to the correct classification of all observations in the sample are derived. Such conditions can take the form of restrictions on group memberships, as typically done in the literature on latent group panel structures \citep{bonhomme_grouped_2015, su_identifying_2016}, or through additional assumptions on the general structure of the outcome vector and/or the available covariates, which is the general strategy that is retained in this paper due to its greater flexibility.
\par
I illustrate those two theoretical contributions with both simulated and real-world data. First, I use simulated data to characterize the finite-sample biases of standard MLE procedures when applied to mixtures of normal, Poisson, and exponential distributions. Then, I use both simulated and real-world data to compare the finite-sample performance of the proposed estimation strategy to standard MLE procedures via the EM algorithm. Both types of data correspond to latent group panel structures with a univariate outcome and $p$ strictly exogenous covariates. Simulation results confirm that estimates obtained from the maximization of the mixture likelihood are more biased and less efficient than those obtained from the proposed estimation strategy under weak regularity conditions. On the contrary, reducing the number of covariates or increasing the overlap between component densities leads to a higher misclassification rate, which results in larger finite-sample biases. Simulation results show that no algorithm outperforms the other when the misclassification rate increases away from zero.
\par
The empirical part of the paper uses the EM algorithm and the proposed estimation algorithm to model individual healthcare expenditure (HCE) from administrative data over time. Both algorithms use a latent group two-part model (LGTPM) for estimation and an identical specification for each component density. Results show that the proposed algorithm leads to a reduction in out-of-sample prediction error of 17.6\% compared to the best result obtained when using the EM algorithm, and of 56.6\% compared to the single-component two-part model. The proposed algorithm also allows the recovery of the true group memberships for almost every observation in the sample, which can be modeled in a second step for forecasting purposes.
\par
The remainder of the paper is as follows. Section \ref{sec1:2} briefly reviews the related literature. Section \ref{sec1:3} shows why maximizing the mixture likelihood leads to large finite-sample biases in simple mixture models under weak regularity conditions. It also details the proposed estimation strategy that is used to reduce the size of such biases. Section \ref{sec1:4} presents simulation scenarios and results, confirming the theoretical insights from the previous section. Section \ref{sec1:5} presents the empirical application and the related results, while Section \ref{sec1:6} concludes.
\section{Related Literature}\label{sec1:2}
\textbf{Finite Mixtures and the EM Algorithm}. 
Although there exist several optimization algorithms that can maximize the mixture likelihood, the EM algorithm is almost always used by practitioners when estimating mixture models due to its simplicity and empirical performance \citep{kwon_global_2024}.\footnote{Note that the EM algorithm is not a maximization algorithm \textit{per se}, but a general strategy for the maximization of any incomplete-data likelihood function. Indeed, it is often desirable to use a numerical optimization method at each M-step of the EM algorithm. See Section \ref{subsec1:53} for an example, and Section 2.4.4 of \cite{fruhwirth-schnatter_finite_2006} for more details on the subject.} The EM algorithm consists of the consecutive repetition of an expectation step (i.e., the E-step) and a maximization step (i.e., the M-step), each conditional on the results obtained from the previous step. The E-step computes assignment probabilities for each observation and each component density, whereas the M-step maximizes the likelihood function conditional on the most recent assignment probabilities. In their seminal paper, \cite{dempster_maximum_1977} showed that such an algorithm never decreases the \textit{incomplete-data likelihood} function between two consecutive iterations, the mixture likelihood being a special case of the former when the mixing distribution has discrete support.
\par
It is also well known that the mixture likelihood is not a convex function of the parameter space \citep{mclachlan_finite_2000}. Hence, the multimodal nature of the mixture likelihood makes it difficult for any optimization algorithm to find its global maximum, which has given rise to various procedures with the explicit goal of reducing the probability of being ``trapped'' in a local spurious maximum \citep{fruhwirth-schnatter_em_2019}. Moreover, it is acknowledged that ``the sample size [...] has to be very large, before asymptotic theory of maximum likelihood applies'', especially when the ``component densities are poorly separated'' \citep{redner_mixture_1984,aragam_uniform_2023}. To provide statistical guarantees, \cite{balakrishnan_statistical_2017} derived sufficient conditions for the initialization of the EM algorithm such that it converges ``geometrically to a fixed point that is within statistical precision of the unknown true parameter'' \citep{balakrishnan_statistical_2017}. However, these conditions do not imply that the true parameters will be close to the global maximum of the mixture likelihood for any fixed sample size and any true mixture density. In fact, I show in the next section that searching for the global maximum of the mixture likelihood can yield highly misleading results in finite samples due to the presence of ``natural'' outliers, which correspond to outliers that are generated by the tails of at least one component density. This intuition is confirmed via simulations that are presented in Section \ref{sec1:4}.
\par
This paper also relates to a series of papers that followed the publication of \cite{balakrishnan_statistical_2017}, and where the authors studied the convergence properties of the EM algorithm when applied to mixtures of Gaussian densities and linear regressions \citep{wu_randomly_2022, kwon_em_2019, kwon_em_2020, kwon_global_2024}. Globally, these papers show that the EM algorithm will asymptotically converge to the true parameter values with high probability -- regardless of the initialization procedure -- for two-component Gaussian and linear regression mixtures with equal mixing weights and known variances. However, the general behavior of the EM algorithm in finite samples remains elusive when mixing weights are unequal or variances are unknown, which is often the case in practice.
\vspace{2mm}
\\
\textbf{K-means and the C-EM Algorithm}. The K-means algorithm is one of the most widely used clustering algorithms in unsupervised machine learning \citep{hastie_elements_2009}. It can be used to estimate Gaussian mixture models just as the EM algorithm, but is less flexible than the EM since it assumes spherical covariance matrices of the outcome(s). Moreover, it is known to yield \textit{inconsistent} estimates of all parameters in the mixture unless the cluster centers are infinitely distant from each other \citep{pollard_strong_1981,bryant_large-sample_1991}. The C-EM algorithm (for Classification-EM) is the likelihood generalization of K-means and is as flexible as the EM, but still leads to inconsistent estimates of the mixture parameters \citep{bryant_asymptotic_1978, celeux_classification_1992}.\footnote{The acronym ``C-EM'' is used to avoid potential confusion with the \textit{Coarsened Exact Matching} (CEM) procedure proposed by \cite{iacus_causal_2012} or with the Component-wise EM algorithm. It also mimics the C-LASSO acronym of \cite{su_identifying_2016}, even though the ``CEM'' acronym was originally employed by \cite{celeux_classification_1992}. Thanks to Giuseppe Moscelli for bringing this point to my attention.} Since the K-means algorithm corresponds to the special case of the C-EM algorithm when all within-cluster errors are independent, normally distributed, and homoskedastic, the rest of the paper focuses on the general and more flexible C-EM algorithm.
\par
Contrary to the EM algorithm, the C-EM algorithm classifies each observation into the group that maximizes its corresponding density value. This practical difference leads to an important theoretical difference, which is that the C-EM algorithm maximizes the C-ML function rather than the mixture likelihood function. The C-ML function is the estimated counterpart of the \textit{complete-data} likelihood function, which is the likelihood function one would get if all group memberships were known \citep{mclachlan_finite_2019}. If this is the case, then it is said that the estimator achieves \textit{oracle efficiency} and all estimated parameters are asymptotically normal and efficient as the corresponding maximum likelihood estimates reach the semiparametric efficiency bound \citep{newey_semiparametric_1990,su_identifying_2016}.
\par
Akin to the EM algorithm, it has been shown that the C-ML function never decreases between two consecutive iterations of the C-EM algorithm.\footnote{This is however only true for a certain class of classifiers. See, for instance, \cite{celeux_classification_1992}, \cite{bottou_convergence_1994}, and the proof of Lemma \ref{lem1:cem} for more details.} Given that the C-ML function is also multimodal, convergence of the C-EM algorithm to the global maximum of the C-ML function is rarely guaranteed, and similar issues are encountered in practice when maximizing either the mixture likelihood or the C-ML function \citep{same_online_2007}. It is also important to stress that global convergence of the estimation algorithm is not very helpful if the global maximum of the objective function does not lie in a shrinking neighborhood of the true parameter values when the sample size remains below a certain threshold. Simulation results presented in section \ref{subsec1:42} shows that such a phenomenon occurs with finite mixtures of parametric densities that overlap significantly.
\vspace{2mm}
\\
\textbf{Latent Group Panel Structures}. Several recent studies in econometrics have used different variants of the K-means algorithm to account for unobserved heterogeneity in panel data through the introduction of grouped fixed-effects (GFE) and/or group-specific coefficients \citep{bonhomme_grouped_2015,bonhomme_distributional_2019,bonhomme_discretizing_2022,okui_heterogeneous_2021,lumsdaine_estimation_2023, liu_identification_2020, wang_panel_2024}. The GFE estimator is also closely related to factor models since it allows time-fixed effects to vary across groups, similar to models with interactive fixed effects \citep{bai_panel_2009}. However, factor models typically assume homogeneity of all the other parameters in the model, whereas the GFE estimator imposes restrictions on group memberships or on the form taken by the unobserved heterogeneity (across units or over time) to achieve consistency \citep{bonhomme_grouped_2015,bonhomme_discretizing_2022, wang_panel_2024}. Contrary to the recent developments on the subject, the estimation procedure proposed in this paper leaves group membership completely unrestricted across units and over time, while still achieving consistency under the ``many covariates'' asymptotics where the number of covariates $p$ is allowed to grow at a strictly slower rate than the sample size \citep{cattaneo_inference_2018, kitamura_nonparametric_2018}.
\par
Several authors also relied on the LASSO device or on binary segmentation algorithms for combining unit-level coefficients into group-level coefficients \citep{su_identifying_2016, su_sieve_2019, qian_shrinkage_2016, wang_homogeneity_2018, wang_identifying_2021}. All estimation strategies presented in this strand of the literature impose certain restrictions on group memberships over time. For instance, in the case of the Classifier-LASSO originally developed by \cite{su_identifying_2016}, leaving group membership completely unrestricted implies the estimation of $NT$ parameters for each time-varying covariate in the sample, which is impractical in regular panel datasets. Contrary to this strand of literature, this paper focuses exclusively on regular finite mixtures where no restriction is imposed on group memberships along any dimension of the dataset.
\section{Maximum Likelihood Estimation of Finite Mixtures}\label{sec1:3}
This section is divided into three parts. Section \ref{subsec1:31} presents the notation that will be used throughout the paper. Section \ref{subsec1:32} then intuitively explains why standard MLE of finite mixtures leads to sometimes heavily biased estimates of all parameters under weak regularity conditions in finite samples. For simplicity, the case that is considered in Sections \ref{subsec1:31} and \ref{subsec1:32} is a cross-section of $N$ independent observations without covariates. Extension of the following results to panel data or time series with strictly exogenous covariates is straightforward. Finally, Section \ref{subsec1:33} describes the proposed estimation strategy and shows under which conditions it will lead to less biased, consistent, and asymptotically efficient estimation of all parameters in the mixture. All proofs are deferred to Appendix \ref{sec1:A}.
\subsection{General Setup and Notation}\label{subsec1:31}
The mixture density function of any observation $y_{i} \in \mathcal{Y} \subseteq \mathbb{R}^{d_y}$ is generally defined as follows
\begin{align}\label{eqn1:1}
	f(y_{i}|\xi) := \sum_{g=1}^G \pi_g f_g(y_{i}|\theta_g) \equiv \sum_{g=1}^G \pi_g f_g(y_{i}|\theta),
\end{align}
where $y_{i}$ is the (possibly multivariate) observed outcome of individual $i$ with $i = 1,...,N$. The function $f: \mathcal{Y} \subseteq \mathbb{R}^{d_y} \rightarrow \mathbb{R}_{>0}$ is the mixture density, defined as a function of $\theta = (\theta_1,...,\theta_g,...,\theta_G)$, which represents the set of \textit{component parameters}, and of $\pi = (\pi_1,...,\pi_G)^{\top}$, the vector of mixing weights, where $\xi = (\theta,\pi)$ represents the whole set of parameters, and where $G \in \mathbb{N}_{>1}$ is the total number of components. The set of component parameters $\theta$ is assumed to lie inside the compact parameter space $\Theta = \Theta_1 \times ... \times \Theta_G \subset \mathbb{R}^{G \times d_{\theta_g}}$, whereas the vector of mixing weights $\pi$ is assumed to lie within the open space $\Pi = (0,1)^{G}$, where the only additional constraint imposed on the mixing weights is that $\sum_{g=1}^G \pi_g =1$. Each component density $f_g(.)$ corresponds to a well-defined parametric density with respect to a $\sigma$-finite measure, generally denoted by $\upsilon(d y_{i})$. All component densities do not necessarily belong to the same family of distributions, although this is common in practice to facilitate the estimation of $\theta$, and can be extended to include (exogenous) covariates provided an appropriate specification of the link between $y_i$ and $x_i$ for each component density.
\par
Throughout the paper, the set of true parameter values is denoted by $ \xi^0 = (\pi^0, \theta^0)$ and lies in the interior of the true parameter space $\Pi \times \Theta\backslash E \equiv \Xi^0$, where $\pi^0 = (\pi^0_1,...,\pi^0_g,...,\pi^0_G)^{\top}$, $\theta^0 = (\theta^0_1,...,\theta^0_g,...,\theta^0_G)$, and where I exclude the set $E = \{\Theta: \cup_{j\ne g}\cup_{g=1}^G (\theta^0_j = \theta^0_g) \}$ from the true parameter space to avoid identification issues. Any dataset generated by this true density is denoted by $\mathbf{y}_N = \{y_{i}\}_{i=1}^N$. Note that the true number of components, $G$, is assumed to be discrete, finite, and known by the econometrician unless stated otherwise. This assumption allows us to rule out identification issues caused by the misspecification of $G$, a topic that has been extensively covered in both statistics and econometrics \citep{redner_note_1981,kasahara_testing_2015,budanova_penalized_2025}.
\par
In the finite mixture framework, each observation is assumed to originate from only one of the $G$ densities. The unobserved, true binary assignment (or grouping) variable, denoted by $z^0_{ig}$, is defined as follows
\begin{align}
	z^0_{ig} := \label{eqn1:2}
	\begin{cases}
		1 \ \ \Leftrightarrow \ y_{i} \sim f_g(\cdot|\theta^0),\\
		0 \ \ \text{otherwise.}
	\end{cases}
\end{align}
Consequently, the true mixing weight of the $g^{th}$ component density, $\pi_g^0$, is defined accordingly
\begin{align}\label{eqn1:3}
	\pi^0_g := \plim_{N \to \infty} \sum_{i=1}^N \frac{z^0_{ig}}{N} = \mathbb{P}[z^0_{ig}=1],
\end{align}
which corresponds to the unconditional probability of any observation to belong to the $g^{th}$ component/group. Note that $z^0_{ig} = 1$ for at least one value of $i \in \{1,2...,N\} = [N]$ for any $g \in \mathbb{G}$ since $\pi^0_g \in (0,1)$ for any $g \in \mathbb{G}$. This also implies that the value $z^0_{ig}$ can be seen as the realization of the random variable $Z_{i}$ drawn from a univariate multinomial distribution with $G$ categories and vector of probabilities $(\pi^0_1,...,\pi_G^0)$ \citep{mclachlan_finite_2019}.
\par
Several objective functions can be written using the above equations. The three most important objective functions are described below in Definition \ref{def1:0}.
\begin{definition}\label{def1:0}
	$ \ $
	Using the previous definitions and equations, I define the following objective functions.
	\begin{enumerate}[label=(\roman*) ]
	\item The \textit{complete-data log likelihood function}~:
	\begin{align}
		l^C(\theta,\pi,\mathbf{z}^0) \equiv l^C(\xi,\mathbf{z}^0) &:= \sum_{i=1}^N \sum_{g=1}^G z^0_{ig} \log(\pi_g f(y_{i}|\theta)),
	\end{align}
	where $\mathbf{z}^0 = (\mathbf{z}^0_1,...,\mathbf{z}^0_g,...,\mathbf{z}^0_G)$, with $\mathbf{z}^0_g = (z^0_{1g},...,z^0_{ig},...,z^0_{Ng})^{\top}$.
	\item The \textit{mixture likelihood} function~:
	\begin{align}
		l^{ML}(\xi) &:= \sum_{i=1}^N \log(\sum_{g=1}^G \pi_g f_g(y_{i}|\theta)).
	\end{align}
	\item The \textit{classification-mixture likelihood} (C-ML) function~:
	\begin{align}
		l^{CM}(\theta, \mathbf{z}) &:= \sum_{i=1}^N \sum_{g=1}^G z_{ig} \log(f_g(y_{i}|\theta)),
	\end{align}
	where $\mathbf{z} = (\mathbf{z}_1,...,\mathbf{z}_g,...,\mathbf{z}_G)$, with $\mathbf{z}_g = (z_{1g},...,z_{ig},...,z_{Ng})^{\top}$, and where each $z_{ig} =\{0,1\}$ with $\sum_{g=1}^G z_{ig} = 1$ (see Definition \ref{def1:1}).
\end{enumerate}
\end{definition}
The complete-data log likelihood function is the objective one would obtain were the true group memberships known for every observation in the sample. Note that maximizing $l^C(\xi,\mathbf{z}^0)$ with respect to $\theta$ is similar to maximizing $G$ distinct log likelihood functions. The vector of mixing weights may then be estimated using Eq. (\ref{eqn1:3}). Since it is assumed that $z^0_{ig}$ is unobserved for all $i \in [N]$, the objective functions shown in Definition \ref{def1:0} $(ii)$ and $(iii)$ can be used to estimate both $\pi^0$ and $\theta^0$.
\par
If $z_{ig} = z^0_{ig}$ for all pairs $(i,g) \in [N] \times \mathbb{G}$, then $l^{CM}(\theta, \mathbf{z}) = l^C(\xi,\mathbf{z}^0) - \sum_{i=1}^N z^0_{ig} \log(\pi_{g})$ and oracle efficiency is achieved for any value of $N$ since maximizing $l^{CM}(\theta, \mathbf{z})$ with respect to $\theta$ will yield estimates as efficient as the ones obtained from maximizing $l^C(\xi,\mathbf{z}^0)$ with respect to $\theta$. Consequently, the estimates $\hat{\theta}^{CM} = \arg \max_{\theta \in \Theta} l^{CM}(\theta, \mathbf{z})$ will be consistent and reach the semiparametric efficiency lower bound asymptotically if $f_g(y_{i}|\theta)$ is regular for any $g \in \mathbb{G}$, meaning that $f_g(y_{i}|\theta)$ has to be a smooth function of $y_i$ for any $\theta \in \Theta$ with nonsingular information matrix (see \cite{newey_semiparametric_1990} for more details). Note also that all three objective functions do not put any restriction on group memberships across units, and that the estimated membership $\hat{z}_{i}(\theta^0):= \arg \max_{g \in \mathbb{G}} f_g(y_i|\theta^0)$ will not always lead to $\hat{\mathbf{z}}(\theta^0) = \mathbf{z}^0$ for any value of $N$ unless all component densities $f_g(\cdot|\theta^0)$ are infinitely distant from each other \citep{bryant_large-sample_1991,celeux_classification_1992}.\footnote{An infinite distance between all component densities implies that the Kullback-Leibler divergence between each pair of different component densities is equal to infinity, meaning that $\int_{\mathcal{Y}} f_g(y_{i}|\theta^0_g) \times f_j(y_{i}|\theta^0_j) \upsilon(dy_{i})$ is arbitrarily close to zero for any pair $(g,j) \in \mathbb{G}\times\mathbb{G}\backslash g$ .}
\par
Although it might not be efficient in finite samples, it is commonly acknowledged that maximizing the mixture likelihood function $l^{ML}(\xi)$ with respect to $\xi$ will yield consistent and asymptotically efficient estimates of the mixture parameters provided appropriate constraints on the parameter space \citep{redner_mixture_1984, chen_consistency_2017, chen_statistical_2023}. However, it remains unclear in the literature on finite mixtures if the estimated parameters $\hat{\xi}^{ML} = \arg \max_{\xi \in \Pi \times \Theta} l^{ML}(\xi)$ feature large biases in finite samples, and what would drive the size of such biases. The next section attempts to shed some light on these aspects.
\subsection{Finite-Sample Bias of MLE of Finite Mixtures}\label{subsec1:32}
This subsection shows why maximizing the mixture likelihood with respect to $\xi$ sometimes leads to large finite-sample biases in all estimated parameters under the following regularity conditions.
\begin{assumption}\label{ass1:1}
	\setstretch{1.15}
	$ \ $
	\begin{enumerate}[label=(\roman*) ]
		\item $\mathbb{E}_{0}[\log(f(y_{i}|\xi))] < \infty$ for any $y_{i} \in \mathcal{Y}$, any $\xi \in \Xi \equiv \Theta \times \Pi$, where $\mathbb{E}_{0}[\cdot]$ stands as the expected value with respect to the true mixture density, $f(\cdot|\xi^0)$, and where all $y_{i}$ are independent and identically distributed according to the true mixture density.
		\item $f_g(y_{i}|\theta_g) > 0$ for any $y_{i} \in \mathcal{Y}$, for any $\theta_g \subset \theta \in \Theta$, and for any $g \in \mathbb{G}$, with $\theta_g = \theta_j \Leftrightarrow g = j$ for any pair $(j,g) \in \mathbb{G}^2$, and where all component densities share the same unbounded support (e.g., $\mathbb{R}, \mathbb{N}, \mathbb{Z}$.).
		\item $l_g(\theta_g) := \sum_{i=1}^N \log(f_g(y_{i}|\theta_g))$ features a unique maximum with respect to $\theta_g$ for any $g \in \mathbb{G}$ and any dataset $\mathbf{y}_N \in \mathcal{Y}^N$.
		\item $l^{ML}(\xi) = l^{ML}(\xi') \Leftrightarrow \xi = \xi'$ up to any permutation in the labels of $\xi$ and $\xi'$, for any pair $(\xi,\xi') \in \Xi^2$ and any dataset $\mathbf{y}_N\in \mathcal{Y}^N$.
		\item $l^{ML}(\xi)$ features a unique global maximum with respect to $\xi$ for any dataset $\mathbf{y}_N\in \mathcal{Y}^N$.
		\item $f(y_{i}|\xi)$ is continuously differentiable with respect to $\xi$, with information matrix $\mathcal{I}(\xi) := \mathbb{E}_0[s_i(\xi) s_i(\xi)^{\top}] < \infty$ where $s_i(\xi) = \frac{\partial \log f(y_i|\xi) }{\partial \xi}$ for any $\xi \in \Xi$ and any $y_{i} \in \mathcal{Y}$.
	\end{enumerate}
\end{assumption}
\setstretch{1.3}
Assumption \ref{ass1:1}$(i)$ assumes that all observations are independently and identically distributed according to the true mixture density, which is standard in the literature. Assumption \ref{ass1:1}$(i)$ however rules out cases where the log likelihood function is not bounded from above. As emphasized recently by \cite{chen_statistical_2023}, this assumption is crucial for the consistency of the standard MLE of finite mixtures. In practice, it rules out standard MLE of mixtures with component densities that belong to the location-scale family \citep{tanaka_strong_2009}. This issue is well-studied in the particular case of a mixture of Gaussian densities, where it is possible to set the mean value of the first component equal to any observation in the sample while making the variance of this same component arbitrarily close to zero, which leads to an infinite density value \citep{li_non-finite_2009}.
\par
To bound the likelihood function from above, \cite{hathaway_constrained_1985} proposes to bound from below the minimum ratio of the variances from any two component densities in the mixture. If the true parameters lie in the constrained parameter space, \cite{redner_note_1981} showed that maximizing the constrained mixture likelihood will yield consistent estimates of all parameters in the mixture. However, constraining the mixture likelihood does not guarantee small biases or estimation variances of the estimated parameters in finite samples. Furthermore, it is not clear how one should determine the exact value of the lower bound for a given sample size \citep{tanaka_strong_2009}. Instead of bounding the minimum ratio of any two variances, \cite{tanaka_strong_2009} and \cite{chen_statistical_2023} suggest using penalized MLE with a sufficiently severe penalty -- with appropriate upper and lower bounds -- to ensure consistency of all parameters in the mixture. Nevertheless, little is known about the finite sample properties of such a strategy given the multimodal nature of the penalized likelihood function.
\par
Assumption \ref{ass1:1}$(ii)$ states that the value taken by every component density is restricted to be strictly positive on the whole support, where this support is assumed to be identical across densities. Note also that this assumption implies that the support of $f_g(\cdot)$ is unbounded, which excludes mixtures of distributions with bounded support, such as the uniform, beta, and Dirichlet distributions. Although this assumption could be relaxed by enforcing bounds on the support of all component densities, these bounds has to be large enough such that there exists outliers $\tilde{y}_i$ that lead to $f_g(\tilde{y}_i|\theta^0_g) > 0$ with $\left|\frac{\tilde{y}_i-\mathbb{E}[y_i]}{\sigma_y}\right| >> 0$ and $\sigma^2_y= Var[y_i]$. Assumption \ref{ass1:1}$(ii)$ also rules out cases where any two component densities have identical parameter values, which prevents the identification issues arising from overspecification of the number of components \citep{budanova_penalized_2025}.
\par
Assumption \ref{ass1:1}$(iii)$ is standard in MLE problems and is satisfied for several families of parametric densities. This assumption guarantees that $\hat{\theta}_{g}:= \arg \max_{\theta \in \Theta} \sum_{i=1}^N \log(f_g(y_{i}|\theta))$ is point identified and always corresponds to a single set of estimated parameter values. Assumption \ref{ass1:1}$(iv)$ is commonly known as generic identifiability of the mixture density. Section 1.3 of \cite{fruhwirth-schnatter_finite_2006} describes the three types of identification issues that can arise when modeling finite mixtures, including generic identifiability. The two other identification issues are characterized as ``weak'' since they can be overcome by using appropriate constraints on the component labels and the number of components. Specifically, \cite{teicher_identifiability_1961} and \cite{yakowitz_identifiability_1968} show that mixtures of Gaussian, Gamma, exponential, Poisson, negative binomial, and binomial distributions (under certain conditions) are generically identifiable provided that the number of components $G$ is correctly specified. Note that the absence of covariates also avoids the \textit{intra-component label switching} issue that arises when label switching in covariates leads to identification issues \citep{grun_finite_2008}.
\par
Assumption \ref{ass1:1}$(v)$ is similar to Assumption \ref{ass1:1}$(iii)$, but applied to the mixture likelihood function. It is different from Assumption \ref{ass1:1}$(iii)$ since it eliminates situations where the mixture likelihood would have two identical global maxima, but at two different locations in $\Xi$ for a given dataset $\mathbf{y}_N$. Finally, Assumption \ref{ass1:1}$(vi)$ is made for convenience and is naturally satisfied for most mixtures of well-defined parametric densities. Although there exists some well-known cases where the Fisher information matrix $\mathcal{I}(\xi)$ is not finite for given values of $\xi \in \Xi$, such pathological cases occur when $\xi$ is located on the boundary of the parameter space and are therefore unlikely to be encountered in most practical settings \citep{li_non-finite_2009}.
\par
If Assumption \ref{ass1:1} is satisfied, it is widely recognized that the standard maximum likelihood estimate $\hat{\xi}^{ML} := \arg \max_{\xi \in \Xi} l^{ML}(\xi)$ is $\sqrt{N}$-consistent, and asymptotically normal and efficient with \citep{chen_statistical_2023}~:
\begin{align*}
	\sqrt{N}(\hat{\xi}^{ML} - {\xi}^0) \xrightarrow{d} \mathcal{N}(0,\mathcal{I}({\xi}^0)).
\end{align*}
However, this distributional behavior is only asymptotic, and it is unclear if the above normal distribution represents a good approximation uniformly across the parameter space $\Xi$ in finite samples. As noted by \cite{redner_mixture_1984}, maximizing the mixture likelihood with respect to $\Xi$ can often be characterized as an \textit{ill-conditioned} problem in the sense that ``its solution is very sensitive to some perturbations in the data while being relatively insensitive to others'' \citep{redner_mixture_1984}, which often makes the asymptotic distribution shown above a very bad approximation of the finite-sample distribution of MLE of finite mixtures.
\par
Such a large discrepancy between the finite-sample and asymptotic behavior of the estimator arises because the global maximum of the mixture likelihood function need not be located near the true parameter values $\xi^0$ for any given sample size $N$ below a certain threshold. This intrinsic yet overlooked feature of the MLE of finite mixtures is exemplified below for a mixture of two normal distributions.
\begin{example}[\textbf{Mixture of normal distributions}]\label{exam:1}
 	Let the mixture density evaluated at the true parameter values be defined as follows
	\begin{align}
		f(y_i|\xi^0) = \pi^0_1 \phi\left(\frac{y_i-\mu^0_1}{\sigma_1^0}\right) + (1-\pi^0_1) \phi\left(\frac{y_i-\mu^0_2}{\sigma_2^0}\right),
	\end{align}
	where $\phi(x)$ is the probability density function (pdf) of the univariate standard normal distribution evaluated at $x$, and where $\xi^0 = (\mu^0_1,\mu^0_2,\sigma^0_1,\sigma^0_2,\pi_1^0)$ correspond to the true parameter values (i.e., means, standard deviations and mixing weight, respectively), which are all treated as constant while $y_i$ is treated as random for any $i \in [N]$. If we assume that $\sigma_1^0 = \sigma_2^0 = 1$ for simplicity, and that the true mean values are separated by a scalar $\eta^0>0$ such that $\mu^0_1 := \mu^0_2 + \eta^0$, then it is possible to rewrite the above density as follows
	\begin{align*}
		f(y_i|\xi^0) &= \phi(y_i-\mu^0_2) \pi_1^0 \left(\exp\left((y_i-\mu^0_2)\eta^0- \frac{(\eta^0)^2}{2}\right) - 1\right)  + \phi(y_i-\mu^0_2),\\
		&\equiv\phi(y_i-\mu^0_2) (1 + \pi_1^0\delta_i(\mu^0_2,\eta^0)),
 	\end{align*}
 	where $\delta_i(\mu_2,\eta) := \exp((y_i-\mu_2)\eta- \eta^2/2) - 1$. If the $j^{th}$ observation $y_j$ is such that $y_j - \mu^0_2 > \eta^0/2$, then $\delta_j(\mu^0_2,\eta^0)>0$ and $f(y_j|\xi^0) > \phi(y_j-\mu^0_2)$. Analogously, if $y_j- \mu^0_2 \le \eta^0/2$, then $\delta_j(\mu^0_2,\eta^0) \le 0$ and $f(y_j|\xi^0) \le \phi(y_j-\mu^0_2)$. Assuming that both $\sigma_1^0$ and $\sigma_2^0$ are known, we can write the corresponding log likelihood function as follows
 	\begin{align*}
 		l^{ML}(\mu_2,\pi_1,\eta) = \sum_{i=1}^N \log(\phi(y_i-\mu_2))+  \sum_{i=1}^N \log(1 + \pi_1\delta_i(\mu_2,\eta)).
 	\end{align*}
 	\par Analytically deriving this expression with respect to $(\mu_2,\pi_1,\eta)$ to obtain the exact MLE is quite difficult due to the second sum in the expression, even for small $N$. However, we know that the first sum is maximized when $\mu_2 = \bar{y} =  N^{-1}\sum_{i=1}^N y_i$. Abstracting away from the improbable case where $\bar{y} = \mu^0_2$, we can analyze separately the two cases where $\bar{y}$ is either below or above $\mu^0_2$ for a given $N$. When $\eta^{0}$ is relatively small and $N$ is fixed, we may observe $\bar{y} < \mu_2^0$ due to the presence of negative outliers in the sample. In this case, we have that
 	\begin{align*}
 		l^{ML}(\mu_2^*(\pi_1,\eta),\pi_1,\eta) > l^{ML}(\bar{y},\pi_1,\eta) > l^{ML}(\mu_2^0,\pi_1,\eta),
 	\end{align*}
 	where $\mu_2^*(\pi_1,\eta) := \arg\max_{\mu_2 \in \mathbb{R}} l^{ML}(\mu_2,\pi_1,\eta)$ for any $\pi_1 \in (0,1)$ and any $\eta > 0$, and where $\mu_2^*(\pi_1,\eta) < \bar{y} < \mu_2^0$. This is verified since $\sum_{i=1}^N \log(\phi(y_i-\bar{y})) >  \sum_{i=1}^N \log(\phi(y_i-\mu_2^0))$ for any $\bar{y} \ne \mu_2^0$, and since $\delta_i(\mu_2,\eta) > \delta_i(\bar{y},\eta) > \delta_i(\mu_2^0,\eta)$ for any $\mu_2 < \bar{y}$ and any $i \in [N]$ when $\bar{y} < \mu_2^0$ and $\eta > 0$. This implies that $\mu_2^*(\pi_1,\eta) \le \bar{y}$ with equality if and only if $\pi_1\eta = 0$, which we are ruling out for the moment. Therefore, we can write that
 	\begin{align}\label{eqn:ineq_bias}
 		\mathbb{E}_0[\mu_2^*(\pi_1,\eta) - \mu_2^0|\bar{y} < \mu_2^0] < \mathbb{E}_0[\bar{y} - \mu_2^0|\bar{y} < \mu_2^0] < 0, 
 	\end{align}
 	where $\mathbb{E}_0[\cdot|\cdot]$ denotes the conditional expected value over the true mixture DGP. Note that $\mathbb{E}_0[\bar{y} - \mu_2^0|\bar{y} < \mu_2^0]$ decreases as $\eta^0$ goes to zero, whereas it increases (up to zero) as $N \to \infty$. By Assumption \ref{ass1:1}$(i)$ and the Central Limit Theorem, we know that the sample mean $\bar{y}$ approximately behaves like a normally distributed variable with $\mathbb{E}_0[\bar{y}] = \pi_1^0\mu_1^0 + \pi_2^0\mu_2^0 = \pi_1^0\eta^0 + \mu_2^0>\mu_2^0$ and $\text{Var}[\bar{y}] = \sigma^2_y/N $ when $N$ is large, with $0 < \text{Var}[y_i] \equiv \sigma^2_y <\infty$. Using standard results on truncated normal distributions, we can thus write that 
 	\begin{align*}
 		\mathbb{E}_0[ \bar{y} - \mu_2^0| \bar{y} < \mu_2^0] \approxeq \pi_1^0\eta^0 - \frac{\sigma_y\phi(\alpha^0)}{\sqrt{N}\Phi(\alpha^0)},
 	\end{align*}
 	where $\alpha^0 := -\sqrt{N}\pi_1^0\eta^0/\sigma_y$, and where $\Phi(\cdot)$ stands as the standard normal cumulative density function (cdf). 
 	\par
 	On the other hand, if $\bar{y} > \mu_2^0$, we still have that $\mu_2^*(\pi_1,\eta) < \bar{y}$, but $\mu_2^*(\pi_1,\eta)$ can either be below or above $\mu^0_2$ for any $\eta > 0$ and any $\pi_1 \in (0,1)$. Using the same logic and the same set of results on truncated distributions as above, we can write that 
 	\begin{align*}
 		\mathbb{E}_0[\mu_2^*(\pi_1,\eta) - \mu_2^0|\bar{y} > \mu_2^0] < \mathbb{E}_0[ \bar{y} - \mu_2^0| \bar{y} > \mu_2^0] \approxeq \pi_1^0\eta^0 + \frac{\sigma_y\phi(\alpha^0)}{\sqrt{N}(1-\Phi(\alpha^0))} > 0.
 	\end{align*}
 	We can now express the overall bias of $\mu_2^*(\pi_1,\eta)$ as follows 
 	\begin{align*}
 		\mathbb{E}_0[\mu_2^*(\pi_1,\eta) - \mu_2^0] &=b_{-}(\mu_2^0,\pi_1,\eta)\times \Phi(\alpha^0) + b_{+}(\mu_2^0,\pi_1,\eta)\times (1-\Phi(\alpha^0)),
 	\end{align*}
 	where $b_{-}(\mu_2^0,\pi_1,\eta) := \mathbb{E}_0[\mu_2^*(\pi_1,\eta) - \mu_2^0|\bar{y} < \mu_2^0]$ refers to the negative part of the bias, $b_{+}(\mu_2^0,\pi_1,\eta) := \mathbb{E}_0[\mu_2^*(\pi_1,\eta) - \mu_2^0|\bar{y} > \mu_2^0]$ refers to the ``positive'' part of the bias (but which may be negative), and where $\Phi(\alpha^0) \equiv \mathbb{P}[\bar{y} < \mu_2^0]$. Due to the symmetry of the univariate normal density, we have that $\lim_{\eta^0 \to 0^+}\mathbb{P}[\bar{y} < \mu_2^0] = \mathbb{P}[\bar{y} > \mu_2^0] =1/2$, which implies that
 	\begin{align*}
 		\lim_{\eta^0 \to 0^+} \mathbb{E}_0[\mu_2^*(\pi_1,\eta) - \mu_2^0] &= \frac{1}{2} (\underbrace{b_{-}(\mu_2^0,\pi_1,\eta)}_{< -c_N}  + \underbrace{b_{+}(\mu_2^0,\pi_1,\eta)}_{< c_N}),
 	\end{align*}
 	where $c_N \approxeq 0.8\sigma_y/\sqrt{N}>0$. Because $\mathbb{E}_0[\mu_2^*(\pi_1,\eta) - \mu_2^0]$ is a smooth function of $\eta^0$ around $\eta^0=0$, the last equation indicates that $\mu_2^*(\pi_1,\eta)$ is negatively biased for small and positive values of $\eta^0$ in finite samples if $\pi_1\eta > 0$, with the bias negatively increasing as $\eta^0$ gets closer to zero. Note that $\eta^*(\pi_1) := \arg\max_{\eta \in \mathbb{R}} \sum_{i=1}^N \log(1 + \pi_1\delta_i(\mu_2^*(\pi_1,\eta),\eta)) > 0$ if $\sum_{i=1}^N \delta_i(\mu_2^*(\pi_1,\eta^*(\pi_1)),\eta^*(\pi_1)) > 0$, which occurs if $\mu_2^*(\pi_1,\eta)$ is negatively biased. Using the same logic, this also leads to $\pi_1^* := \arg\max_{\pi_1 \in (0,1)} \sum_{i=1}^N \log(1 + \pi_1\delta_i(\mu_2^*(\pi_1,\eta^*(\pi_1)),\eta^*(\pi_1)))$ being strictly positive if $\sum_{i=1}^N \delta_i(\mu_2^*(\pi^*_1,\eta^*(\pi^*_1)),\eta^*(\pi^*_1)) > 0$, As a result, the MLE $\mu_2^*(\pi_1^*,\eta^*(\pi_1^*))$ is negatively biased in finite samples (or positively biased if $\mu_2^0 > \mu_1^0$ due to label switching). This finite-sample bias is confirmed by Monte Carlo simulations that are described in Section \ref{sec1:4}. Finally, note hat that $\mathbb{E}_0[\mu_2^*(\pi_1^*,\eta^*(\pi_1^*)) - \mu_2^0]$ converges to $b_{+}(\mu_2^0,\pi_1^*,\eta^*(\pi_1^*)) \le \pi_1^0\eta^0 > 0$ as whether $N$ or $\eta^0$ tends to infinity, which is uninformative about the presence of any bias in either $\mu_2^*(\pi_1^*,\eta^*(\pi_1^*))$, $\eta^*(\pi_1^*)$, or $\pi_1^*$.
\end{example}
\par
Example \ref{exam:1} shows that the MLE of finite mixtures can mistake apparent contamination for $\epsilon$-contamination in finite samples \citep{huber_robust_1964}, at least for mixtures of normal distributions. $\epsilon$-contamination refers to the situation where the observed data originates from the following ``contaminated'' density 
\begin{align*}
	y_i \sim \epsilon f_{\epsilon}(\cdot|\theta_{\epsilon}) + (1-\epsilon)f(\cdot|\theta),
\end{align*}
where $\epsilon > 0$ is close to zero and where $f_{\epsilon}(\cdot|\theta_{\epsilon})$ represents the ``contamination'' distribution, which is governed by the set of parameters $\theta_{\epsilon}$. In practice, mixture models are well-suited to estimate the parameters of an $\epsilon$-contaminated density, but can also lead to higher likelihood values in finite sample by fallaciously treating natural outliers from well-balanced mixtures as a product of $\epsilon$-contamination. Contrary to true cases of $\epsilon$-contamination, natural outliers will asymptotically be ``smoothed out'' by the tails of the mixture density. Therefore, the probability of observing a large difference between the estimates and the true parameter values will depend on the sample size and on the tails of the true density~: mixture densities with ``fatter'' tails will generate more extreme outliers in finite samples, and more extreme outliers will lead to a larger probability of observing a large difference between the estimated parameters and the true values. Bounding the component variances from below only puts a mild restriction on the number of outliers included in the contamination density.
\par
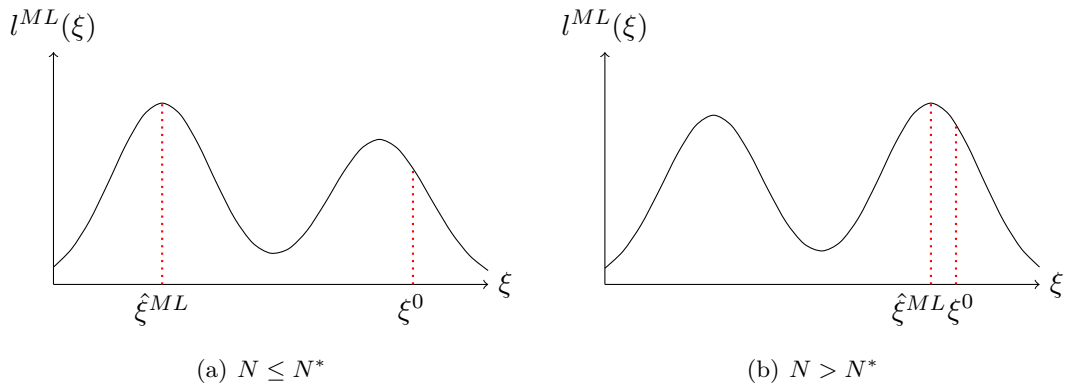
\begin{figure}[t!]
	\setstretch{1.15}
	\centering
	\subfigure[$N \le N^*$]{
		\begin{tikzpicture}[scale=2.2]
			\draw[->] (-1.3, -1.2) -- (1.3, -1.2) node[right] {$\xi$};
			\draw[->] (-1.3, -1.2) -- (-1.3, 0.2) node[above] {$l^{ML}(\xi)$};
			\draw[scale=1, domain=-1.3:1.3, smooth, variable=\x, black] plot ({\x}, 	{(1.5/(2*pi*0.3)^0.5*exp(-0.5*((\x+0.65)^2/0.3^2)) +1.2/(2*pi*0.3)^0.5*exp(-0.5*((\x-0.65)^2/0.3^2)))-1.2});
			\draw[scale=1, thick, dotted, red] (0.85, -1.2) -- (0.85, -0.49);
			\draw[scale=1, thick, dotted, red] (-0.65, -1.2) -- (-0.65, -0.11) ;
			\node at (0.84,-1.35) {$\xi^0$};
			\node at (-0.65,-1.35) {$\hat{\xi}^{ML}$};
		\end{tikzpicture}
	}
	\subfigure [$N > N^*$]{
		\begin{tikzpicture}[scale=2.2]
			\draw[->] (-1.3, -1.2) -- (1.3, -1.2) node[right] {$\xi$};
			\draw[->] (-1.3, -1.2) -- (-1.3, 0.2) node[above] {$l^{ML}(\xi)$};
			\draw[scale=1, domain=-1.3:1.3, smooth, variable=\x, black] plot ({\x}, 	{(1.4/(2*pi*0.3)^0.5*exp(-0.5*((\x+0.65)^2/0.3^2)) +1.5/(2*pi*0.3)^0.5*exp(-0.5*((\x-0.65)^2/0.3^2)))-1.2});
			\draw[scale=1, thick, dotted, red] (0.80, -1.2) -- (0.80, -0.22);
			\draw[scale=1, thick, dotted, red] (0.65, -1.2) -- (0.65, -0.105) ;
			\node at (0.58,-1.35) {$\hat{\xi}^{ML}$};
			\node at (0.83,-1.35) {$\xi^0$};
		\end{tikzpicture}
	}
	\caption{\label{fig1:1} Hypothetical representation of the mixture likelihood $l^{ML}(\xi)$ when $N$ is small (Panel (a)) and $N$ is sufficiently large (Panel (b)).}
\end{figure}
\par
The effect of these ``natural'' outliers on the mixture likelihood is illustrated in Figure \ref{fig1:1}. For values of $N$ that are below a certain threshold $N^*$, the true parameters $\xi^0$ can be arbitrarily far away from the global maximum for any $N \le N^*$, depending on the location and the number of outliers generated by the true mixture density, as shown in Panel (a) of Figure \ref{fig1:1}. As $N$ remains below $N^*$, the $\epsilon$-contamination model may better represent the observed data than a mixture model evaluated near the true parameters. As $N$ increases and surpasses $N^*$, natural outliers are ``smoothed out'' and the consistency of $\hat{\xi}^{ML}$ lead to a displacement in the location of the global maximum of $l^{ML}(\xi)$, which will then become much closer to the true parameter values, as shown in Panel (b) of Figure \ref{fig1:1}. Therefore, the consistency property of $\hat{\xi}^{ML}$, combined with Assumption \ref{ass1:1}$(v)$, leads to an asymptotic global maximum of $l^{ML}(\xi)$ that is located in a shrinking neighborhood of the true parameter values. However, this questions the uniform consistency of $\hat{\xi}^{ML}$ over $\Xi^0$ since $\hat{\xi}^{ML}$ may converge at an arbitrarily slow rate as $\xi^0$ approaches the boundary of the parameter space $\Xi^0$. Validating this last intuition goes beyond the scope of this paper.
\par
In summary, the previous example and the above discussion demonstrated that it is impossible to guarantee that the bias of $\hat{\xi}^{ML}$ will shrink toward zero as $N$ increases but remains below $N^*$ for a given $\xi^0 \in \Xi^0$. The bias could even increase with the sample size, as shown in Figure \ref{fig1:2} of Section \ref{subsec1:42}. Note that the value of the threshold $N^*$ is data-dependent and typically increases with the total degree of overlap between the true component densities \citep{redner_mixture_1984}. Given the analytical complexities related to the finite-sample distribution of MLE of finite mixtures, formally deriving the analytical expression of such a bias is left to future research.
\subsection{Bias-Reduced Estimation of Finite Mixtures}\label{subsec1:33}
Although it is asymptotically unbiased under standard regularity conditions, it is well known that the maximum likelihood estimator exhibits finite-sample bias even for well-defined, single-component densities. Formally, the bias of the MLE can be generally represented as follows \citep{firth_bias_1993}
\begin{align}\label{eq1:bias}
	b(\beta) = \mathbb{E}_0[\hat{\beta}- \beta] = \frac{b_1(\beta)}{N}+\frac{b_2(\beta)}{N^2}+ \frac{b_3(\beta)}{N^3}+ O(N^{-4}),
\end{align}
for a general set of true parameters denoted by $\beta$, and where $\frac{b_d(\beta)}{N^d}$ with $d \in \mathbb{N}_{>0}$ represents the $d^{th}$-order term of the bias. Several bias-reduction methods exist in order to get rid of the first-order term $\frac{b_1(\beta)}{N}$~: direct bias-correction, jackknifing, bootstrap, etc.; see \cite{kosmidis_bias_2014} for an introductory review.
\par
The bias of the MLE of finite mixtures is different. As argued in the previous subsection, the finite-sample bias of the MLE of finite mixtures might be represented by Eq. (\ref{eq1:bias}) only once $N$ is above a certain threshold $N^*$ due to the multimodal nature of the mixture likelihood. Given that $N^*$ is data-dependent, it is not clear if this finite-sample bias necessarily reduces when $N$ increases but remains below $N^*$. It is also not clear how traditional bias-reduction techniques can be directly applied to finite mixtures estimated by traditional MLE techniques, such as the EM algorithm.
\par
The strategy that I am proposing to reduce the bias in the MLE of finite mixtures relies on the correct classification of almost every observation in the sample. By correct classification, I mean the estimation of the true group memberships $\mathbf{z}^0$ such that the estimated memberships $\hat{\mathbf{z}} = \mathbf{z}^0$ up to an appropriate permutation in the group labels for any given sample size $N$. Once this is achieved, the finite-sample bias can be generally represented by Eq. (\ref{eq1:bias}) for each group separately. A second bias-reduction step may then be applied to the within-group estimates using traditional bias-reduction methods. The rest of this article focuses exclusively on the first step of this two-step procedure.
\par
The proposed bias-reduction strategy relies on the availability of a consistent unsupervised classification procedure. More precisely, I show that the use of the Mahalanobis distance leads to a consistent classifier such that, under the assumptions stated below, its misclassification rate goes down to zero as $p$ increases, where $p$ can either represent the size of the set of available (exogenous) covariates $x_{i}$ if the density of interest is the conditional mixture density $f(y_i|x_i;\xi)$, or the size of the outcome vector when $\dim(y_i) > 1$. In this section, I consider the second case where $y_i$ is a $p$-sized vector with no covariate for notational simplicity. All results can however be easily extended to the first setup with $f(y_i|x_i;\xi)$, where $p = \dim(x_i)$ and $x_i$ is used instead of $y_i$ for classification (see Definition \ref{def1:5}). In either case, additional information is required to reduce the misclassification rate and the estimation bias in finite samples. If no additional information is available, then traditional MLE can still be applied -- namely via the EM algorithm -- but this can lead to severe biases if the sample size is not sufficiently large or if the true component densities feature a strong degree of overlap. Another option would be to restrict the group memberships across observations, but it is unclear which restrictions should be applied to guarantee consistency of the estimates in cross-sectional data.
\begin{assumption}\label{ass1:2}
	\setstretch{1.15}
	$ \ $
	\begin{enumerate}[label=(\roman*)]
		\item $\mathbb{E}_{0,g}[\log(f_g(y_{i}|\theta_g))] < \infty$ for any $y_i \in \mathcal{Y}$ and any $\theta_g \in \Theta_g$, where $\mathbb{E}_{0,g}[\cdot]$ stands as the expected value with respect to the $g^{th}$ true mixture density $f_g(\cdot|\theta^0_g)$, and where all $y_{i}$ are independent and identically distributed according to the true mixture density $f(\cdot|\xi^0)$.		
		\item $l_g(\theta_g) = l_j(\theta_j) \Leftrightarrow \theta_g = \theta_j$ for any pair $(g,j) \in \mathbb{G} \times \mathbb{G}$ and any dataset $\mathbf{y}_N \in \mathcal{Y}^{N}$, where $l_g(\theta_g)$ is defined as in Assumption \ref{ass1:1}$(iii)$.
		\item $f_g(y_i|\theta_g)$ is twice continuously differentiable with respect to $\theta_g$, with information matrix $\mathcal{I}_g(\theta_g) := \mathbb{E}_0[s_i(\theta_g) s_i(\theta_g)^{\top}] < \infty$ where $s_i(\theta_g) = \frac{\partial \log f_g(y_i|\theta_g) }{\partial \theta_g}$ for any $\theta_g \in \Theta_g$ and any $y_{i} \in \mathcal{Y}$.
		\item $\mathbb{E}[y_i] = \boldsymbol{\mu}^0_g = (\mu^0_{g1},...,\mu^0_{gp})^{\top}$ and $Var[y_i] = \Sigma^0_g$ if and only if $z^0_{ig} = 1$, where $||\boldsymbol{\mu}^0_g||^2 < \infty$ for any $g \in \mathbb{G}$, where $\Sigma^0_g$ is a $p \times p$ positive-definite matrix with diagonal elements $0 < \sigma^2_{g,ll} < \infty$, where $Cov[y_{i},y_{j}] = \mathbf{0}$ for any $j \ne i$ and $\mathbb{E}[(y_{il})^4] < \infty$ for any $l \in \{1,2...,p\}$.
		\item $W_g \ne W_j$ for any pair $(g,j) \in \mathbb{G}\times \mathbb{G}\backslash g$, and for a nonvanishing proportion of the elements in each covariance matrix as $p \to \infty$, where $W_g$ is a lower triangular matrix such that $\{\Sigma^0_g\}^{-1} = W_gW_g^\top$ for any $g\in \mathbb{G}$.
	\end{enumerate}
\end{assumption}
\setstretch{1.3}
Assumption \ref{ass1:2}$(i)$ is the analog of Assumption \ref{ass1:1}$(i)$ but adapted to the C-ML objective, where the expected log likelihood of the $g^{th}$ component density is bounded from above. It is worth noting that this assumption is less restrictive than Assumption \ref{ass1:1}$(i)$ since there is no value of $\theta_g \in \Theta_g$ and $y_i \in \mathcal{Y}$ such that $f_g(y_{i}|\theta_g) = \infty$ if $f_g(\cdot|\theta_g)$ is a well-defined, single-component density.
\par
Assumption \ref{ass1:2}$(ii)$ is analogous to Assumption \ref{ass1:1}$(iv)$ but expressed at the level of the components. This assumption implies that the C-ML objective will be generically identifiable if each $\theta_g$ is point identified. Contrary to the mixture likelihood, overspecifying the number of components in the C-ML objective does not lead to any identification issue since there is no mixing weight involved in the function. In practice, it is possible to test for the presence of a ``useless'' component by testing if $\theta_g = \theta_j$ for any pair $(g,j) \in \mathbb{G}\times \mathbb{G}\backslash g$. Contrary to standard MLE of finite mixtures, it is possible to do so with a likelihood ratio test without facing any non-standard asymptotics provided that $\theta_g$ lies in the interior of $\Theta_g$ \citep{chen_likelihood_2014}. Testing the presence of a ``useless'' component in this context is also facilitated by the notion that estimators based on classification objectives tend not to mix observations from different groups when $G$ is overspecified \citep{liu_identification_2020}. Assumption \ref{ass1:2}$(iii)$ is the component-level analog of Assumption \ref{ass1:1}$(vi)$, and is slightly weaker than Assumption \ref{ass1:1}$(vi)$ given the absence of any mixing weight in the density $f_g(y_i|\theta_g)$ and the score function $s_i(\theta_g)$.
\par
Assumption \ref{ass1:2}$(iv)$ states that the true mean values and variance-covariance matrix of $y_{i}$ are identical within each group, and that only the variance-covariance matrix has to differ across groups. This is not stronger than assuming that $y_i$ is distributed according to $f_g(\cdot|\theta^0)$ if and only if $z^0_{ig} = 1$, which has already been assumed in Section \ref{subsec1:31}. This assumption however states that each  $\Sigma^0_g$ is positive-definite, which implies that no element of $y_i$ is perfectly collinear with any other set of outcomes within each group. This is crucial to guarantee that $\Sigma^0_g$ is invertible for any $g \in \mathbb{G}$. The same assumption also implies the absence of cross-sectional dependence between vectors of outcome values, which could be relaxed at the expense of tedious theoretical complications (see the proof of Corollary \ref{th1:2}). It also assumes that the fourth moment of each element in $y_{i}$ is bounded from above, which is naturally satisfied for most parametric densities.
\par
Finally, Assumption \ref{ass1:2}$(v)$ implies that the proportion of elements that differ between each lower triangular matrix $W_g$, where $W_g$ is such that $\{\Sigma^0_g\}^{-1} = W_gW_g^\top$ for any $g\in \mathbb{G}$, does not go to zero as $p \to \infty$. This assumption puts a relatively weak restriction on each true density $f_g(\cdot|\theta^0_g)$ since it can be satisfied even when a relatively small and fixed number of elements in the covariance matrix $\Sigma^0_g$ are different across groups. However, the magnitude of the differences in $\Sigma^0_g$ across groups will influence classification error in finite samples depending on the chosen classifier (or group membership indicator), which is generally defined below.
\par
\begin{definition}\label{def1:1}
	Let Assumptions \ref{ass1:1}(ii)-(iii) and \ref{ass1:2} hold. Then the classifier $z_{ig}(\theta)$ is generally defined as follows
	\begin{align*}
		z_{ig}(\theta) := 
		\begin{cases}
			1 \ \ \text{if} \ g = \arg\underset{j\in \mathbb{G}}\max \ h_j(y_i|\theta_j),\\
			0 \ \ \text{otherwise,}
		\end{cases}
	\end{align*}
	where $h_j: \mathbb{R}^{p} \to \mathbb{R}$ is a discriminant function and which is continuous with respect to $\theta_j$ for any $j \in \mathbb{G}$.
\end{definition}
Although they are ruled out by assumption \ref{ass1:2} $(ii)$, eventual ties between values of $h_j(y_i|\theta_j)$ require a tie-breaking rule in order to complete the above definition, but it is left unspecified for simplicity. Using Definition \ref{def1:1}, the bias of any classifier $z_{ig}(\theta)$ is defined as follows.
\begin{definition}\label{def1:2}
	Let Assumptions \ref{ass1:1}(ii)-(iii) and \ref{ass1:2} hold. A classifier $z_{ig}(\theta)$ is said to be \textit{unbiased} if and only if
	\begin{align*}
		\arg\underset{j\in \mathbb{G}}\max \ \mathbb{E}_0[h_j(y_i|\theta_j^0)] = z^0_{i},
	\end{align*}
	for any $i \in [N]$ and any value of $g \in \mathbb{G}$, where $z^0_{i} = \{1,2...,G\}$ is such that $z^0_{i} = g$ if and only if $z^0_{ig} = 1$ after a suitable permutation in the group labels.
\end{definition}
\noindent
\textbf{Remark 1.} An unbiased classifier \textit{does not} correspond to a classifier such that $\mathbb{E}_0[z_{ig}(\theta^0)] = \mathbb{E}_0[\arg\underset{j\in \mathbb{G}}\max \ h_j(y_{i}|\theta^0_j)] = z^0_{ig}$, which actually corresponds to a uniformly consistent classifier (see Definition \ref{def1:3}). Instead, the bias of any classifier refers to the capacity of its discriminant function to correctly classify any observation after averaging $h_j(\cdot|\theta^0_j)$ over an infinite number of random draws from $f(y_i|\theta^0_j)$.
\vspace{2mm}
\par
A \textit{consistent} classifier is then defined as follows.
\begin{definition}\label{def1:4}
	A consistent classifier is a classifier $z_{ig}(\theta)$ such that
	\begin{align}\label{eqn1:10}
		\hat{E}_{Np}(\theta^0) := &\sum_{i=1}^N \sum_{g=1}^G \frac{\mathbbm{1}[ \hat{z}_{Np,i(g)}(\theta^0) \ne z^0_{ig}]}{2N} \xrightarrow{p} 0,
	\end{align}
	as $N \to \infty$, where $\mathbbm{1}[\cdot]$ denotes the indicator function, $\hat{z}_{Np,i(g)}(\theta)$ is the estimated value of $z_{ig}(\theta)$ based on a sample of $N$ $p$-sized vector $y_i$ and a given value of $\theta$, $\hat{E}_{Np}(\theta)$ is the misclassification rate of $\hat{z}_{Np,i(g)}(\theta)$, and where $(g)$ is a permutation from $g \in \mathbb{G} \to j \in \mathbb{G}$ that minimizes $\hat{E}_{Np}(\theta)$ for a given value of $\theta$.
\end{definition}
\par
Akin to \cite{su_identifying_2016}, I now define a classifier that is \textit{uniformly consistent}.
\begin{definition}\label{def1:3}
	A uniformly consistent classifier is a classifier $z_{ig}(\theta)$ such that
	\begin{align*}
		\hat{E}^u_{Np}(\theta^0):= \mathbb{P}[\cup_{i=1}^N \cup_{g=1}^G ( \hat{z}_{Np,i(g)}(\theta^0) \ne z^0_{ig})] \xrightarrow{a.s.} 0,
	\end{align*}
	as $N \to \infty$, where $\hat{z}_{Np,i(g)}(\theta)$ is defined as in Definition \ref{def1:4}, $\hat{E}^u_{Np}(\theta)$ is the probability of misclassifying at least one observation in the sample, and where $(g)$ is a permutation from $g \in \mathbb{G} \to j \in \mathbb{G}$ that minimizes $\hat{E}^u_{Np}(\theta)$ for a given value of $\theta$.
\end{definition}	
\noindent
\textbf{Remark 2.} A uniformly consistent classifier is a classifier that groups every observation into its true group with probability approaching 1 as $N \to \infty$~: it is consistent uniformly over all possible realizations of $y_{i}$. Note that a uniformly consistent classifier will not necessarily yield a misclassification rate that is equal to zero in finite samples. Analogously, a misclassification rate equal to zero in finite samples does not necessarily imply that the employed classifier is uniformly consistent. If the rate of convergence of a uniformly consistent classifier is (very) fast, then it should lead to an estimated misclassification rate $\hat{E}_{Np}(\theta)$ that is (very) close to zero when $\theta$ is close to or equal to its true value $\theta^0$.
\vspace{2mm}
\par
Unlike \cite{su_identifying_2016}, the latter definition does not distinguish between Type I and Type II errors since they can be controlled simultaneously in finite samples if no observation is left unclassified, which is implicitly assumed throughout the paper due to Definition \ref{def1:1}. I now define three different classifiers~: the joint density classifier, the Euclidean distance classifier, and the Mahalanobis distance classifier.
\begin{definition}\label{def1:5}
	Let Assumptions \ref{ass1:1}(ii)-(iii) and \ref{ass1:2} hold, and let the classifier $z_{ig}(\theta)$ be defined as in Definition \ref{def1:1}.\\
	\begin{enumerate}[label=(\roman*)]
	\item The joint density classifier $z^D_{ig}(\theta)$ is defined such that
	\begin{align*}
		z^D_{ig}(\theta): h_{j}(y_{i}|\theta_j) = f_{j}(y_{i}|\theta_j),
	\end{align*}
	where $f_{j}(y_{i}|\theta_j)$ represents the $p$-variate density of the $i^{th}$ observation for the $j^{th}$ component density.\\ 
	\item The Euclidean distance classifier $z^E_{ig}(\boldsymbol{\mu})$ is defined such that
	\begin{align*}
		z^{E}_{ig}(\boldsymbol{\mu}): h_{j}(y_{i}|\theta_j) = - ||y_{i} -\boldsymbol{\mu}_j||^2.
	\end{align*}
	\item The Mahalanobis distance classifier $z^M_{ig}(\boldsymbol{\mu},\Sigma)$ is defined such that
	\begin{align*}
		z^{M}_{ig}(\boldsymbol{\mu},\Sigma): h_{j}(y_{i}|\theta_j) = -d^2(y_{i}, \boldsymbol{\mu}_j, \Sigma_j),
	\end{align*}
	where $d^2(y_{i}, \boldsymbol{\mu}_j, \Sigma_j) := (y_{i} - \boldsymbol{\mu}_j)^{\top}\Sigma_j^{-1}(y_{i} - \boldsymbol{\mu}_j)$ denotes the squared Mahalanobis distance.
	\end{enumerate}
\end{definition}
Using the joint density classifier is a natural choice when an appropriate distributional assumption is made on the outcome vector. This classifier is known as the Bayes' classification rule and reduces to a naive Bayes' classifier when all covariance elements in $\Sigma = (\Sigma_1,...,\Sigma_G)$ are assumed to be null \citep{hastie_elements_2009}. If the density of interest is the conditional mixture density $f(y_i|x_i;\xi)$ where $x_i$ is multivariate (and may be composed of continuous and/or discrete elements), then it is necessary to assume that Assumption \ref{ass1:2}$(iv)$-$(v)$ also applies to the vector of covariates $x_i$. In this case, the discriminant function of the joint density classifier is written as $h_{j}(y_{i},x_i|\theta_j,\psi_j) = f_j(y_i|x_i;\theta_j)p_j(x_i|\psi_j)$, where $p_j(\cdot|\psi_j)$ refers to the covariates' density for the $j^{th}$ component, and where $\psi = (\psi_1,...,\psi_G) \in \Psi$ represents the set of parameters that governs each $p_j(\cdot)$ density.
\par
The Euclidean distance classifier is a nonparametric classifier that relies on the first moment of $y_i$ and is the one that is normally used with the K-means algorithm. The Mahalanobis distance classifier is also nonparametric but relies on the first and second moments of $y_i$ to classify each observation. Several other classifiers exist and are used in practice, but those three classifiers are among the most prevalent in the literature on unsupervised classification and clustering techniques.
\par
The next two lemmas show that all three classifiers are unbiased under specific conditions.
\begin{lemma}\label{lem1:34}
	Let Assumptions \ref{ass1:1}(ii)-(iii) and \ref{ass1:2} hold. Then all three classifiers defined in Definition \ref{def1:5} are unbiased if $\boldsymbol{\mu}^0_j = \boldsymbol{\mu}^0_g \Leftrightarrow j = g$.
\end{lemma}
\begin{lemma}\label{lem1:35}
	Let Assumptions \ref{ass1:1}(ii)-(iii) and \ref{ass1:2} hold. Then the Mahalanobis distance classifier is unbiased if $p = \dim(y_i)$ is sufficiently large.
\end{lemma}
Lemma \ref{lem1:34} shows that all three classifiers are unbiased if the vector of true mean values $\boldsymbol{\mu}^0_g$ is different across all groups, whereas Lemma \ref{lem1:35} shows that the Mahalanobis distance classifier can be unbiased even if the vectors of true mean values are identical across components. This makes this classifier more robust to cases where the component densities feature strong overlap between each other, compared to the Euclidean distance classifier. Even though the joint density classifier is unbiased if $\boldsymbol{\mu}^0_j = \boldsymbol{\mu}^0_g \Leftrightarrow j = g$, it is generally not possible to show that it is (uniformly) consistent without specifying the distribution of the joint density. Nonetheless, Theorem \ref{th1:1} and Theorem \ref{th1:2} show that the Mahalanobis distance classifier is uniformly consistent if the size of $y_i$ (or $x_i$) grows at a strictly faster rate than the number of observations for a fixed value of $G$. 
\begin{theorem}\label{th1:1}
	Let Assumptions \ref{ass1:1}(ii)-(iii) and \ref{ass1:2} hold, and define the Mahalanobis distance classifier as in Definition \ref{def1:5}$(iii)$ with a fixed number of groups $G$. Then
	\begin{align*}
		\mathbb{P}[\cup_{g=1}^G (z^{M}_{i(g)}(\boldsymbol{\mu}^0,\Sigma^0) \ne z^0_{ig})] = O(p^{-1}),
	\end{align*}
	for any pair $i \in [N]$, where $(g)$ corresponds to a suitable permutation in the labels of the groups.
\end{theorem}
\begin{theorem}\label{th1:2}
	Let Assumptions \ref{ass1:1}(ii)-(iii) and \ref{ass1:2} hold, and define the Mahalanobis distance classifier as in Definition \ref{def1:5}(iii) with a fixed number of groups $G$. Then $z^{M}_{ig}(\boldsymbol{\mu}^0,\Sigma^0)$ is a uniformly consistent classifier if $p/N \to \infty$ as $N,p \to \infty$.
\end{theorem}
The conclusion of Theorem \ref{th1:2} does not necessarily imply that $p > N$ is necessary to correctly classify every observation in the sample; this will however be required as $N$ tends to infinity. In this case, each estimated matrix $\hat{\Sigma}_g$ will not be invertible, and it will not be possible to compute $z^{M}_{ig}(\hat{\boldsymbol{\mu}},\hat{\Sigma})$ for any $g \in \mathbb{G}$. However, Corollary \ref{cor1:34} shows that the misclassification rate given by the Mahalanobis distance classifier will go to zero as $p \to \infty$ if the number of covariates grows at a strictly faster rate than the number of groups, regardless of the rate at which the sample size increases. This means that $p$ could be much smaller than $N$ in finite samples, and the Mahalanobis distance classifier would still be able to correctly classify every observation in the sample under Assumption \ref{ass1:1}$(ii)$-$(iii)$ and Assumption \ref{ass1:2}. When $N < \infty$ is fixed, this corollary also indicates that $p$ has to be larger than $G$ to ensure that the misclassification rate $\hat{E}_{Np}(\theta^0)$ remains close to or equal to zero.
\begin{corollary}\label{cor1:34}
	Let Assumptions \ref{ass1:1}(ii)-(iii) and \ref{ass1:2} hold, and define the Mahalanobis distance classifier as in Definition \ref{def1:5}(iii). Then $z^{M}_{ig}(\boldsymbol{\mu}^0,\Sigma^0)$ is a consistent classifier if $p/G \to \infty$ as $N,p,G \to \infty$, regardless of the rate at which $N$ increases.
\end{corollary}
\textbf{Remark 3.} As emphasized by \cite{dzemski_convergence_2021}, uniform consistency of a classifier is not necessary to obtain consistent estimates of $\theta^0$ when estimating latent group panel structures or finite mixtures. Such a statement is confirmed below in Theorem \ref{the1:35} where consistency of the classifier is sufficient to yield consistent estimates of $\theta^0$.
\vspace{2mm}
\par
Corollary \ref{cor1:35} shows that the estimates obtained from the maximization of the C-ML objective, when equipped with the Mahalanobis distance classifier, will be identical to those obtained from the maximization of the complete-data log likelihood $l^{C}(\theta,\pi,\mathbf{z}^0)$ when the size of $y_i$ (or $x_i$) tends to infinity. This corollary is a direct consequence of Theorem \ref{th1:2} when $p \to \infty$ and $N$ is fixed, which makes the two objective functions $l^{CM}(\theta,\mathbf{z}^M(\boldsymbol{\mu}^0,\Sigma^0))$ and $l^{C}(\xi,\mathbf{z}^0)$ identical up to a constant factor, where $\mathbf{z}^M(\boldsymbol{\mu}^0,\Sigma^0) = (\mathbf{z}^M_1(\boldsymbol{\mu}^0,\Sigma^0),...,\mathbf{z}^M_G(\boldsymbol{\mu}^0,\Sigma^0))$, with $\mathbf{z}^M_g(\boldsymbol{\mu}^0,\Sigma^0) = (z^M_{1g}(\boldsymbol{\mu}^0,\Sigma^0),...,z^M_{Ng}(\boldsymbol{\mu}^0,\Sigma^0))^{\top}$. 
As a result, the finite-sample bias of the estimated parameters will decrease linearly with $N$, as long as $p/N \to \infty$ such that no observation is misclassified.
\begin{corollary}\label{cor1:35}
	Let Assumptions \ref{ass1:1}(ii)-(iii) and \ref{ass1:2} hold, and define the estimated parameters
	\begin{align*}
		\hat{\theta}^{CM}(\mathbf{z}^M(\boldsymbol{\mu}^0,\Sigma^0)) &:= \arg\max_{\theta \in \Theta} l^{CM}(\theta,\mathbf{z}^M(\boldsymbol{\mu}^0,\Sigma^0)),\\
		\hat{\pi}_g^{CM}(\mathbf{z}^M(\boldsymbol{\mu}^0,\Sigma^0)) &:= N^{-1}\sum_{i=1}^N z_{ig}^M(\boldsymbol{\mu}^0,\Sigma^0),
	\end{align*}
	for any $g \in \mathbb{G}$. Then 
	\begin{align*}
		&\hat{\theta}^{CM}(\mathbf{z}^M(\boldsymbol{\mu}^0,\Sigma^0)) \xrightarrow{a.s.} \hat{\theta}^{C}, &\hat{\pi}^{CM}(\mathbf{z}^M(\boldsymbol{\mu}^0,\Sigma^0)) \xrightarrow{a.s.} \hat{\pi}^{C},
	\end{align*}
	as $p \to \infty$ with $N$ and $G$ fixed, where $(\hat{\theta}^{C},\hat{\pi}^{C}) = \arg\max_{\theta \in \Theta,\pi \in \Pi} l^{C}(\theta,\pi,\mathbf{z}^0)$, and where $\hat{\pi}^{CM}(\mathbf{z}^M(\boldsymbol{\mu}^0,\Sigma^0)) = (\hat{\pi}_1^{CM}(\mathbf{z}^M(\boldsymbol{\mu}^0,\Sigma^0)),...,\hat{\pi}_G^{CM}(\mathbf{z}^M(\boldsymbol{\mu}^0,\Sigma^0)))$. In this case, we have that
	\begin{align*}
		&b(\hat{\theta}^{CM}(\mathbf{z}^M(\boldsymbol{\mu}^0,\Sigma^0))) = O(N^{-1}), &b(\hat{\pi}^{CM}(\mathbf{z}^M(\boldsymbol{\mu}^0,\Sigma^0))) = O(N^{-1}),
	\end{align*}
	as $p/N \to \infty$ with $G$ fixed, where $b(\hat{\beta}) := \mathbb{E}[\hat{\beta} - \beta^0]$ for any estimator $\hat{\beta}$ of the true parameter values $\beta^0$.
\end{corollary}
In practice, small values of $p$ will lead to larger biases in both $\hat{\theta}^{CM}(\mathbf{z}^M(\cdot,\cdot))$ and $\hat{\pi}_g^{CM}$ compared to their infeasible counterparts $(\hat{\theta}^{C},\hat{\pi}^{C})$. However, results from the simulations presented in Section \ref{subsec1:421} tend to show that the size of $p$ does not have to be very large to yield relatively small finite-sample biases and low variances in the estimated parameters when Assumption \ref{ass1:2} is satisfied.
\par
The next theorem and corollary show that the Euclidean distance classifier does not offer the same guarantee as the Mahalanobis distance classifier under Assumption \ref{ass1:2}. In fact, the Euclidean distance classifier may misclassify a nonvanishing proportion of the observations in the sample if the distance between each vector $\boldsymbol{\mu}^0_j$ does not become sufficiently large as $p \to \infty$. This is because the Euclidean distance is based only on the mean values, which is not sufficient to guarantee that any single observation in the sample is correctly classified as $p \to \infty$ under Assumption \ref{ass1:2}. For instance, if a fixed number of true mean values differ across groups, Theorem \ref{the1:33} and Corollary \ref{cor1:36} show that the ratio $\frac{||\boldsymbol{\mu}^0_g - \boldsymbol{\mu}^0_{j}||^2}{\text{tr}(\Sigma^0_{g})}$ will converge to 0 as $p \to \infty$, which will make the upper bound on $\mathbb{P}[\cup_{g=1}^G (z^{E}_{i(g)}(\boldsymbol{\mu}^0) \ne z^0_{ig})]$ converge to one as $p \to \infty$. This upper bound may still be sharpened under the assumption that each observation is randomly assigned to any given group when all true mean values are identical across groups.\footnote{For simplicity, I assumed that identical distances across groups necessarily imply misclassification, although this depends on the chosen tie-breaking rule. Note also that identical true mean values do not violate Assumption \ref{ass1:2}$(ii)$ as long as Assumption \ref{ass1:2}$(v)$ is satisfied.} On the contrary, the Mahalanobis distance classifier is robust to identical true mean values across groups, and is a consistent classifier as long as a sufficiently large number of elements in the covariance matrix $\Sigma^0_g$ differ across groups.
\begin{theorem}\label{the1:33}
	Let Assumptions \ref{ass1:1}(ii)-(iii) and \ref{ass1:2} hold, and define the Euclidean distance classifier as in Definition \ref{def1:5}$(ii)$ with a fixed number of groups $G$. Then
	\begin{align*}
		\mathbb{P}[\cup_{g=1}^G (z^{E}_{i(g)}(\boldsymbol{\mu}^0) \ne z^0_{ig})] \le 1 - \prod_{j \ne z^0_{i}} (1- {c}_{jz^0_{i},p}),
	\end{align*}
	for any $i \in [N]$, where ${c}_{jg,p} = \left(1+\frac{p\sqrt{M}} {2\text{tr}(\Sigma^0_{g})}\right) \bigg/ \left(1 + \frac{||\boldsymbol{\mu}^0_{g} - \boldsymbol{\mu}^0_{j}||^2}{\text{tr}(\Sigma^0_{g})} \right)$ with $0 < M < \infty$ and $tr(\cdot)$ denoting the trace operator, $z^0_{i}$ is defined as in Definition \ref{def1:2}, and where ${c}_{jg,p} \to 1$ as $\boldsymbol{\mu}^0_g \to \boldsymbol{\mu}_j^0$ for any pair $(g,j) \in \mathbb{G}^2$ and any $p>0$.
\end{theorem}
\begin{corollary}\label{cor1:36}
	Let Assumptions \ref{ass1:1}(ii)-(iii) and \ref{ass1:2} hold, and define the Euclidean distance classifier as in Definition \ref{def1:5}$(ii)$ with a fixed number of groups $G$. Then $z^{E}_{ig}(\boldsymbol{\mu}^0)$ is a consistent classifier if the ratio $\frac{||\boldsymbol{\mu}^0_g - \boldsymbol{\mu}^0_{j}||^2}{\text{tr}(\Sigma^0_{g})} \to \infty$ as $p\to \infty$ for any pair $(g,j) \in \mathbb{G}\times \mathbb{G}\backslash g$.
\end{corollary} 
\noindent
\textbf{Remark 4.} When all elements in $y_{i}$ are normally distributed, the joint density classifier can be expressed as a function of the Mahalanobis distance given that the squared Mahalanobis distance is embedded into the exponential part of the multivariate normal distribution. In this case, the joint density classifier is both unbiased and consistent under the same conditions as the ones stated in Lemma \ref{lem1:35} and Corollary \ref{cor1:34}. Simulations confirm that the joint density classifier leads to a smaller misclassification rate than the Mahalanobis distance classifier in finite samples when the normality assumption is verified (results not shown).
\vspace{2mm}
\begin{algorithm}[t]
	\caption{The C-EM algorithm}\label{alg:CEM}
	\setstretch{1.3}
	Let $z_{ig}(\theta)$ be any \textit{consistent} classifier as defined in Definition \ref{def1:4}. Given initial parameters $\theta^{(0)}$ and initial assignment values $\mathbf{z}(\theta^{(0)}) = (\mathbf{z}_{1}(\theta^{(0)}),...,\mathbf{z}_{g}(\theta^{(0)}),...,\mathbf{z}_{G}(\theta^{(0)}))$ with $\mathbf{z}_{g}(\theta^{(0)}) = (z_{1g}(\theta^{(0)}),...,z_{ig}(\theta^{(0)}),...,z_{Ng}(\theta^{(0)}))^{\top}$, the C-EM algorithm consists of the consecutive repetition of the two following steps until $\mathbf{z}(\hat{\theta}^{(k)}) = \mathbf{z}(\hat{\theta}^{(k+1)})$~:
	\begin{enumerate}
		\item The Maximization Step (the M-step)~:\\
		Compute $\hat{\theta}^{(k+1)} = \arg \max_{\theta \in \Theta} \sum_{i=1}^N \sum_{g=1}^G z_{ig}(\hat{\theta}^{(k)}) \log( f_g(y_{i}|\theta))$ conditionally on $z_{ig}(\hat{\theta}^{(k)})$.
		\item The Classification Step (the C-step)~:\\
		 Compute $z_{ig}(\hat{\theta}^{(k+1)})$ conditionally on $\hat{\theta}^{(k+1)}$ for all $i \in [N]$ and all $g \in \mathbb{G}$.
	\end{enumerate}
\end{algorithm}
\par
If the normality assumption is not verified, the Mahalanobis distance classifier offers a very good nonparametric alternative to the joint density classifier. Extending the use of the Mahalanobis distance classifier to a fully nonparametric estimation strategy is straightforward via the use of a kernel density estimator for estimating each $f_g(y_{i}|\theta_g)$ at each maximization step of the C-EM algorithm (see Algorithm \ref{alg:CEM}). Note that the proposed C-EM algorithm is more general than the CEM algorithm of \cite{celeux_classification_1992} since it leaves the classifier $z_{ig}(\theta)$ unspecified, as long as it is consistent. Note also that the C-EM algorithm can be adapted to any conditional component density of the form $f_g(y_i|x_i;\theta_g)$, where the chosen classifier may be based on $y_i$ and/or $x_i$. For instance, the Mahalanobis distance classifier could be applied to the vector of covariates $x_i$, where the maximization step would have to also include the estimation of $\hat{\psi}^{(k+1)} = \arg \max_{\psi \in \Psi} \sum_{i=1}^N \sum_{g=1}^G z^M_{ig}(\hat{\psi}^{(k)}) \log(p_g(x_{i}|\psi))$.
\par
I now derive the asymptotic distribution of the estimates given by Algorithm \ref{alg:CEM} under the following assumptions.
\begin{assumption}\label{ass1:3}
	\setstretch{1.15}
	$ \ $
	\begin{enumerate}[label=(\roman*)]
		\item $z_{ig}(\cdot)$ is a classifier that is unbiased and consistent as $N,p \to \infty$.
		\item $\plim_{N \to \infty} \arg \max_{\theta \in \Theta} l^{CM}(\theta,\mathbf{z}^0) = \theta^0$.
		\item $\lim_{N \to \infty} N^0_g = \infty$ for every $g \in \mathbb{G}$, where $N^0_g = \sum_{i=1}^N z^0_{ig}$.
		\item $\mathcal{I}_g := \mathbb{E}_{0} \left[ \left\{s_{i}(\theta^0_g)\right\}\left\{s_{i}(\theta^0_g)\right\}^{\top} \right] < \infty$, where $s_{i}(\theta_g)$ is defined as in Assumption \ref{ass1:2}$(iii)$, with $\frac{1}{N^0_g} \sum_{i=1}^N z^0_{ig} \left\{s_{i}(\theta^0_g)\right\}\left\{s_{i}(\theta^0_g)\right\}^{\top} \xrightarrow{p} \mathcal{I}_g$ for every $g \in \mathbb{G}$.
		\item $\mathcal{H}(\theta_g) := \mathbb{E}_{0}[\frac{\partial}{\partial \theta_g } s_{i}(\theta_g)]$ where $\mathcal{H}_g \equiv \mathcal{H}(\theta_g^0)$ is finite and non-singular for every $g \in \mathbb{G}$, and where $\frac{1}{N^0_g} \sum_{i=1}^N z^0_{ig} \frac{\partial}{\partial \theta_g} s_{i}(\theta_g) \xrightarrow{p} \mathcal{H}(\theta_g)$ uniformly in $\theta_g$ in an open neighbourhood of $\theta_g^0$ for all $g \in \mathbb{G}$ as $N \to \infty$.
	\end{enumerate}
\end{assumption}
\setstretch{1.3}
Assumptions \ref{ass1:3}$(i)$-$(ii)$ are necessary to ensure that the algorithm leads to consistent estimates of $\theta$. Those assumptions can be adapted to the case of asymptotically biased estimators, such as MLE applied to nonlinear fixed effects models, by using a bias-corrected estimator. Assumption \ref{ass1:3}$(iii)$ implies that the number of observations within each group goes to infinity as the sample size increases. Assumption \ref{ass1:3}$(iv)$ is a special case of Assumption \ref{ass1:2}$(iii)$ and is standard in most MLE problems. Finally, Assumptions \ref{ass1:3}$(v)$ states that the true Hessian matrix $\mathcal{H}_g$ is finite and non-singular for each $g \in \mathbb{G}$ and that the Hessian matrix $\mathcal{H}(\theta_g)$ can be computed by using the derivative of the score vector with respect to $\theta_g$ for each group separately.
\par
The next lemma and two theorems follow from Assumption \ref{ass1:3} and Algorithm \ref{alg:CEM}, and can be adapted to panel data and time series by modifying the score function, the Information matrix, and the Hessian matrix appropriately. Lemma \ref{lem1:cem} shows that the C-ML function never decreases between two consecutive iterations of the C-EM algorithm when the joint density classifier is employed. Theorem \ref{the1:35} shows that the C-EM algorithm will yield consistent estimates of all parameters in the mixture after a finite number of iterations provided appropriate initial parameter values, and a sufficiently large size of $y_i$ (or $x_i$) and sample size. If those conditions are satisfied, Theorem \ref{the1:34} shows that each estimated component parameter $\hat{\theta}^{(k)}_g$ will be normally distributed asymptotically.
\begin{lemma}\label{lem1:cem}
	Let Assumptions \ref{ass1:1}(ii)-(iii), \ref{ass1:2} and \ref{ass1:3} hold. Then the C-ML function will never decrease between two consecutive iterations of the C-EM algorithm if $h_j(y_i|\theta_j) = f_j(y_i|\theta_j) \ \forall j \in \mathbb{G}$.
\end{lemma}
\begin{theorem}\label{the1:35}
	Let Assumptions \ref{ass1:1}(ii)-(iii), \ref{ass1:2} and \ref{ass1:3} hold. Given a fixed number of groups $G$, initial values $\theta^{(0)}$ sufficiently close to $\theta^0$, and sufficiently large values of $N$ and $p$, the C-EM algorithm equipped with a consistent classifier will converge to $\hat{\theta}^{(k+1)} = \hat{\theta}^{(k)}$ after $k+1$ iterations, where $\hat{\theta}^{(k)}$ is such that $\hat{\theta}^{(k)} \xrightarrow{p} \theta^0$ as $N,p \to \infty$, and where $k$ is a positive, discrete, finite number.
\end{theorem}
\noindent
\textbf{Remark 5.} It is worth noting that Theorem \ref{the1:35} refers to the asymptotic behavior of $\hat{\theta}^{(k)}$ \textit{conditional} on the previous set of estimates $(\theta^{(0)},\hat{\theta}^{(1)},...,\hat{\theta}^{(k-1)})$. This can be very different from the asymptotic behavior of the whole sequence of estimated parameters, which implies that $\hat{\theta}^{(1)}$ is such that $\mathbf{z}({\theta}^{(0)}) = \mathbf{z}(\hat{\theta}^{(1)})$ asymptotically given a consistent classifier and initial parameter values that are sufficiently close to the true values. On the other hand, it may be that $\hat{\theta}^{(k)}$ does not converge in probability to the true parameter values if we fix the $k-1$ sets of previous estimates, even if the algorithm is initialized at the true parameter value. This will happen if the entire sequence $(\theta^{(0)},\hat{\theta}^{(1)},...,\hat{\theta}^{(k)})$ converges, in finite samples, to a global maximum of the C-ML function that does not coincide with the global maximum asymptotically. This is why Theorem \ref{the1:35} states that $N$ and $p$ have to be sufficiently large to guarantee the consistency of $\hat{\theta}^{(k)}$ due to the conditional nature of the estimated parameters.
\vspace{2mm}
\begin{theorem}\label{the1:34}
	Let Assumptions \ref{ass1:1}(ii)-(iii), \ref{ass1:2} and \ref{ass1:3} hold, and let $\hat{\theta}^{(k)} \xrightarrow{p} \theta^0$ as $N,p \to \infty$ as described in Theorem \ref{the1:35}. Then
	\begin{align*}
		\sqrt{n_g(\hat{\theta}^{(k)})}(\hat{\theta}_{g}^{(k)} - \theta^0_{g}) \xrightarrow{d} \mathcal{N}(0, \mathcal{H}_g^{-1} \mathcal{I}_g \mathcal{H}_g^{-1}),
	\end{align*}
	as $N,p \to \infty$ for any $g \in \mathbb{G}$, where $n_g(\hat{\theta}^{(k)}) = \sum_{i=1}^N z_{ig}(\hat{\theta}^{(k)})$.
\end{theorem}
\noindent
\textbf{Remark 6.} The convergence of $\hat{\theta}^{(k)}$ is not uniform over $\pi^0$ since arbitrarily small values of $\pi^0_g$ can make the convergence of $\hat{\theta}_g^{(k)}$ arbitrarily slow as well. This is why all convergence results presented here are ``group-wise'' under Assumption \ref{ass1:3}(iii). It is possible to make $\hat{\theta}^{(k)}$ converge to $\theta^0$ uniformly over $\pi^0_g$ by modifying Assumption \ref{ass1:3}(iii) such that $\lim_{N \to \infty} N^0_g/N = \alpha_g$, where $\alpha_g$ is bounded away from zero for every $g \in \mathbb{G}$, which is a stronger assumption than the original one and might not be appropriate depending on the context. Note also that the estimates provided by the C-EM algorithm will asymptotically reach the semiparametric efficiency bound even if the employed classifier is not uniformly consistent since $\hat{\theta}^{(k)}$ will converge to the same distribution regardless of the employed classifier, as long as it is consistent.
\vspace{2mm}
\par
As in standard MLE problems, misspecification of the component densities will make the C-EM algorithm converge to a pseudo-true value $\bar{\theta} \ne \theta^0$ that minimizes the Kullback-Leibler divergence between the ``working'' and the true component densities \citep{gourieroux_pseudo_1984}. Misspecification of the component densities might also lead to the inconsistency of the joint density classifier if the working density is too far from the truth. On the other hand, the Mahalanobis distance classifier remains consistent under Assumption \ref{ass1:1}(ii)-(iii) and Assumption \ref{ass1:2} even if the component densities are misspecified. This is one of the main benefits of using the Mahalanobis distance classifier compared to the joint density classifier.
\vspace{2mm}\\
\textbf{Remark 7.} Under misspecification of the component densities and/or lack of efficiency, estimation of $\text{Var}[\hat{\theta}^{(k)}_{g}]$ can be done using the sample counterpart of the asymptotic variance given in Theorem \ref{the1:34}. This estimated variance is defined as follows
\begin{align}\label{eqn1:11}
	\widehat{\text{Var}[\hat{\theta}^{(k)}_{g}]} = \left\{s'(\hat{\theta}_g^{(k)}) \right\}^{-1}  \sum_{i=1}^N \left[\left\{s_{i}(\hat{\theta}_g^{(k)}) \right\}\left\{s_{i}(\hat{\theta}_g^{(k)})\right\}^{\top} \right]\left\{ s'(\hat{\theta}_g^{(k)}) \right\}^{-1},
\end{align}
where $s'(\hat{\theta}_g^{(k)}) = \sum_{i=1}^N \frac{\partial}{\partial \hat{\theta}^{(k)}_{g}} s_{i}(\hat{\theta}_g^{(k)})$, and where $\hat{\theta}^{(k)}$ is the set of parameters that maximize the C-ML function after searching for the global maximum of the objective function. Note that the expressions for both the asymptotic variance and its estimated counterpart will not account for the finite-sample uncertainty in group memberships. This uncertainty can be taken into account by the use of resampling techniques, such as the bootstrap, or by constructing confidence sets that account for group memberships \citep{dzemski_confidence_2024, higgins_bootstrap_2024}. Implementing such methods goes however beyond the scope of this paper.
\vspace{2mm}
\par
The next section confirms the theoretical intuitions and results presented in this section via simulations. Those theoretical intuitions and results are summarized as follows~: First, standard MLE of finite mixtures yields estimates that can be heavily biased in finite samples, with the size of the bias increasing with the overlap between the component densities. Second, maximizing the C-ML function equipped with a consistent classifier will yield consistent (as $N,p\to \infty$) and less biased estimates in finite samples compared to estimates obtained from standard MLE. Finally, the C-EM algorithm presented above can be used to maximize the C-ML function, which will yield consistent estimates provided that the algorithm is initialized with parameter values that are sufficiently close to their true counterparts.
\section{Monte Carlo Simulations}\label{sec1:4}
\subsection{Data-generating Processes and Estimation Strategies}\label{subsec1:41}
In this section, I describe the objectives and details of two distinct simulation exercises. The main objective of the first simulation exercise is to confirm the presence of a large finite-sample bias in the MLE of finite mixtures for relatively small values of $N$. The main objective of the second simulation exercise is to compare the finite-sample performance of the EM and the C-EM algorithms in the context of a mixture of linear panel data, where each algorithm is used to maximize the mixture likelihood and the C-ML function, respectively.
\subsubsection{Finite-Sample Bias of Standard MLE}\label{subsec1:411}
This first simulation exercise uses two-component mixtures of various distributions to show that maximizing the mixture likelihood function $l^{ML}(\xi)$ may yield heavily biased estimates of all parameters in the mixture. All datasets are generated using the following DGP~:
\begin{align*}
	y_i \sim z^0_{i1} f(\cdot|\theta_1^0) + (1-z^0_{i1}) f(\cdot|\theta_2^0), 
\end{align*}
where $y_i$ is univariate and $f(\cdot)$ represents the normal, Poisson, or exponential distribution, and where each $z^0_{i1} = \{0,1\}$ is set arbitrarily such that the set of true mixing weights $\pi^0 = (\pi^0_1, 1-\pi_1^0)$ corresponds to $(N^{-1}\sum_{i=1} z^0_{i1},N^{-1}\sum_{i=1} (1-z^0_{i1}))$.
\par
To ensure the consistency of the obtained estimates in the case of the mixture of normals, a penalized version of $l^{ML}(\xi)$ is maximized with respect to $\xi$ via the EM algorithm of \cite{dempster_maximum_1977}.\footnote{The EM algorithm is similar to the C-EM algorithm shown in Algorithm \ref{alg:CEM} but with $z_{ig}(\theta)$ replaced by $\tau_{ig}(\theta)$ in both steps, the latter corresponding to the probability that the $i^{th}$ observation belongs to the $g^{th}$ group. There are several ways to compute this probability, but the most popular and the one that I use in this paper is the following~: $\tau_{ig}(\theta) = \frac{\pi_gf_g(y_i|\theta)}{\sum_{g=1}^G\pi_gf_g(y_i|\theta)}$, where $\pi_g$ is estimated by averaging $\tau_{ig}(\theta)$ (from the previous iteration) over $i$.} The penalized mixture likelihood for the mixture of normal distributions is described as follows
\begin{align*}
	\tilde{l}^{ML}(\xi) &= \sum_{i=1}^N \log\left(\pi_1 \phi\left(\frac{y_i-\mu_1}{\sigma_1}\right) +(1-\pi_1)\phi\left(\frac{y_i-\mu_2}{\sigma_2}\right)\right)+ p_N(\sigma^2_1,\sigma^2_2),
\end{align*}
where
\begin{align*}
	p_N(\sigma^2_1,\sigma^2_2) &= - N^{-1/2}\left(\sigma^{-2}_1+\log(\sigma^2_1)+\sigma^{-2}_2+\log(\sigma^2_2)\right),
\end{align*}
which corresponds to the recommendation of \cite{chen_statistical_2023} for mixtures of normal distributions in large samples. The other simulations based on mixtures of Poisson and exponential distributions maximize the standard mixture likelihood $l^{ML}(\xi)$ since they do not suffer from the same consistency problem as with the mixture of normal distributions.
\par
Finally, to find the global maximum of ${l}^{ML}(\xi)$ (or $\tilde{l}^{ML}(\xi)$ for the mixture of normals), I use the following initialization and maximization strategy for each simulated dataset~:
\begin{enumerate}
	\setstretch{1}
	\item Build the set of quantiles $Q = (0.9999,0.9995,0.999,0.995,0.99,0.98,0.97,0.95)$.
	\item Separate the generated dataset in two, where the first dataset $\mathbf{y}_{N1,j} = \{y_i: y_i < Q_{j}\}$ while the second $\mathbf{y}_{N2,j} = \{y_i: y_i \ge Q_{j}\}$ with $Q_j$ being the $j^{th}$ element in $Q$.
	\item Use $\mathbf{y}_{N1,j}$ to generate the initial set of parameters $\theta^{(0)}_{1,j}$ and $\mathbf{y}_{N2,j}$ to generate the initial set of parameters $\theta^{(0)}_{2,j}$.
	\item Perform the EM algorithm using the initial parameters values $\theta^{(0)}_{1,j}$, $\theta^{(0)}_{2,j}$, and $\pi^{(0)}_{1,j} := |\mathbf{y}_{N1,j}|/N$, where $|\mathbf{y}_{N1,j}|$ corresponds to the size of $\mathbf{y}_{N1,j}$.
	\item Save the resulting value of ${l}^{ML}(\xi)$ (or $\tilde{l}^{ML}(\xi)$) once the algorithm has converged (i.e., ${l}^{ML}(\xi^{(k)})-{l}^{ML}(\xi^{(k-1)}) < 1\text{e}-10$).
	\item (For mixture of normals only) Repeat steps 2 to 5 but using the new set of quantiles $Q^c = 1-Q$.
	\item Perform the EM algorithm using the true values $\xi^0$ as initial values and save the resulting value of ${l}^{ML}(\xi)$ (or $\tilde{l}^{ML}(\xi)$).
	\item Select the estimates that yield the highest value of ${l}^{ML}(\xi)$ (or $\tilde{l}^{ML}(\xi)$) among all initial parameter values and identify them as the final estimates for this particular replication.
\end{enumerate}
The goal of this strategy is to compare the results provided by the maximization of ${l}^{ML}(\xi)$ (or $\tilde{l}^{ML}(\xi))$ via the EM algorithm for initial parameter values that are either equal to the true values, or equal to values that are influenced by the presence of natural outliers in the dataset. Since it is well-known that the EM algorithm does not decrease the value of the mixture likelihood between two consecutive iterations, initializing the algorithm with the true parameter values $\xi^0$ should yield estimated parameters that are close to these true values unless other parameter values can yield a higher likelihood value. According to the discussion in Section \ref{subsec1:32}, the parameter values that might yield a higher likelihood value are the ones where one component density is strongly influenced by a small set of outliers while the other component density is strongly influenced by every other observation. Dividing the sample into two sets of observations such that the deemed outliers (of the same sign) are all in the same set leads to the computation of initial values that are likely to identify such parameters, while also yielding a relatively high value of $\tilde{l}^{ML}(\xi)$.
\par
To identify the bias related to the maximization of the mixture likelihood, 1,000 replications with different simulated data were performed for each type of mixture and each set of true parameters. Label switching across replications was also avoided by sorting the component densities by their estimated mean values (or estimated standard deviations when $\mu_1^0 = \mu_2^0$) for each replication. The sample size was also allowed to vary between $N=100$ and $N=10,000$ for each set of true parameter values to confirm the idea according to which there exists a value $N^*$ for each simulation scenario such that the total estimation bias and mean squared error of $\hat{\xi}^{ML}$ drop when $N > N^*$. Convergence of the EM algorithm for each scenario was assumed to be reached when the value of the log likelihood function was less than 1e-10 between two consecutive iterations, or when the number of iterations reaches 100.
\subsubsection{Latent Group Linear Panel Structure}\label{subsec1:412}
I consider here the typical case in latent panel structure where the conditional density of the outcome value follows a normal distribution for all values of $g \in \mathbb{G}$. The goal of this simulation exercise is to determine which one of the two objective functions, $l^{ML}(\xi)$ or $l^{CM}(\theta,\mathbf{z})$, yields the best finite-sample results when these functions are maximized by the EM and the C-EM algorithms, respectively. In this simulation exercise, $p$ refers to the number of time-varying covariates that are used for classification jointly with the conditional density of the univariate outcome. 
\par
The exact DGP used to simulate the outcome variable is described as follows
\begin{align}\label{eqn1:13}
	y_{it} &= x_{it,1}\beta_{z^0_{it}} + \bar{x}_{i,1}\gamma_{z^0_{it}} + \delta_{tz^0_{it}} + \alpha_{iz^0_{it}} + \epsilon_{it},\\
	&= X_{it,1}\tilde{\beta}_{z^0_{it}} + \alpha_{iz^0_{it}} + \epsilon_{it}, \nonumber
\end{align}
where $z^0_{it}$ represents the true group membership of the $it^{th}$ observation and is generated by a categorical distribution where the vector of probabilities follows a Markovian process,\footnote{The transition matrix that is used to create such a process is generated via a Dirichlet distribution that varies across replications. This Markovian process is coincidental and is not central to further developments.} $\tilde{\beta}_{g} = (\beta_{g}^{\top}, \gamma_{g}^{\top},\delta_{1g},...,\delta_{Tg})^{\top}$, $X_{it,1} = (x_{it,1}, \bar{x}_{i,1},\mathbbm{1}[t=1],....,\mathbbm{1}[t=T])$, $x_{it,1}$ is a time-varying covariate and stands as the first element in the vector of covariates $x_{it}$, $\bar{x}_{i,1}^{\top} = T^{-1} \sum_{t=1}^T x_{it,1}$ is the time-averaged value of $x_{it,1}$, $\beta_{g}$ and $\gamma_{g}$ are both scalar parameters to be estimated for each value of $g \in \mathbb{G}$, $\alpha_{ig} \sim N(0,\sigma^2_{\alpha,g})$ where $\sigma^2_{\alpha,g} = g$, and where $\delta_{tg}$ is a time-fixed effect that is normally distributed with mean $\bar{x}_{tg,1} := \frac{\sum_{i=1}^N\mathbbm{1}[z^0_{it}=g]x_{it,1}}{\sum_{i=1}^N\mathbbm{1}[z^0_{it}=g]}$ and unit variance. For simplicity, I specify $\epsilon_{it} \sim N(0,\sigma^2_{\epsilon})$ with $\sigma^2_{\epsilon} = 1$, and also $\mathbb{E}[\epsilon_{it}\epsilon_{ls}] = 0$ for any pair $(l,s) \ne (i,t)$.
\par
The vector of covariate $x_{it}$ is generated according to
\begin{align}\label{eqn1:14}
	x_{it} &\sim \mathcal{N}_p(\boldsymbol{\mu}^0_{z^0_{it}}, \Sigma^0_{z^0_{it}}),
\end{align}
where $\mathcal{N}_p$ denotes the $p$-variate normal distribution, $\boldsymbol{\mu}^0_{g}$ is the vector of true mean values for the $g^{th}$ group where the $j^{th}$ element $\mu^0_{gj}$ is generated by a standard normal distribution for every $j \in \{1,2...,p\}$, and where $\Sigma^0_{g}$ is generated according to the following equation
\begin{align*}
	\Sigma^0_{g} &= P_gP_g^{\top},
\end{align*}
with
\begin{align*}
	P_g &= \begin{bmatrix}
		1 & \omega_{g,12} &  \cdots & \omega_{g,1p}\\
		0 & 1 & \cdots & \omega_{g,2p}\\
		\vdots & \vdots & \ddots & \vdots \\
		0 & 0 & \cdots &  1 \\
	\end{bmatrix},
\end{align*}
where $\omega_{g,ij} \sim N(0,1)$ for all triples $(g,i,j) \in \mathbb{G}\times \{1,2...,p\}^2$. Both $\boldsymbol{\mu}^0_{g}$ and $\Sigma^0_{g}$ vary across replications for every $g \in \mathbb{G}$. Defining the matrix $P_g$ as such guarantees that $\Sigma^0_{g}$ will always be positive-definite for each $g \in \mathbb{G}$. Finally, the parameters $\beta_g$ and $\gamma_g$ are drawn from a standard normal distribution for all values of $g \in \mathbb{G}$. Note that only the first element of $x_{it}$ is included as a regressor in Eq. (\ref{eqn1:13}) for simplicity. Including the other generated covariates in the specification of Eq. (\ref{eqn1:13}) does not alter significantly the conclusions of the simulation exercise, but makes the results less interpretable due to a larger number of elements in $\beta_g$.
\par
Eq. (\ref{eqn1:13}) corresponds to the Mundlak specification where the time-average values of the covariates plus unit-random effects are used in place of unit-specific fixed effects. It has been shown that it is equivalent to the \textit{least-square dummy variable} (LSDV) estimator \citep{wooldridge_two-way_2021} but with the advantage of being more computationally tractable since it greatly reduces the number of parameters to estimate. Moreover, using the Mundlak approach allows the non-random part of the unit-fixed effects, $\bar{x}_{i,1}^{\top}\gamma_{z^0_{it}}$, to vary across groups for any given individual. The LSDV estimator also allows for such flexibility, but each unit has to be assigned to the same group for at least two periods in order to avoid identification issues when using the C-EM algorithm. The Mundlak specification also facilitates cross-validation since the non-random part of the unit-fixed effect can be predicted for any unit in the test set. This framework also avoids the biases generated by the incidental parameter problem encountered in nonlinear panels with unit-fixed effects and fixed $T$.
\par
As mentioned earlier, estimation is carried out using both the EM and the C-EM algorithms for comparison. The C-EM algorithm uses the joint density classifier with normal distributions where the discriminant function $h_{j}(y_{it},x_{it}|\theta_j,\psi_j) = f_{j}(y_{it}|x_{it};\theta_j)p_j(x_{it}|\psi_j)$. To make sure that the C-ML function never decreases between two consecutive iterations of the C-EM algorithm, I use $l^{CM}(\theta,\psi,\mathbf{z}) = \sum_{i=1}^N\sum_{t=1}^T\sum_{g=1}^G z^D_{itg}(\theta,\psi) \log(f_{j}(y_{it}|x_{it};\theta_j)p_j(x_{it}|\psi_j))$ as the objective function (see the proof of Lemma \ref{lem1:cem} for a justification). All models are correctly specified, including the use of normal distributions for the random-unit effect $\alpha_{ij}$, the error term $\epsilon_{it}$, and the density of $x_{it}$. To ease computational burden, 250 replications were performed with only 25 distinct sets of randomly-generated initial parameter values per replication. Comparison of the finite-sample performance between the EM and the C-EM algorithms was performed by using the same set of initial values for both algorithms. All covariates, true parameters, and random terms are allowed to vary across replications according to their respective distribution or generating process. More details regarding the two estimation procedures for this second simulation exercise are given in Appendix \ref{sec1:B}.
\par
Since the random part $\alpha_{iz^0_{it}} + \epsilon_{it}$ is normally distributed, the penalized likelihood function $\tilde{l}^{ML}(\xi)$ is used instead of the regular mixture likelihood $l^{ML}(\xi)$ to ensure consistency. Non-singularity of the estimated $\hat{\Sigma}^{(k)}_g$ is obtained by ensuring that all eigenvalues of $\hat{\Sigma}^{(k)}_g$ remain above 1e-8 for every $g \in \mathbb{G}$ at each iteration. The maximum number of iterations for every replication is set to 100. Convergence of the EM algorithm is assumed to be reached when the relative change between two consecutive values of $\tilde{l}^{ML}(\xi)$ is less than 0.01\%. Convergence of the C-EM algorithm is achieved when the estimated group memberships do not change between two consecutive iterations. All simulations are performed using Python 3.11.10 with NumPy version 2.2.0. Replication codes for the two simulation exercises are available on GitHub \href{https://github.com/raph1177/Finite_mixture}{here}. Additional simulation results for mixtures of binary choice models (Probit) are also available from the author upon request.
\subsection{Simulation Results}\label{subsec1:42}
\subsubsection{Finite-Sample Bias of Standard MLE}\label{subsec1:421}
\begin{figure}[t!]
	\setstretch{1.15}
	\centering
	\begin{subfigure}[Estimation biases for different sample sizes]{
			\centering
			\includegraphics[width=14.5cm]{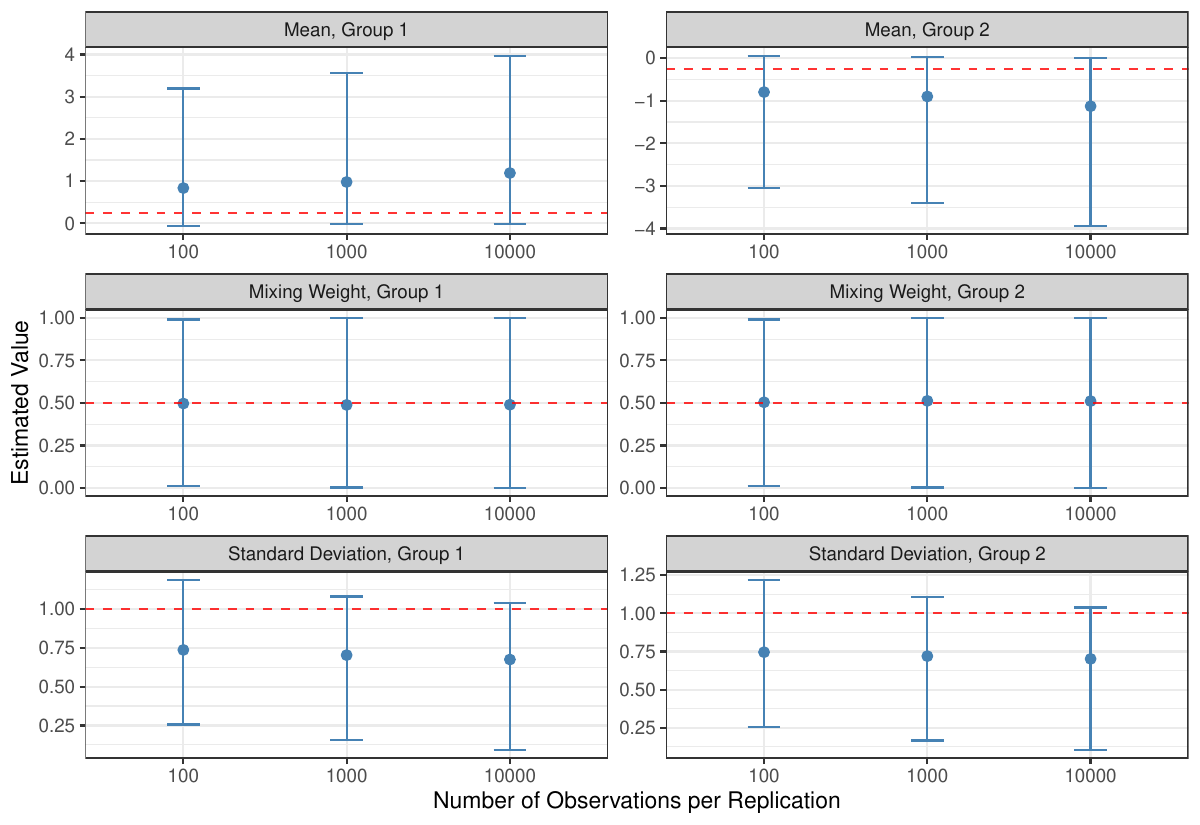}
	}\end{subfigure}
	\begin{subfigure}[Estimated parameters for N=10,000]{
			\centering
			\includegraphics[width=14.5cm]{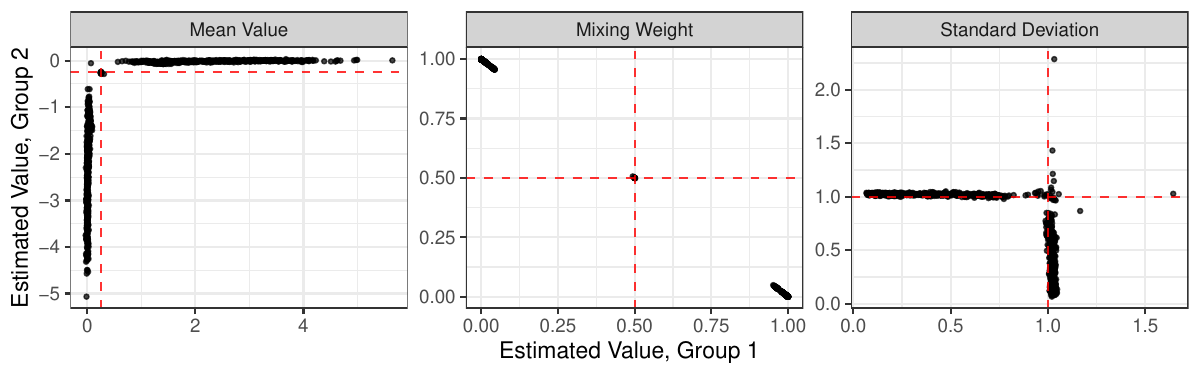}
	}\end{subfigure}
	\caption{Simulation results for the mixture of normal distributions with true values $\boldsymbol{\mu}^0 = (0.25,-0.25)$, $\sigma^0 = (1,1)$, and $\pi^0_1 = 0.5$. The blue dots correspond to the estimated value averaged across 1,000 replications, while the error bars represent the $2.5^{th}$ and $97.5^{th}$ percentiles of the estimated parameter's empirical distribution. The red dashed lines correspond to the true parameter values.\label{fig1:2}}
\end{figure}
\setstretch{1.3}
Figure \ref{fig1:2} shows the results of the first simulation exercise for the two-component mixture of normal distributions with true mean values $\boldsymbol{\mu}^0 = (0.25,-0.25)$, true standard deviations $\sigma^0 = (1,1)$, and true mixing weight $\pi^0_1 = 0.5$. Panel (a) of Figure \ref{fig1:2} shows that the estimation bias for both the mean values and the standard deviations, which corresponds to the distance between the blue dots and the red dashed line, tends to \textit{increase} with the sample size. Note that the estimated mixing weights are unbiased in this case, but the estimation variance (due to the unbiasedness) is very high because some estimated mixing weights are very close to the boundary of the parameter space $\Pi$. Note also that the upper graphs of Figure \ref{fig1:2} confirm the formal intuition described in Example \ref{exam:1}, which is that $\hat{\mu}_1$ and $\hat{\mu}_2$ are positively and negatively biased in finite samples, respectively, when the distance between the true mean values is small and ${\mu}^0_1 > {\mu}^0_2$.
\par
The results shown in panel (b) of Figure \ref{fig1:2} also confirm that, for several simulated samples, one of the two sets of estimated parameters that maximize $\tilde{l}^{ML}(\xi)$ is strongly influenced by a small number of outliers. Indeed, panel (b) shows that, when $N=10,000$, several replications feature a set of maximizers $\hat{\xi}$ that are heavily influenced by outliers, as indicated by the strange estimated parameter values across several replications. Only a few number of estimated parameters feature a very small level of bias compared to the others, indicating that not all simulated datasets generated sufficiently large outliers (or a sufficiently large number of outliers) to drastically change the location of the global maximum of $\tilde{l}^{ML}(\xi)$.
\begin{figure}[t!]
	\setstretch{1.15}
	\centering
	\begin{subfigure}[Estimation biases for different sample sizes]{
			\centering
			\includegraphics[width=14.5cm]{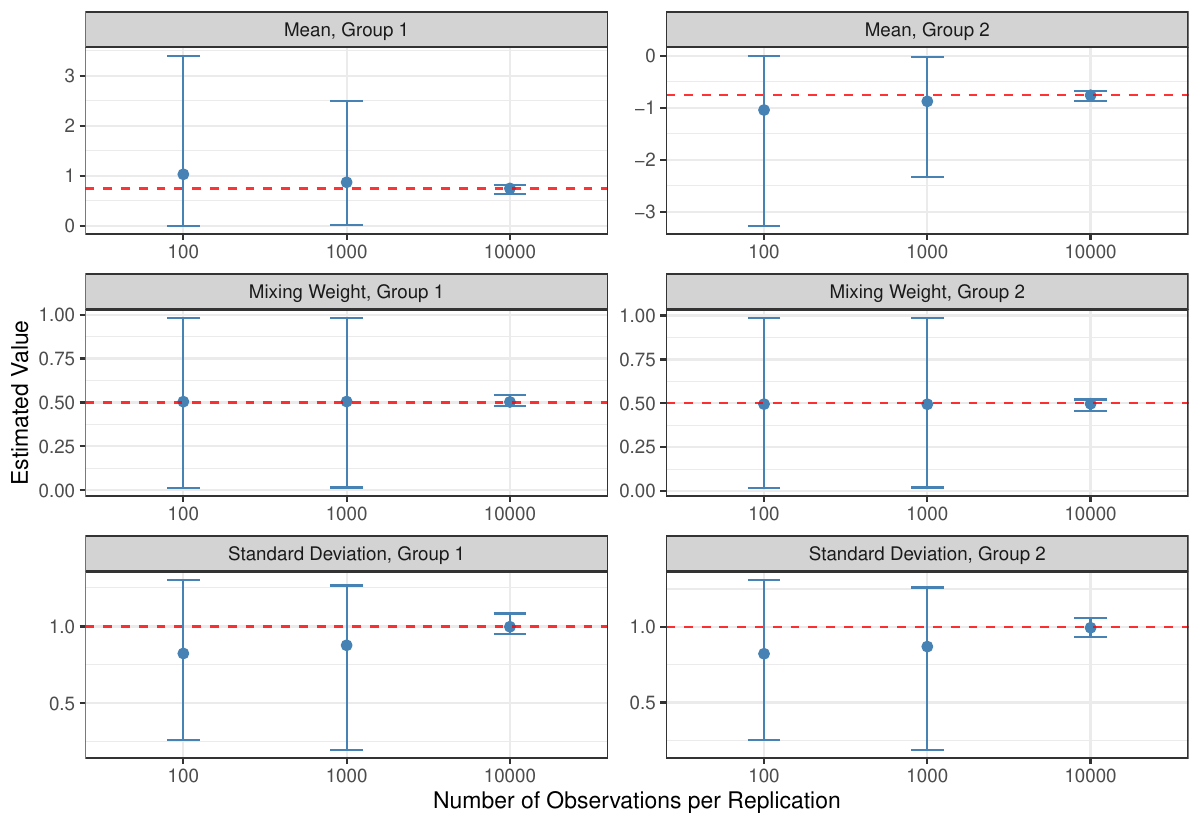}
	}\end{subfigure}
	\begin{subfigure}[Estimated parameters for N=10,000]{
			\centering
			\includegraphics[width=14.5cm]{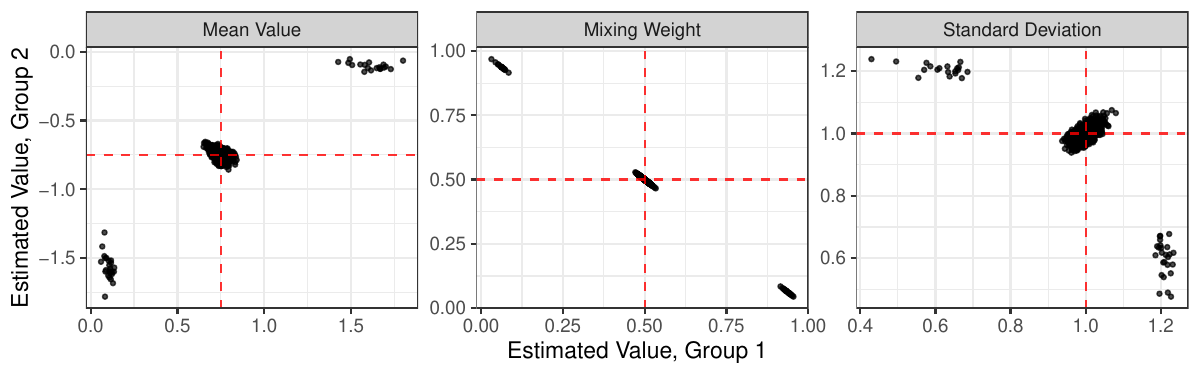}
	}\end{subfigure}
	\caption{Simulation results for the mixture of normal distributions with true values $\boldsymbol{\mu}^0 = (0.75,-0.75)$, $\sigma^0 = (1,1)$, and $\pi^0_1 = 0.5$. The blue dots correspond to the estimated value averaged across 1,000 replications, while the error bars represent the $2.5^{th}$ and $97.5^{th}$ percentiles of the estimated parameter's empirical distribution. The red dashed lines correspond to the true parameter values.\label{fig1:3}}
\end{figure}
\setstretch{1.3}
\par
Figure \ref{fig1:3} shows the results for the two-component mixture of normal distributions with true parameter values that are similar to those used to produce Figure \ref{fig1:2}, except for the mean values where $\boldsymbol{\mu}^0 = (0.75,-0.75)$. Increasing the distance between the true mean values leads to a reduction in the degree of overlap between the two component densities, which translates into a smaller influence of outliers on the estimated parameters. Figure \ref{fig1:3} also shows that increasing the sample size from 1,000 to 10,000 leads to a sharp reduction in estimation biases and/or mean squared errors for all parameters. Specifically, the results portrayed in Figure \ref{fig1:3} confirm the intuition behind Figure \ref{fig1:1}, which is that there exists a threshold value $N^*$ such that when $N> N^*$, the location of $\hat{\xi}^{ML}$ then becomes much closer to the true parameter values $\xi^0$. In the case illustrated by Figure \ref{fig1:3}, it is clear that this threshold lies somewhere between $1,000$ and $10,000$ for most simulated datasets. Determining the location of this threshold as a function of $\xi^0$ and characteristics of the data goes beyond the scope of this paper but is an interesting avenue for future research.
\par
Other simulation results with a mixture of normal distributions where standard deviations and mixing weights are allowed to vary are shown in Appendix \ref{subsec1:C1}. For instance, Figure\ref{fig1:C1} 
shows that the value of the threshold $N^*$ is, on average, lower when the true standard deviations differ across components, and that this value may differ across groups. Indeed, Figure \ref{fig1:C1}
shows that the reductions in estimation biases and mean squared errors are captured primarily by the first group of parameters given the larger value of the true mixing weight $\pi_1^0$. However, the differences in standard deviations and mixing weights across groups tend to matter less as the distance between the true mean values increases, as shown in Figure \ref{fig1:C2}. Figure \ref{fig1:C3} also shows that large differences in standard deviations alone are sufficient to reduce estimation biases and mean squared errors with a moderately large sample size. In fact, this is the only scenario where no component density is ever influenced by any outlier when $N = 10,000$ (see Panel (b) of Figure \ref{fig1:C3}).
\par
Finally, Figures \ref{fig1:C4} to \ref{fig1:C7} of Appendix \ref{subsec1:C1} show similar results to the previous figures, but for mixtures of Poisson and exponential distributions. Since these distributions are both governed by a single parameter, the estimation biases and mean squared errors are less salient than for mixtures of normal distributions. Nonetheless, Figures \ref{fig1:C4} 
and \ref{fig1:C6} show that both the estimated mean and mixing weight parameters from a mixture of Poisson or exponential distributions can be heavily biased in small- to medium-sized samples when the true mean values are too close to each other.
\begin{figure}[t!]
	\setstretch{1.15}
	\centering
	\subfigure{
		\includegraphics[width=14.5cm]{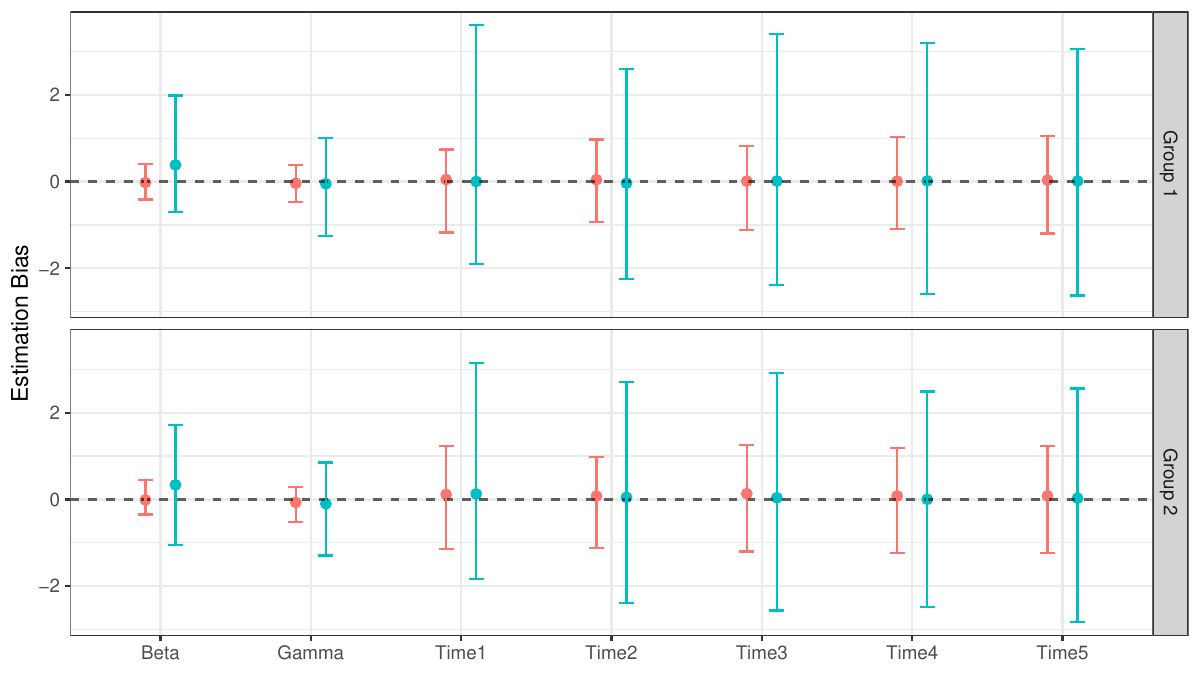}
	}
	\subfigure{
		\includegraphics[width=14.5cm]{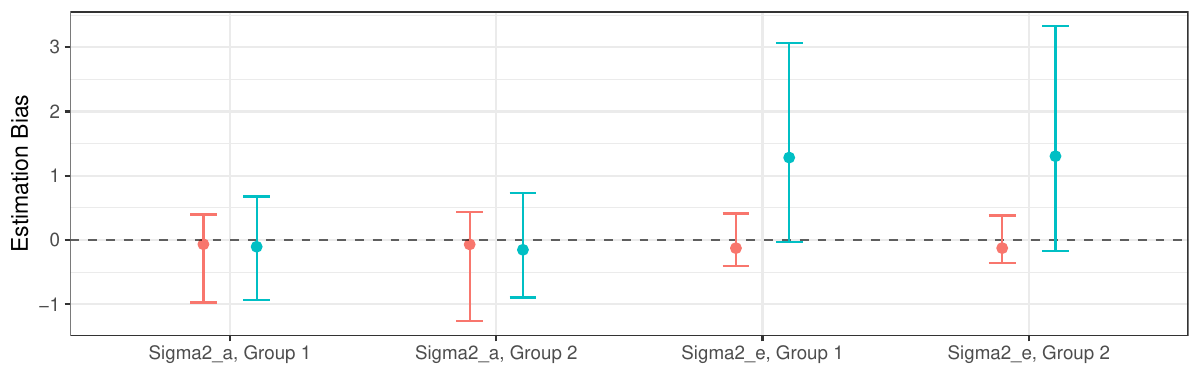}
	}
	\subfigure{
		\includegraphics[width=14.5cm]{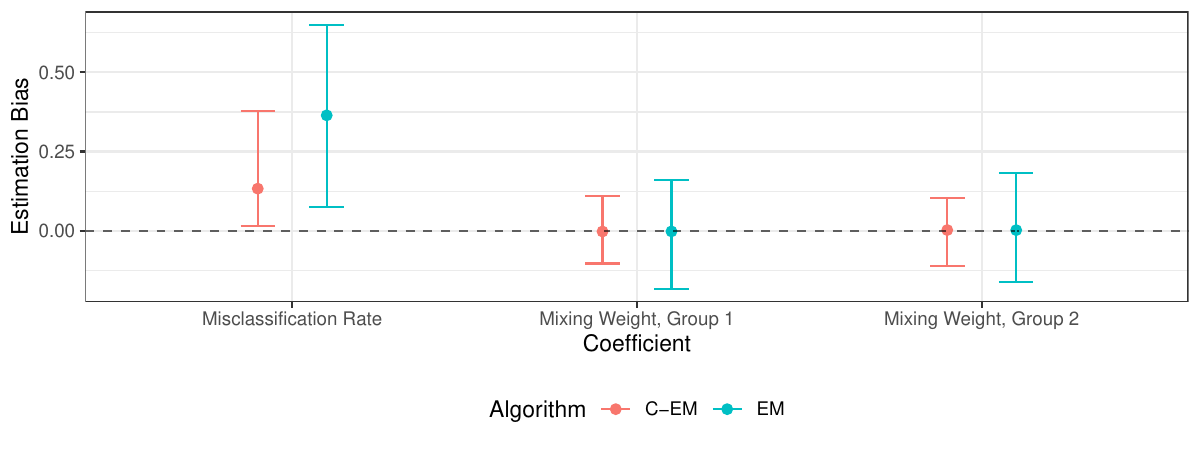}
	}
	\caption{Estimation bias for all parameters in the mixture when $N = 500$, $T=5$, $G=2$, and $p=1$. The dots correspond to the estimated bias over 250 replications, while the error bars represent the $2.5^{th}$ and $97.5^{th}$ percentiles of the empirical distribution of the bias. The coefficients $Time1,....,Time5$ represent the time-fixed effects, with $Sigma2\_a = \sigma^2_{\alpha,g}$ and $Sigma2\_e = \sigma^2_{\epsilon}$.\label{fig1:4}}
\end{figure}
\setstretch{1.3}
\subsubsection{Latent Group Linear Panel Structure}\label{subsec1:422}
The four graphs of Figure \ref{fig1:4} show the estimation biases of the mean parameters (i.e., $\beta_g$, $\gamma_g$, and $\delta_{tg}$ for all $g \in \mathbb{G}$ and $t \in \{1,2,...,T\}$), variance parameters, and mixing weights for each group and each algorithm when $N=500$, $T=5$, $G=2$, and the number of variables $p$ is equal to one. This implies that the joint density classifier is based only on two variables~: the outcome $y_{it}$ (conditional on the linear predictor $X_{it,1}\tilde{\beta}_g$) and the scalar covariate $x_{it,1}$. The two upper graphs of Figure \ref{fig1:4} show that the C-EM algorithm yields mean parameter estimates that have similar biases -- except for $\beta_g$ -- but much smaller mean squared errors compared to those produced by the EM algorithm. The two other graphs of Figure \ref{fig1:4} also show that the variance terms and the mixing weights estimated by the C-EM algorithm feature less variability and smaller or equal bias than those produced by the EM algorithm. This is because the C-EM algorithm, even with $p$ as low as 1, leads to a misclassification rate that is, on average, lower than the one produced by the EM algorithm. Note that the misclassification rate for the EM algorithm is based on the maximum posterior probability at convergence, and it should be equal to or close to zero in order to limit estimation biases and variability.
\par
\begin{figure}[t!]
	\setstretch{1.15}
	\centering
	\subfigure{
		\includegraphics[width=14.5cm]{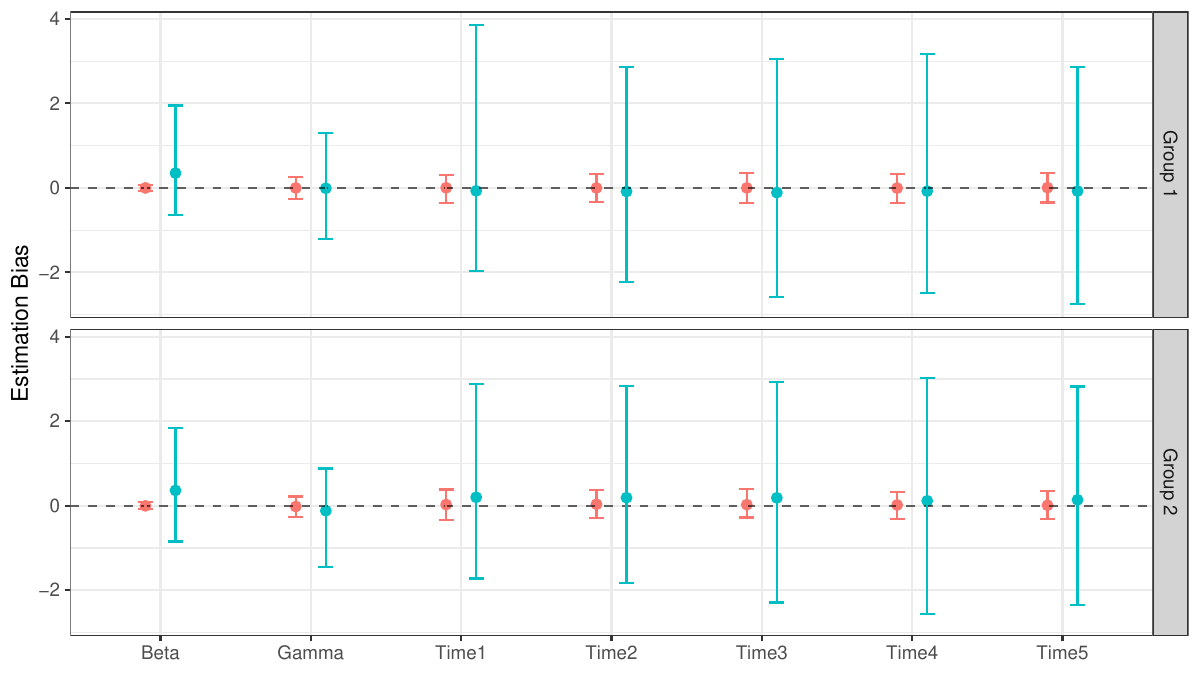}
	}
	\subfigure{
		\includegraphics[width=14.5cm]{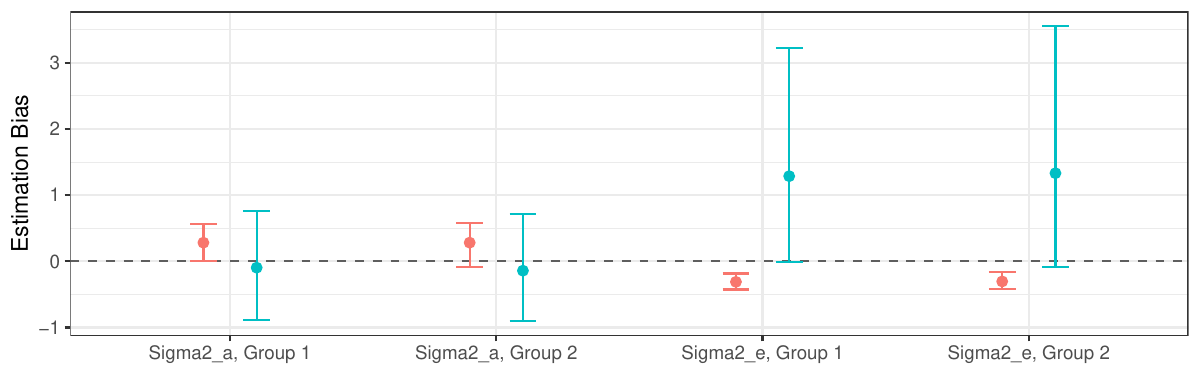}
	}
	\subfigure{
		\includegraphics[width=14.5cm]{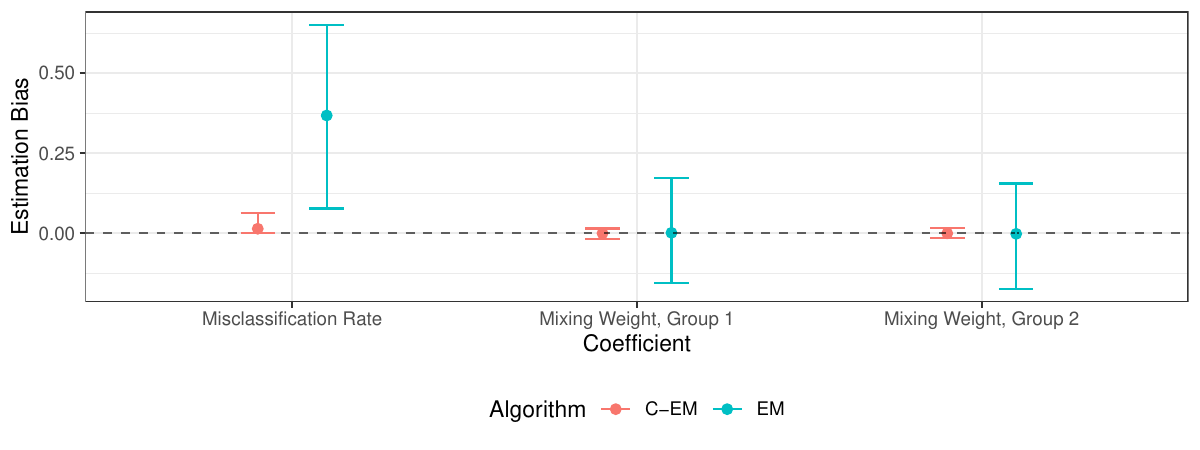}
	}
	\caption{Estimation bias for all parameters in the mixture when $N = 500$, $T=5$, $G=2$, and $p=5$. The dots correspond to the estimated bias over 250 replications, while the error bars represent the $2.5^{th}$ and $97.5^{th}$ percentiles of the empirical distribution of the bias. The coefficients $Time1,....,Time5$ represent the time-fixed effects, with $Sigma2\_a = \sigma^2_{\alpha,g}$ and $Sigma2\_e = \sigma^2_{\epsilon}$. \label{fig1:5}}
\end{figure}
\begin{figure}[t!]
	\setstretch{1.15}
	\centering
	\subfigure{
		\includegraphics[width=14.5cm]{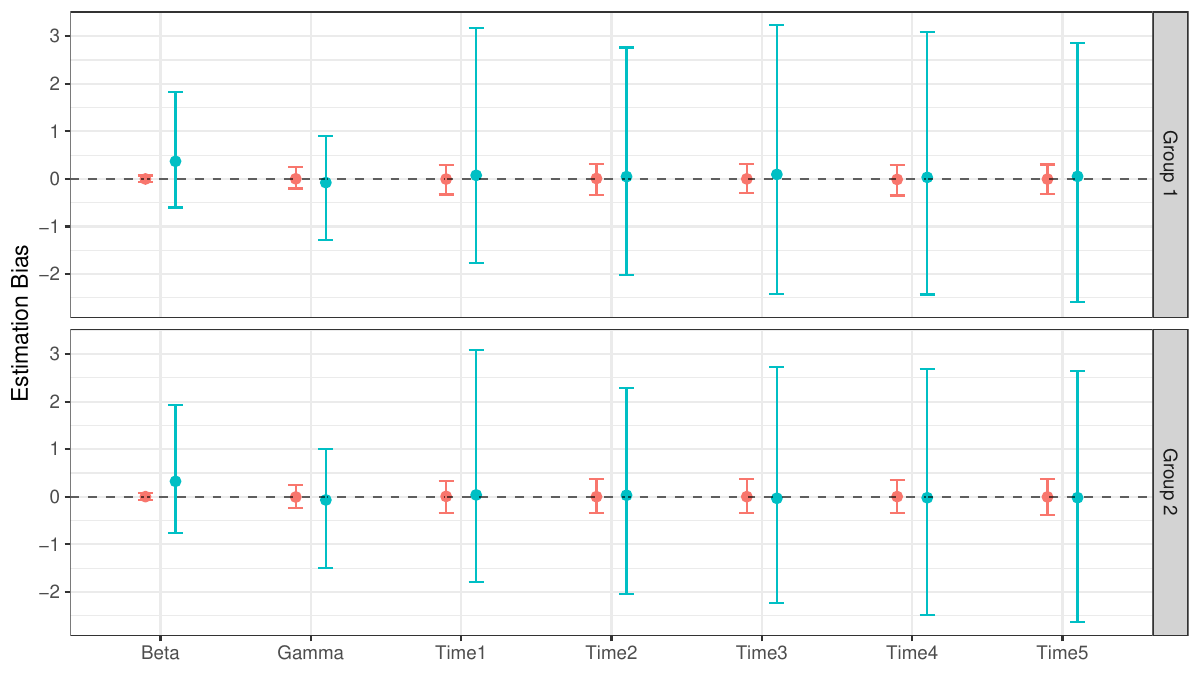}
	}
	\subfigure{
		\includegraphics[width=14.5cm]{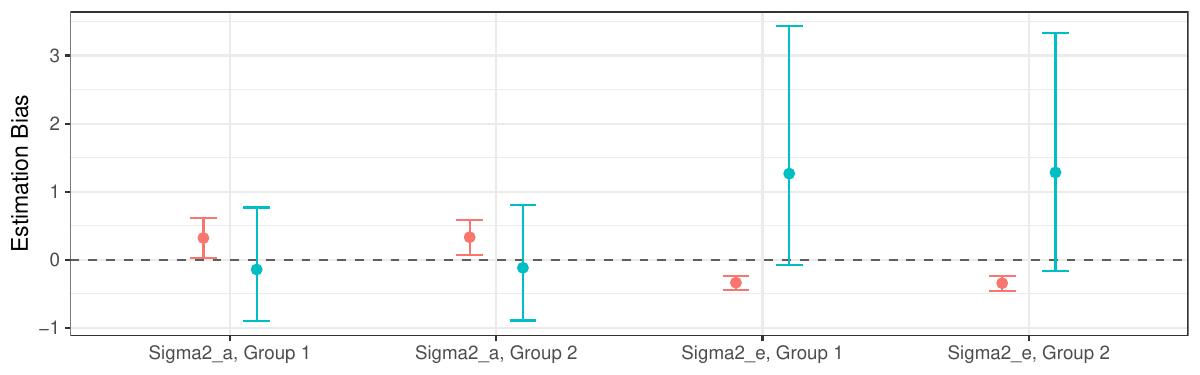}
	}
	\subfigure{
		\includegraphics[width=14.5cm]{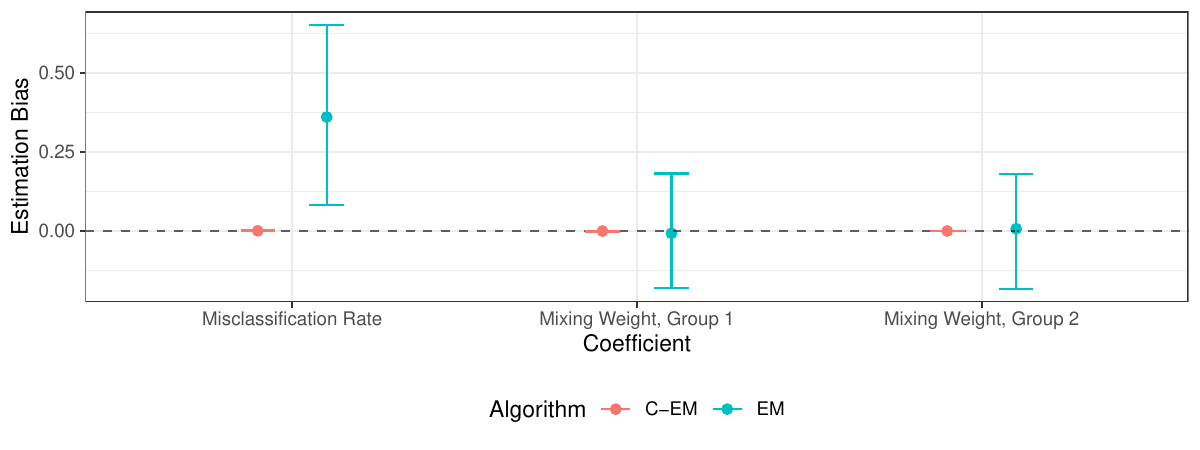}
	}
	\caption{Estimation bias for all parameters in the mixture when $N = 500$, $T=5$, $G=2$, and $p=10$. The dots correspond to the estimated bias over 250 replications, while the error bars represent the $2.5^{th}$ and $97.5^{th}$ percentiles of the empirical distribution of the bias. The coefficients $Time1,....,Time5$ represent the time-fixed effects, with $Sigma2\_a = \sigma^2_{\alpha,g}$ and $Sigma2\_e = \sigma^2_{\epsilon}$.\label{fig1:6}}
\end{figure}

\setstretch{1.3}
Figures \ref{fig1:5} and \ref{fig1:6} show similar results, but when the number of covariates $p$ is equal to 5 and 10, respectively. When $p$ goes from 1 to 5, there is a substantial reduction in mean squared errors for all parameters estimated by the C-EM algorithm, including the variance terms and the mixing weights. The misclassification rate is also close to zero for the vast majority ($\ge 95\%$) of the replications. As expected, the overall performance of the EM algorithm is not influenced by the additional covariates that are not included in the outcome model shown in Eq. (\ref{eqn1:13}). It is worth noting that introducing additional covariates in the outcome model will not improve the general performance of the EM algorithm unless the associated coefficients largely differ across groups, but this will also improve the performance of the C-EM algorithm in finite samples. This intuition is confirmed by additional simulation results that are available from the author upon request.
\par
When the number of covariates is equal to 10, the results illustrated in Figure \ref{fig1:6} show that the C-EM performs even better than with $p=5$. However, the marginal reduction in biases and mean squared errors diminishes as the number of covariates increases, which is due to the C-EM reaching the oracle efficiency lower bound for this sample size. The lower graph of Figure \ref{fig1:6} indeed shows that the misclassification rate produced by the C-EM algorithm is equal to zero in (at least) 19 replications out of 20. In this case, the C-EM algorithm produces mean parameter and mixing weights estimates with very low biases and mean squared errors, while the estimates produced by the EM algorithm remain slightly biased but highly variable. Note also that the bias and precision of the variance terms estimated by the C-EM algorithm are similar between $p=5$ and $p=10$ due to the short length of the panel.
\setstretch{1.3}
\par
Figures \ref{fig1:C8} to \ref{fig1:C10} in Appendix \ref{subsec1:C2} show the set of results obtained from the same simulation scenarios as the ones used to produce Figures \ref{fig1:4} to \ref{fig1:6}, but with $N=750$ and $T=8$. When $p=1$, Figure \ref{fig1:C8} shows that increasing the sample size does not necessarily lead to less biased and/or less variable estimates for either algorithm. However, when $p$ increases from 1 to 10, the C-EM algorithm produces less biased estimates with lower mean squared errors than with the smaller sample size, which is true for all parameters, including the variance terms. Figures \ref{fig1:C8} to \ref{fig1:C10} also show that there is no clear improvement in the performance of the EM algorithm after increasing the sample size from $NT = 2,500$ to $NT = 6,000$. Such a poor finite-sample performance can be explained by an insufficient increase in the sample size, or perhaps by the relatively small number of initial parameter values that were used to identify the global maximum of each objective function. In any case, the results obtained with the C-EM algorithm demonstrate that it is not always necessary to use a larger number of initial parameter values to yield estimated parameters that are close to true values.
\par
Finally, Figures \ref{fig1:C11} to \ref{fig1:C13} in Appendix \ref{subsec1:C2} show the set of results associated with similar simulation scenarios as the three shown above, but with an additional group (i.e., $N=500$, $T=5$, and $G=3$). For each value of $p$, increasing the number of groups leads to higher misclassification rates for the C-EM algorithm compared to when $G=2$, which then leads to higher mean squared errors for most estimated parameters. These results also show that increasing the number of groups $G$ entails similar effects on the parameters estimated by the C-EM algorithm as if the number of covariates $p$ had been reduced with $G$ fixed. This validates the conclusion of Corollary \ref{cor1:34}, which states that the ratio $p/G$ has to go to infinity for the classification procedure to be consistent. Therefore, increasing the number of groups $G$ while keeping $p$ constant increases the misclassification rate and thus leads to larger biases and/or mean squared errors overall.
\par
If the general form of the joint density $f(y_{it},x_{it}|\theta,\psi)$ is known, the results presented in this section show that the joint density classifier generally outperforms the results produced by the EM algorithm Assumption \ref{ass1:2} is satisfied, even when the numbers of covariates employed to improve classification is as small as 1. To note, the EM algorithm produces values of $\hat{\beta}_g$ that are always positively biased across all simulation scenarios, while the C-EM algorithm always yields mean parameters with very low biases and mean squared errors when $p \ge 5$, although this may be an artifact of the employed DGPs. 
\par
If the general form of the joint density is unknown and no assumption on this general form is being made, then the Mahalanobis distance classifier can be used as a nonparametric substitute for every classification step of the C-EM algorithm. However, the Mahalanobis distance classifier typically leads to higher misclassification rates compared to the joint density classifier (when the parametric assumptions are correct) and might result in a decrease of the C-ML function between two consecutive iterations. Using the Mahalanobis distance classifier with a slightly larger number of covariates will however lead to similar performances as with the joint density classifier in the context of the current simulation exercise (results not shown).
\section{Empirical Application}\label{sec1:5}
\subsection{Objectives}\label{subsec1:51}
The main objective of the empirical analysis is to group within the same component all the observations that share the same latent, unobserved individual characteristics. In this context, it is assumed that the unobserved characteristics explain a non-null proportion of the observed variation in individual healthcare expenditures (HCE), and that the unobserved characteristics can also influence the relationships between the outcome and the observed individual characteristics. For instance, individuals who are more frail than others, which is unobserved, are more likely to develop adverse health outcomes and become more dependent on the healthcare system after experiencing a minor injury. Such poor underlying characteristics will be reflected in the data by higher individual HCE in the periods following the minor injury.
\par
This approach is similar to health state modeling in the literature on hidden Markov models (HMMs) \citep{luo_bayesian_2021,komariah_health_2019} but using a two-step approach where the first consistently estimates the group membership (i.e., the ``health state'') and the second step models the dynamic behavior of the group membership. Unlike this literature, I will refer to health groups rather than ``health state'' given that the unobserved individual characteristics explaining HCE can be related to non-health factors (e.g. accessibility to healthcare services, peer effects, etc.).
\par
The second objective of the empirical analysis is to compare the results coming from both algorithms (C-EM and EM) when using real-world data with missing information. The predictive performance of the results obtained from each algorithm is compared using cross-validation, as detailed in Section \ref{subsec1:55}.
\subsection{Data Sources and Characteristics}\label{subsec1:52}
The employed dataset is the Quebec portion of the Canadian Emergency departments Team Initiative (CETI). The CETI research program on mobility and aging is a Canadian clinical program ``which aims to improve emergency department (ED) care for older adults with minor injuries'' \citep{provencher_decline_2015}. In the province of Quebec, the CETI research program followed 1,391 patients with a medical consultation at the ED of one of three hospitals (Hôpital du Sacré-Coeur de Montréal, Hôpital de l’Enfant-Jésus, and Hôpital du Saint-Sacrement) after a minor injury that occurred between 2009 and 2015. Individuals were included in the cohort if they were aged 65 and older and suffered from a minor injury that did not lead to hospitalization. A detailed description of the eligibility criteria and monitoring schedules can be found in \cite{provencher_decline_2015}.
\par
The health administrative data for each participant has been extracted from Quebec's physician claims database, which is managed by the Régie de l’Assurance-Maladie du Quebec (RAMQ). Each claim included a unique patient identification number, billing codes corresponding to the services rendered, the diagnostic code according to the International Classification of Diseases, 9th Revision (ICD-9-CM), dates and locations of the services provided, an identification code for the physician making the claim, and the amount paid to the physician. The period covered by the dataset goes from January 1, 2008, to June 30, 2016. Note that physician claims do not include all the HCE that an individual could incur in a given period. However, physician claims are strongly correlated with total HCE while being usually more precise than other cost measures (e.g., hospitalization costs, costs of treatment episodes, etc.).
\par
An unbalanced panel dataset containing 1,330 individuals and seven three-month periods per individual was created using the information described above. Some individuals died before the end of the last period, hence leading to 64 observations with no information. A weight of zero was attributed to those 64 observations at every classification/expectation step of both algorithms. Consequently, the total number of observations with non-null weights in the dataset is 9,246. The initial ED visit occurred at the end of the second period, while two follow-up visits occurred at the end of the third and fourth period,s respectively. Price effects (inflation) were accounted for by adjusting all costs with Quebec’s all-item consumer price index (CPI) between 2008 and 2016.
\subsection{Estimation Strategy}\label{subsec1:53}
The employed model is a mixture of two-part models, where the first part corresponds to a Probit binary choice model and the second part uses a lognormal distribution. The lognormal distribution accounts for the heavy-tailed distribution that is usually observed in individual HCE \citep{manning_estimating_2001}. Conceptually speaking, the first part models the ``decision'' of the $i^{th}$ participant to incur a strictly positive amount of HCE at the $t^{th}$ period, whereas the second part models the total amount, if any, that was incurred at this period by the participant. Two-part models have been extensively used in econometrics and health economics to model the decision-making of agents that lead to the generation of a strictly positive, continuous outcome depending on the initial decision of the agent \citep{norton_specification_2008, neelon_bayesian_2011}.
\par
The specification used within each part of the model is similar to the one used for the second simulation exercise, as formulated by Eq. (\ref{eqn1:13}). For simplicity, it is assumed that all covariates follow a normal distribution with means and covariance matrices varying across components. The general form of the $g^{th}$ joint component's density for the $it^{th}$ observation is written as follows
\begin{align}\label{eqn1:19}
	f_g(y^c_{it},x_{it}|\theta_g,\psi_g) &= \left[\left(1-\Phi(\eta^b_{itg})\right) f_N(x^b_{it}|\boldsymbol{\mu}^b_g,\Sigma^b_g) \right]^{1-d_{it}} \left[\Phi(\eta^b_{itg}) \phi\left(\eta_{itg}\right) f_N(x_{it}|\boldsymbol{\mu}_g,\Sigma_g) \right]^{d_{it}},
\end{align}
with $\eta^b_{itg} = X_{it}^{b}\tilde{\beta}^{b}_{g}$ and $\eta_{itg} = \frac{X_{it}\tilde{\beta}_g-y^c_{it}}{\sigma^{2}_{\alpha+\epsilon,g}}$, where $f_N(\cdot|\boldsymbol{\mu}_g,\Sigma_g)$ corresponds to the multivariate normal pdf with mean $\boldsymbol{\mu}_g$ and variance-covariance matrix $\Sigma_g$, $\Phi(\cdot)$ and $\phi(\cdot)$ are defined as in Example \ref{exam:1}, $y^c_{it} = \log(y_{it})$ stands as the log value of the HCE for the ${it}^{th}$ observation (and is set to zero if $y_{it} = 0$), $d_{it} = \mathbbm{1}[y_{it}>0]$ is equal to one if $y_{it}>0$ and zero otherwise, and where $\theta_g = (\tilde{\beta}_g,\tilde{\beta}^{b}_g, \sigma^{2}_{\alpha+\epsilon,g})$ and $\psi_g=(\boldsymbol{\mu}^b_g,\Sigma^b_g,\boldsymbol{\mu}_g,\Sigma_g)$ for any $g \in \mathbb{G}$. All other coefficients have the same interpretation as in Section \ref{subsec1:412}, with $X_{it}$ including all covariates and fixed effects, and with the $b$ superscript denoting the estimates of the binary part. Note that all covariates and parameters associated with the binary part need not be equal to those of the continuous part, even if the true distributions of the common elements in $x^b_{it}$ and $x_{it}$ are identical. More details on the covariates included in each part are provided in the next subsection.
\par
Estimation of the model is carried out using similar approaches to the ones used in the second simulation exercise. Two differences are however worth noting. First, each M-step for both binary parts (one for each algorithm) corresponds to a single Newton-Raphson step to update the estimates $\tilde{\beta}_g^b$. Second, the C-EM algorithm uses the joint density classifier $z^{D}_{itg}(\theta,\psi)$ at each C-step of the algorithm, where $h_{j}(y^c_{it},x_{it}|\theta,\psi) = f_j(y^c_{it},x_{it}|\theta_j,\psi_j)$, whereas the EM algorithm uses the posterior probabilities $\tau_{itg}(\theta,\psi)$ at each E-step of the algorithm, where $\tau_{itg}(\theta,\psi) = \frac{f_g(y^c_{it},x_{it}|\theta_g,\psi_g)}{\sum_{j=1}^{G} f_{j}(y^c_{it},x_{it}|\theta_j,\psi_j)}$. Such posterior probabilities avoid the need to choose between the binary and the continuous density at each E-step of the EM algorithm, while leading to values of $\tau_{itg}(\theta,\psi)$ that are closer to either 0 or 1 on average. Not introducing any mixing weight in the RHS of the posterior probability $\tau_{itg}(\theta,\psi)$ also makes the two algorithms more comparable to each other. Such a formulation also suggests that the covariate densities $f_N(x^b_{it}|\boldsymbol{\mu}^b_g,\Sigma^b_g)$ and $f_N(x_{it}|\boldsymbol{\mu}_g,\Sigma_g)$ are used as prior grouping information for computing $\tau_{itg}(\theta,\psi)$.
\par
For simplicity, the unit-random effects of the binary and the continuous parts are assumed to be independent of each other. It has been shown that this assumption is likely to introduce bias in the continuous part of the model \citep{su_bias_2009}. Note that this bias is different from the sample selection bias often encountered in econometrics \citep{heckman_sample_1979}. If the probability for every individual of generating a non-null amount of HCE is never equal to zero over time, then the sample selection bias will vanish as $T$ increases (provided a representative sample). On the other hand, the bias introduced by the independence assumption between the two parts of the model may be maintained even in the absence of any sample selection bias. However, the finite mixture setup naturally accounts for cross-part correlation that might be caused by correlated group-specific intercepts/time-fixed effects. Introducing correlated random effects is also possible using Bayesian estimation methods or simulated maximum likelihood, but using such methods goes beyond the scope of this paper. Inference on the estimated parameters is performed using a cluster-robust variance estimator that is described in Appendix \ref{sec1:B}.
\par
If the independence assumption described above is satisfied, then consistent estimation of the parameters can be done by estimating each part of the model separately. This implies that only the observations with a strictly positive outcome value are used to estimate the continuous parts of the model. The estimation of the binary parts is performed using an iterative weighted Newton-Raphson procedure that uses all the observations in the sample, where the weights correspond to the joint density classifier (for the C-EM) and the posterior probabilities (for the EM) described above. Convergence of the whole estimation procedure is assumed to be achieved when the relative change between two consecutive log likelihood values is less than 0.01\%. A maximum number of 100 iterations is enforced for each iterative procedure.
\subsection{Covariates}\label{subsec1:54}
The list of covariates included in $X^b_{it}$ and $X_{it}$ is shown in Appendix \ref{sec1:D}. The included covariates are the same across all component densities. To proxy for frailty, I use the Elder's Risk Assessment (ERA) index, which predicts the hospitalization and health risks among elders \citep{crane_use_2010}. Frailty is known to be an important predictor of medical resource consumption and individual HCE over time \citep{sirven_cost_2017}. The composition of the ERA index is shown in Appendix \ref{sec1:E} and slightly differs from the original index due to data availability issues.
\par
In addition to frailty, the global amount of comorbidity is also known to be an important predictor of individual HCE \citep{charlson_charlson_2008}. Consequently, the Charlson index has been used to control for the overall burden of comorbidities in both parts of the model. The composition of the Charlson index is shown in Appendix \ref{sec1:E}. Given that the ERA and Charlson indices have comorbidities in common, the overlapping covariates were removed from the Charlson index to limit collinearity issues. Note that only the time-averaged Charlson index was used in every specification due to strong collinearity between the time-varying and the time-averaged Charlson indices.
\par
Finally, the Continuity of Care Index (COCI) was also introduced in the model to account for the peculiar relationships between each patient and the healthcare system. Continuity of care is defined as how one patient experiences care over time as coherent and linked \citep{bice_quantitative_1977, haggerty_experienced_2013}. Values of the index range from zero to one, with a zero value referring to a total absence of continuity of care (i.e., each visit is associated with a different provider), whereas a value of one represents perfect continuity of care (i.e., all visits are associated with the same unique provider). Several studies have shown that the COCI is strongly associated with individual HCE and that even modest variations in the index value are associated with large variations in medical costs \citep{chu_continuity_2012,hussey_continuity_2014}. The computation of the COCI relies on the identification of the billing physician at each visit, which cannot be done if $y_{it}=0$ for a given pair $(i,t) \in [N]\times \{1,2,...,T\}$. Although the missing COCIs could be proxied from other information in the dataset, I include the time-varying COCI only in the continuous part of the model for simplicity.\footnote{To maintain comparability with the other coefficients, the COCI has been rescaled from 0 to 10.}
\subsection{Selection of the Initial Parameter Values and Number of Groups}\label{subsec1:55}
Between 300 and 2,000 different sets of initial parameters were assessed to estimate the model for each algorithm and each value of $G$.\footnote{The number of initial parameter values is different for each algorithm since it is more frequent for the C-EM algorithm to experience convergence issues due to the presence of empty component(s). The exact number of initial parameter values used for each value of $G$ is indicated in the Python code provided by the author upon request.} The selection of the ``optimal'' initial parameter values and number of groups $G$ was performed using the maximum value of the corresponding likelihood function and cross-validation (CV). If some observations are misclassified, then maximizing the likelihood might select parameter estimates that may be quite far from the true parameters. This is why CV was also used to select the optimal set of initial parameter values and the number of groups.
\par
The goal of the CV procedure is to estimate the out-of-sample prediction error associated with the estimated parameters and the chosen number of groups. Following the recommendations of \cite{zhang_cross-validation_2015}, a 2-fold CV procedure with 10 different data splittings was performed for each value of $G \in \{2,6\}$ and each set of estimates associated with the 15 highest likelihood values.\footnote{The 15 highest likelihood values were chosen to reduce the computational burden of the CV procedure for the selection of the initial parameter values.} All data splittings randomly allocated each unit to the training set or the test set, which maintains the correlation structure among units in the training set. To limit the probability of being ``trapped'' in a local maximum of the objective function, the estimated parameter values of the complete dataset were used as the initial values for each repetition during the CV procedure.
\par
The predicted values of the test set for both algorithms are computed as follows
\begin{align*}
	\hat{y}^c_{it} = \widehat{\log(y_{it})} = \sum_{g=1}^{{G}} w^{(k)}_{itg} \Phi(X^b_{it}\hat{\tilde{\beta}}^{b,(k)}_{g}) X_{it}\hat{\tilde{\beta}}^{(k)}_g,
\end{align*}
where $w^{(k)}_{itg}$ is defined as in Section \ref{subsec1:53}, and where $\hat{\tilde{\beta}}^{b,(k)}_{g}$ and $\hat{\tilde{\beta}}^{(k)}_g$ are the respective estimated analogs of $\tilde{\beta}^{b}_g$ and $\tilde{\beta}_g$ after convergence of the algorithm. The cross-validated RMSEs were computed by combining the prediction errors of all observations for each test set and each data splitting. Note that all $\hat{\tilde{\beta}}^{(k)}_g$ correspond to semi-elasticities and that retransformation to the original scale is not necessary to perform the CV procedure.
\subsection{Results}\label{subsec1:56}
\subsubsection{Selection of the Initial Parameter Values and Number of Groups}\label{subsec1:561}
\begin{figure}[t!]
	\begin{center}		
		\includegraphics[width=15.5cm]{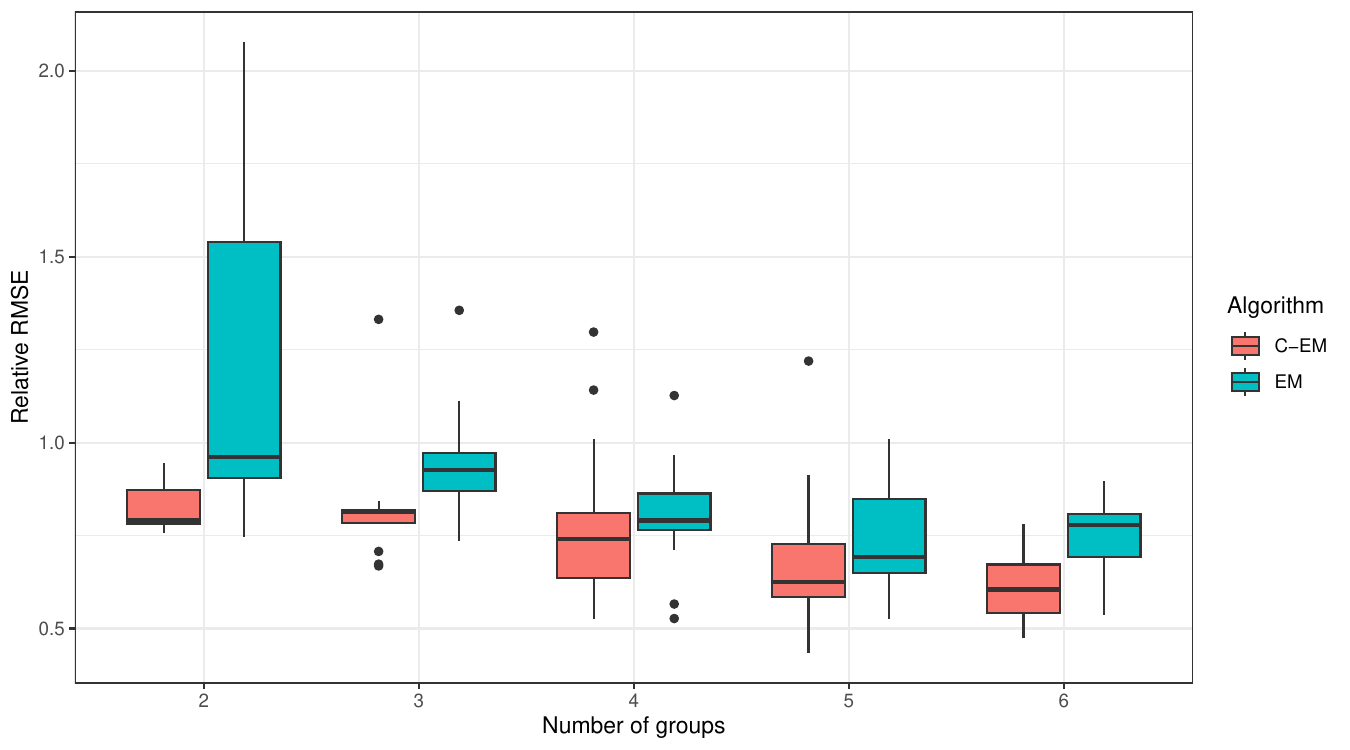}
		\caption{\label{fig1:7} Relative cross-validated RMSE values obtained by repeated 2-fold cross-validation (with 10 repetitions) for each one of the 15 highest likelihood values obtained by random initialization.}
	\end{center}
\end{figure}
\par
Figure \ref{fig1:7} shows the relative, cross-validated RMSE values obtained by the CV procedure described in Section \ref{subsec1:55} for each algorithm, each value of $G$, and each set of initial parameter values that produced the 15 highest likelihood values. The values shown in Figure \ref{fig1:7} are relative to the RMSE computed when $G=1$, which corresponds to a value of 2.05.
\par
The results illustrated in Figure \ref{fig1:7} show that the EM algorithm minimizes the cross-validated RMSE when $G=4$ while the C-EM algorithm minimizes the cross-validated RMSE when $G=5$. Several repetitions of the CV procedure have been performed to ensure the stability of the lowest RMSE values for each algorithm (results not shown). The lowest RMSE value obtained from the C-EM algorithm is equal to 0.89, which is 17.6\% smaller than the lowest RMSE value obtained from the EM algorithm. This also corresponds to a reduction of 56.6\% compared to the standard, single-component two-part model (2.05 vs. 0.89). Figure \ref{fig1:7} also shows that the out-of-sample prediction errors obtained from the C-EM algorithm are generally lower and less variable than those obtained from the EM algorithm, especially when $G=2$. For parsimony, the subsequent subsections will focus on the set of parameter estimates that yielded the lowest RMSE among all values.
\subsubsection{Groups' Analysis}\label{subsec1:562}
Table \ref{tab1:6} shows the estimated mean and variance values of each covariate for each group of the optimal set of estimates according to the CV procedure. Note that the distributions of the covariates in each group are likely to overlap with each other, as denoted by the large within-group variance values. Although each group contains observations that are homogeneous with respect to their unobserved characteristics, this means that all groups might contain observations that appear very different from each other based on their \textit{observed} characteristics. Because of this \textit{observed} heterogeneity, it is hard to broadly characterize the observations contained within each group. This is why the expression ``mostly feature'' is used below for the description of each group.
\begin{table}[t!]
	\begin{center}
		\caption{\centering Descriptive statistics of the covariates contained within each group created by the optimal set of estimates \label{tab1:6}}	
		\begin{tabular}{ c  c cccccc} \toprule\toprule
			\multirow{3}{1.4cm}{Group Number}& \multirow{3}{1.4cm}{Mixing Weights}  & \multirow{3}{1cm}{Male} &  \multirow{3}{1cm}{ERA} & {Time-} &\multirow{3}{1cm}{COCI} &{Time-}& {Time-}  \\
			& & & & averaged& & averaged&averaged\\
			& & & & ERA&& COCI&Charlson\\
			&\multicolumn{1}{c}{(1)} & \multicolumn{1}{c}{(2)} & \multicolumn{1}{c}{(3)} &\multicolumn{1}{c}{(4)} &\multicolumn{1}{c}{(5)} &\multicolumn{1}{c}{(6)} &\multicolumn{1}{c}{(7)} \\ \midrule
			\multirow{2}{1cm}{All groups} & \multirow{2}{1cm}{1.000} &0.38&2.05&2.05&3.19&4.00&1.35  \\
			&&(0.23)&(2.96)&(2.61)&(9.81)&(4.30)&(1.34) \\ [1mm]
			\multirow{2}{1cm}{$\mathbf{1}$} &\multirow{2}{1cm}{0.224} &0.38&1.46&1.57&1.34&4.01&1.50 \\
			& &(0.24)&(1.28)&(1.41)&(0.60)&(5.33)&(0.94) \\ [1mm]
			\multirow{2}{1cm}{$\mathbf{2}$} &\multirow{2}{1cm}{0.192} &0.39&3.90&3.68&3.06&3.25&2.41 \\
			& &(0.24)&(3.29)&(2.85)&(3.19)&(2.42)&(2.32)\\ [1mm]
			\multirow{2}{1cm}{$\mathbf{3}$} & \multirow{2}{1cm}{0.232} &0.40&2.03&2.06&0.99&3.49&0.93 \\
			& &(0.24)&(2.64)&(2.38)&(0.31)&(3.66)&(0.63) \\  [1mm]
			\multirow{2}{1cm}{$\mathbf{4}$}&\multirow{2}{1cm}{0.201}  &0.35&1.22&1.30&3.49&1.30&0.82 \\
			& &(0.23)&(0.86)&(0.91)&(1.39)&(2.86)&(0.32) \\  [1mm]
			\multirow{2}{1cm}{$\mathbf{5}$} &\multirow{2}{1cm}{0.151} &0.37&1.83&1.80&10.00&5.67&1.18 \\	  
			& &(0.23)&(2.46)&(2.08)&(0.00)&(3.65)&(1.00) \\ \bottomrule\bottomrule   
		\end{tabular}\\
		\vspace{1.2mm}
		\small \textbf{Note}~: Variance estimates are shown between parentheses.
	\end{center}
\end{table}
\par
The first group contains observations that mostly feature a moderately high number of comorbidities, as indicated by columns (4) and (7), and low continuity of care. The second group contains observations that mostly feature a larger number of comorbidities compared to the first group, but also a higher contemporaneous level of continuity of care, as indicated by column (5). The third group contains observations that mostly feature very low continuity of care and a moderately high number of comorbidities. Note that this group is the largest in size and also has the largest proportion of males. The fourth group contains observations that mostly feature a very low number of comorbidities and a relatively high continuity of care. This group also features the smallest proportion of males. Finally, the fifth group contains exclusively observations that feature perfect continuity of care. It is also the smallest group in size, and observations contained in this group mostly feature a relatively low number of comorbidities. Given that one covariate is constant within this group, this creates a perfect collinearity issue that can be dealt with by removing the group's intercept.
\par
Note that there exist large differences between the time-varying and the time-averaged COCI in most groups. For instance, the first and third groups contain observations that mostly feature low continuity of care at the current period, but high continuity of care on average over time. This is the opposite for the fourth and fifth groups. Therefore, it is not unreasonable to think that the level of continuity of care, both contemporaneously and over time, is an important factor in determining group memberships in the sample.
\par
\begin{figure}[t!]
	\begin{center}
		\includegraphics[width=14.5cm]{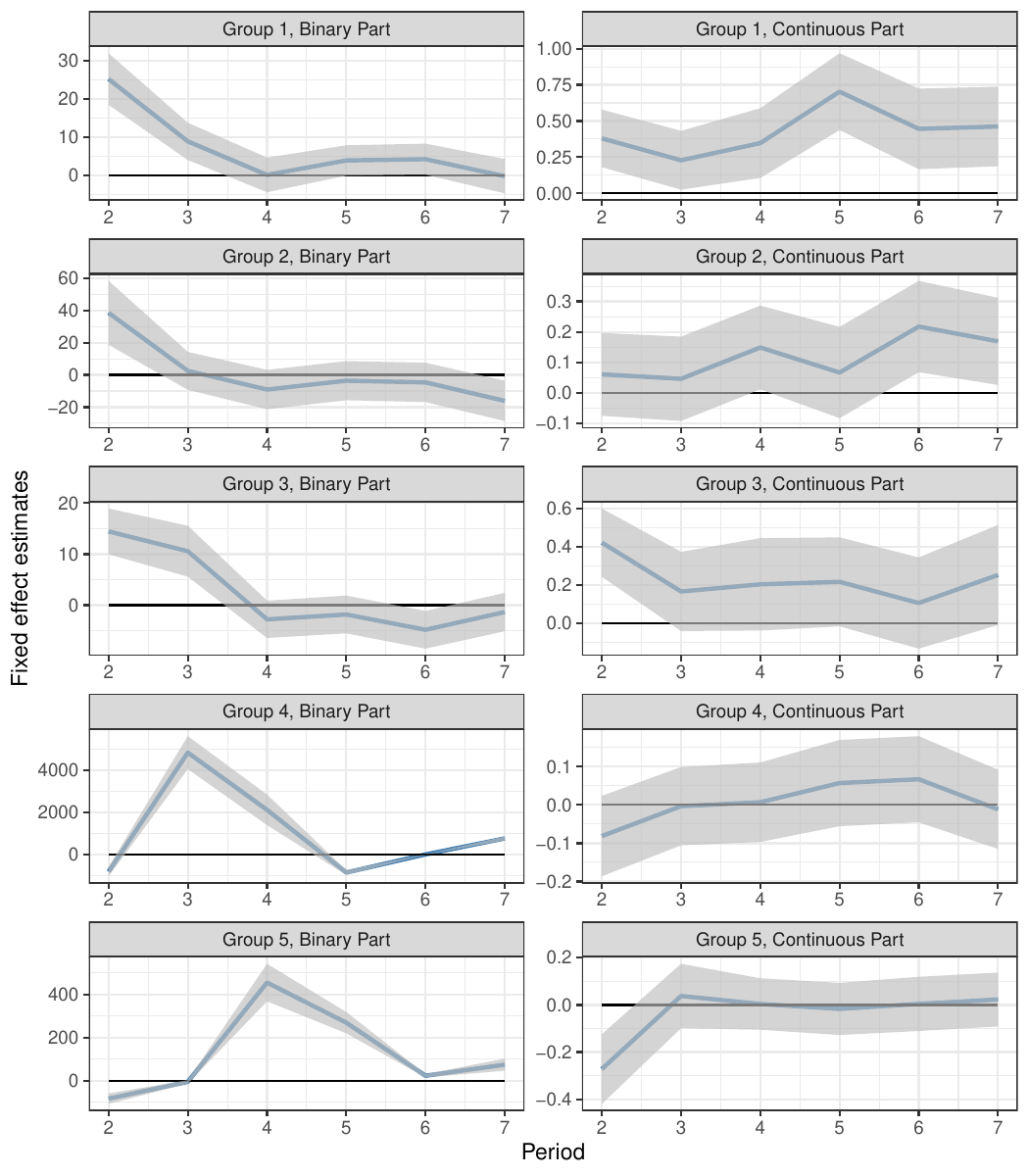}
	\end{center}
	\caption{\label{fig1:8} Time-fixed effects of the optimal set of estimates according to the CV procedure. The value of the first time-fixed effect is equal to zero and is set as the reference value. The shaded areas correspond to the 95\% cluster-robust confidence interval and do not account for uncertainty in group memberships.}
\end{figure}
Figure \ref{fig1:8} shows the estimated time-fixed effects for each group and each part of the model that are associated with the same set of estimates. Confidence intervals on the time-fixed effects (and all other coefficients) are computed using Eq. (\ref{eqn1:11}). The first period, ranging from 6 months to 3 months before the initial ED visit, is set as the reference value for the other time-fixed effects. The shaded areas correspond to the 95\% cluster-robust confidence intervals. Given that the continuous part of the model is in the log scale, each mean value in the graphs on the right of Figure \ref{fig1:8} corresponds to the average, relative increase in individual HCE (in \%) for each period. For instance, the observations in Group 1 feature an average increase of 70\% in their individual HCE at $t=5$ compared to $t=1$, which is explained by the low levels of continuity of care and the poor health of the observations contained within this group.
\par
The marginal effects of the time-fixed effects in the binary parts can be obtained using the estimated coefficient values shown in Appendix \ref{sec1:F1} and the covariates' average values presented in Table \ref{tab1:6}. Nonetheless, the sign of the mean values of the time-fixed effects on the LHS of Figure \ref{fig1:8} still indicates the direction of each effect on the probability of consuming medical resources. The very large time-fixed effects depicted in Group 4 at $t=3$ and $t=4$ indicate that every or almost every observation in this group consumed a strictly positive amount of medical resources during those two periods. Note that the initial ED visit after the minor injury did not necessarily lead to a visit to the physician, thus explaining why some time-fixed effects at $t=2$ are significantly negative in both parts of the model.
\par
\begin{figure}[t!]
	\begin{center}
		\includegraphics[width=14.5cm]{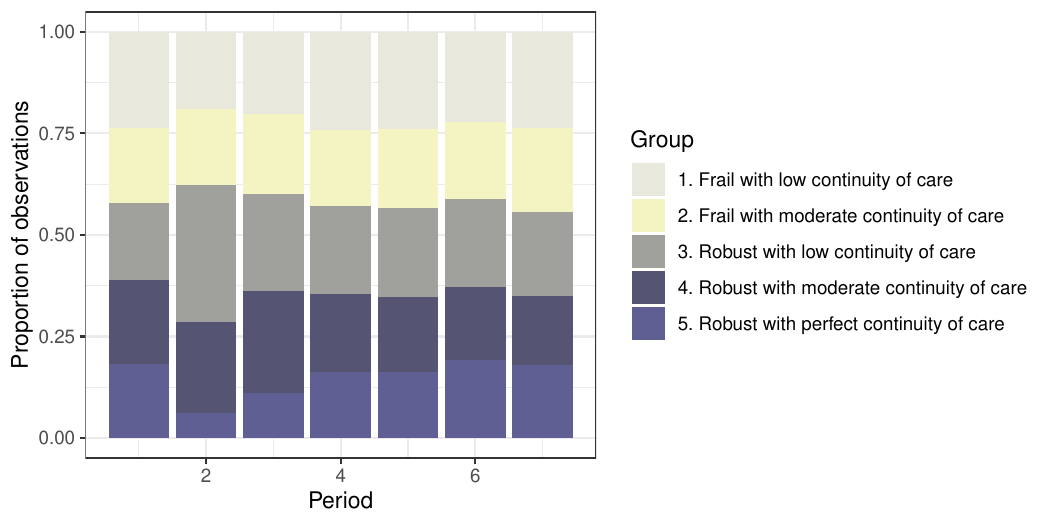}
	\end{center}
	\caption{\label{fig1:9} Proportions of the total number of observations in each group at each period.}
\end{figure}
The analogous set of time-fixed effects that are associated with the ``best'' set of parameters generated by the EM algorithm is shown in Appendix \ref{sec1:F3}. Trying to make formal connections between the estimates obtained from each algorithm is, according to me, hazardous given that the two methods do not treat the observed information similarly, and can produce results that are substantially different from each other. Such differences can also be appreciated by examining all other parameters estimated by each algorithm.
\subsubsection{Transition Between Groups}\label{subsec1:57}
Using the results from Table \ref{tab1:6} and Figure \ref{fig1:8}, it is possible to broadly characterize each group. This is shown below in Figure \ref{fig1:9}, along with the proportions of the total number of observations contained in each group at each period. This ``naming'' exercise is highly subjective and does not rely on a comprehensive analysis of all observed characteristics and results. The proposed ``names'' for the groups rely exclusively on frailty and continuity of care, which are known to be good predictors of individual HCE \citep{sirven_cost_2017, chu_continuity_2012, hussey_continuity_2014}.\footnote{More precisely, it was assumed that a group is composed of frail individual-periods if at least one of the time-fixed effects of the continuous part is significantly positive after the initial injury, and if the group mean value for the ERA or the Charlson index is larger than its global average. It was assumed to be composed of robust individual-periods if no time-fixed effect of the continuous part is significantly different from zero after the initial injury. Overall continuity of care was based exclusively on the average value of the time-varying COCI covariate, as shown in column (6) of Table \ref{tab1:6}, where the attribution of the ``type'' of continuity of care (i.e., low, moderate, and perfect) is obvious. Note that the overlap between the distributions of the COCI covariate for each type of continuity of care is very small compared to the other covariates, which advocates for the use of this covariate to broadly define the groups.} Note that the group numbers have been previously arranged in descending order so that the first group corresponds to the ``high-cost'' group, whereas the fifth group corresponds to the ``low-cost'' group.
\par
Figure \ref{fig1:9} shows that the initial ED visit substantially reduced the number of units in the group with perfect continuity of care at $t=2$. This is not surprising since the initial ED visit was associated with an unexpected minor trauma, therefore increasing the chance for the patient's regular healthcare provider to be unavailable at this precise moment. On the other hand, the initial ED visit seemed to have substantially increased the number of patients in Group 3. This can be explained by the fact that (robust) individuals who rarely consume medical resources can easily switch from perfect continuity to low continuity of care if they see a different healthcare provider each time they consume medical resources.
\par
Figure \ref{fig1:9} also shows that Group 4 experienced the largest reduction in size with a decrease of 3.4 percentage points between period 1 and period 7, while Group 2 is the group whose size increased the most with 2.3 additional percentage points between period 1 and period 7. This can be easily explained by a natural transition from robustness to frailty as time passes after the minor injury.
\par
\begin{figure}[t!]
	\begin{center}
		\includegraphics[width=14.5cm]{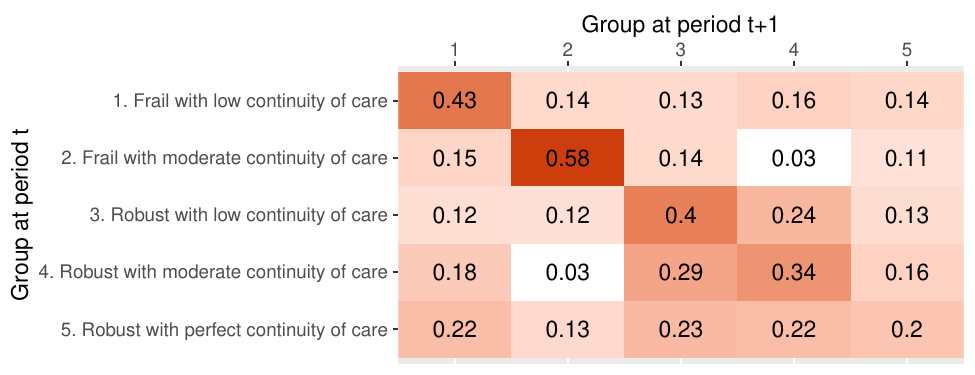}
		\caption{\label{fig1:10} Estimated transition matrix of the group memberships based on the optimal set of estimates according to the CV procedure.}
	\end{center}
\end{figure}
Figure \ref{fig1:10} depicts the estimated transition matrix of the group memberships over time. This transition matrix shows that robustness and frailty are persistent characteristics since transitions from a robust group to another robust group account for 74\% of all transitions out of a robust group on average, whereas transitions from a frail group to another frail group account for 65\% of all transitions out of a frail group on average. Therefore, robustness appears more persistent than frailty, which is in line with the fact that frailty is reversible provided appropriate care and healthy lifestyle habits \citep{kolle_reversing_2023}.
\par
Such a transition matrix can also be used to predict the group membership at period $t+1$ given the group membership at period $t$. Using the Bayes' rule of categorical assignment and the probabilities shown in Figure \ref{fig1:10}, group membership at period $t+1$ is correctly predicted 3,226 times out of 7,980 total memberships, which corresponds to a ``success'' rate of 40.4\%. This success rate could be easily improved upon using more sophisticated approaches, such as dynamic multinomial logit models, namely by including additional observed factors that might affect future group memberships.
\section{Conclusion}\label{sec1:6}
This paper showed that maximizing the likelihood function of a mixture density, as defined generally by Eq. (\ref{eqn1:1}), leads to estimates that can be severely biased in finite samples under weak regularity conditions. It is worth emphasizing that the resulting biases \textit{do not} depend on the chosen algorithm to maximize the mixture likelihood if the global maximum of the mixture likelihood is indeed reached. The motivation behind the widespread use of the EM algorithm is that it maximizes the mixture likelihood function just as any other numerical optimization method, but is easier to implement in practice. Still, it is true that for the same initial parameter values, different algorithms might converge to different local maxima. This, however, does not invalidate any of the claims made in this paper.
\par
Rather than relying on the mixture likelihood function, this paper showed that maximizing the classification-mixture likelihood function combined with a consistent classifier leads to (much) less biased and consistent estimates of all parameters in the mixture. As shown in Section \ref{subsec1:33}, consistency of the classifier is essential to obtain consistent estimates of the component parameters and the mixing weights. Although uniform consistency of the chosen classifier is a desirable property, Theorem \ref{the1:35} showed that it is not necessary to obtain consistent estimates of all parameters in the mixture. Nonetheless, Corollary \ref{th1:2} showed that the Mahalanobis distance classifier is uniformly consistent if the number of covariates goes to infinity at a faster rate than the sample size for a fixed number of groups, which is however problematic for estimation in large samples.
\par
Consistency of the Mahalanobis distance classifier (and, by extension, of the joint density classifier under normality) heavily hinges on the validity of Assumptions \ref{ass1:2}$(iv)-(v)$, which assumes that all the elements in the outcome vector $y_i$ and/or the covariates vector $x_i$ share the same group structure. To what extent the violation of those assumptions reduces the performance, and the usefulness of the proposed estimation strategy is an empirical question~: stronger violations of those assumptions are likely to lead to worse performance of the C-EM algorithm and vice versa. A strong violation would be a case where the group structure of the conditional outcome density is completely different from that of the covariates' density. Because it is not difficult to rule out such a possibility before analyzing the data, it is a good practice to use both the EM and the C-EM algorithms to compare the predictive performance of each approach, as done in Section \ref{sec1:5} of this paper.
\par
Contrary to the recent papers on latent group panel structures, the estimation strategy proposed in this paper has the benefit of leaving group membership completely unrestricted. This has important implications from a policy perspective given that transitions between groups over time allow for the prediction of future group memberships, which can then be used to improve decision-making in healthcare. Achieving such an objective implies that group membership has to be consistently estimated in the first step. The dynamic behavior of the group membership can then be modeled in the second step using a consistent estimator for categorical outcomes. Such a two-step procedure is advocated by \cite{bonhomme_distributional_2019} in the context of matched employer-employee panel data analysis, while being more in line with the literature on finite mixtures, where no restriction is imposed on group memberships to estimate the mixture parameters.
\par
A general recommendation concerning the estimation of any kind of finite mixture models (e.g., Gaussian mixture models, mixture of experts, latent group panel structures, etc.) with unrestricted group membership can thus be formulated~: instead of using the EM algorithm, one should always use the C-EM algorithm combined with a consistent classifier with as many covariates as possible. If the large number of covariates makes the computational burden too heavy, then one could remove from the conditional mixture density $f(y_i|x_i;\xi)$ (if it is the density of interest) the covariates that are deemed to be irrelevant for predicting the outcome $y_i$. If this is not sufficient, then a simple and effective procedure for variable selection at every classification step would be to remove covariates that feature the lowest in-sample variances. Indeed, covariates that feature a low variance do not contain a substantial amount of information in order to guide the classification process. Therefore, removing these covariates from all classification steps should not significantly increase the misclassification rate while easing computational burden.
\par
Results from the simulation exercises and a real-world application showed that the estimation procedure leads to less biased and more stable estimates than those obtained by standard MLE procedures such as the EM algorithm. Results from the empirical analysis also show that the proposed estimation strategy identified five heterogeneous health groups that account for a large part of the unobserved heterogeneity in the sample. The use of the proposed estimation strategy reduced the out-of-sample prediction error by more than 55\% compared to the single-component model and by 17.6\% compared to the best results obtained from standard, widely used MLE procedures.
\par
Compared to other algorithms, the proposed estimation strategy contrasts with HMMs, where the dynamic behavior of the latent variable is estimated simultaneously with the parameters of each component density. Although this method has been proven to yield consistent estimates \citep{douc_asymptotics_2001}, the two-step estimation procedure proposed in this paper seems to be less biased and more efficient than HMMs since the latter is often estimated using special cases of the EM algorithm such as the Baum-Welch or the Viterbi algorithm \citep{bishop_pattern_2006}. Confirming this intuition goes beyond the scope of this paper, but it is an interesting avenue for future research. Introduction of endogeneity in the covariates of the conditional mixture density, namely through feedback effects, is also another avenue for future research \citep{chamberlain_feedback_2022}.
\newpage
\setstretch{1.15}
\bibliographystyle{chicago}
\bibliography{main}

@incollection{grun_finite_2008,
	address = {Heidelberg},
	title = {Finite {Mixtures} of {Generalized} {Linear} {Regression} {Models}},
	isbn = {978-3-7908-2064-5},
	url = {https://doi.org/10.1007/978-3-7908-2064-5_11},
	doi = {10.1007/978-3-7908-2064-5_11},
	booktitle = {Recent {Advances} in {Linear} {Models} and {Related} {Areas}: {Essays} in {Honour} of {Helge} {Toutenburg}},
	publisher = {Physica-Verlag HD},
	author = {Grün, Bettina and Leisch, Friedrich},
	year = {2008},
	pages = {205--230},
}

@article{huber_robust_1964,
	title = {Robust {Estimation} of a {Location} {Parameter}},
	volume = {35},
	issn = {0003-4851},
	url = {https://www.jstor.org/stable/2238020},
	number = {1},
	urldate = {2025-12-08},
	journal = {The Annals of Mathematical Statistics},
	author = {Huber, Peter J.},
	year = {1964},
	pages = {73--101},
}

@article{kosmidis_bias_2014,
	title = {Bias in parametric estimation: reduction and useful side‐effects},
	volume = {6},
	copyright = {http://creativecommons.org/licenses/by-nc/3.0/},
	issn = {1939-5108, 1939-0068},
	shorttitle = {Bias in parametric estimation},
	url = {https://wires.onlinelibrary.wiley.com/doi/10.1002/wics.1296},
	doi = {10.1002/wics.1296},
	language = {en},
	number = {3},
	urldate = {2025-09-14},
	journal = {WIREs Computational Statistics},
	author = {Kosmidis, Ioannis},
	year = {2014},
	pages = {185--196},
}

@article{firth_bias_1993,
	title = {Bias {Reduction} of {Maximum} {Likelihood} {Estimates}},
	volume = {80},
	issn = {0006-3444},
	url = {https://www.jstor.org/stable/2336755},
	doi = {10.2307/2336755},
	number = {1},
	urldate = {2025-09-09},
	journal = {Biometrika},
	author = {Firth, David},
	year = {1993},
	pages = {27--38},
}

@article{newey_semiparametric_1990,
	title = {Semiparametric {Efficiency} {Bounds}},
	volume = {5},
	url = {https://people.eecs.berkeley.edu/~jordan/courses/210B-spring06/readings/newey.pdf},
	number = {2},
	urldate = {2025-07-09},
	journal = {Journal of Applied Econometrics},
	author = {Newey, Whitney K.},
	year = {1990},
	pages = {99--135},
}

@article{wu_randomly_2022,
	title = {Randomly initialized {EM} algorithm for two-component {Gaussian} mixture achieves near optimality in \${O}({\textbackslash}sqrt\{n\})\$ iterations},
	volume = {4},
	issn = {2520-2316, 2520-2324},
	url = {https://ems.press/doi/10.4171/msl/29},
	doi = {10.4171/msl/29},
	number = {3},
	urldate = {2026-01-23},
	journal = {Mathematical Statistics and Learning},
	author = {Wu, Yihong and Zhou, Harrison H.},
	year = {2022},
	pages = {143--220},
}

@article{kwon_global_2024,
	title = {Global {Optimality} of the {EM} {Algorithm} for {Mixtures} of {Two}-{Component} {Linear} {Regressions}},
	volume = {70},
	issn = {1557-9654},
	url = {https://ieeexplore.ieee.org/document/10614292/},
	doi = {10.1109/TIT.2024.3435522},
	number = {9},
	urldate = {2025-08-23},
	journal = {IEEE Transactions on Information Theory},
	author = {Kwon, Jeongyeol and Qian, Wei and Chen, Yudong and Caramanis, Constantine and Davis, Damek and Ho, Nhat},
	year = {2024},
	pages = {6519--6546},
}

@misc{kwon_em_2020,
	title = {The {EM} {Algorithm} gives {Sample}-{Optimality} for {Learning} {Mixtures} of {Well}-{Separated} {Gaussians}},
	url = {http://arxiv.org/abs/2002.00329},
	urldate = {2022-09-27},
	publisher = {arXiv},
	author = {Kwon, Jeongyeol and Caramanis, Constantine},
	year = {2020},
	note = {arXiv preprint:2002.00329 [cs, math, stat]},
}

@inproceedings{kwon_minimax_2021,
	title = {On the {Minimax} {Optimality} of the {EM} {Algorithm} for {Learning} {Two}-{Component} {Mixed} {Linear} {Regression}},
	url = {https://proceedings.mlr.press/v130/kwon21b.html},
	language = {en},
	volume = {130},
	urldate = {2022-09-27},
	booktitle = {Proceedings of {The} 24th {International} {Conference} on {Artificial} {Intelligence} and {Statistics}},
	author = {Kwon, Jeongyeol and Ho, Nhat and Caramanis, Constantine},
	year = {2021},
	pages = {1405--1413},
}

@misc{kwon_em_2019,
	title = {{EM} {Converges} for a {Mixture} of {Many} {Linear} {Regressions}},
	url = {http://arxiv.org/abs/1905.12106},
	urldate = {2022-09-27},
	publisher = {arXiv},
	author = {Kwon, Jeongyeol and Caramanis, Constantine},
	year = {2019},
	note = {arXiv preprint:1905.12106 [cs, stat]},
}

@article{bonhomme_grouped_2015,
	title = {Grouped {Patterns} of {Heterogeneity} in {Panel} {Data}},
	volume = {83},
	issn = {0012-9682},
	url = {http://www.jstor.org/stable/43616962},
	number = {3},
	urldate = {2022-09-27},
	journal = {Econometrica},
	author = {Bonhomme, Stéphane and Manresa, Elena},
	year = {2015},
	pages = {1147--1184},
}

@article{bonhomme_discretizing_2022,
	title = {Discretizing {Unobserved} {Heterogeneity}},
	volume = {90},
	issn = {1468-0262},
	url = {http://onlinelibrary.wiley.com/doi/abs/10.3982/ECTA15238},
	doi = {10.3982/ECTA15238},
	language = {en},
	number = {2},
	urldate = {2022-09-27},
	journal = {Econometrica},
	author = {Bonhomme, Stéphane and Lamadon, Thibaut and Manresa, Elena},
	year = {2022},
	pages = {625--643},
}

@article{okui_heterogeneous_2021,
	title = {Heterogeneous structural breaks in panel data models},
	volume = {220},
	issn = {03044076},
	url = {https://linkinghub.elsevier.com/retrieve/pii/S0304407620301287},
	doi = {10.1016/j.jeconom.2020.04.009},
	language = {en},
	number = {2},
	urldate = {2022-09-27},
	journal = {Journal of Econometrics},
	author = {Okui, Ryo and Wang, Wendun},
	year = {2021},
	pages = {447--473},
}

@article{su_identifying_2016,
	title = {Identifying {Latent} {Structures} in {Panel} {Data}},
	volume = {84},
	issn = {1468-0262},
	url = {http://onlinelibrary.wiley.com/doi/abs/10.3982/ECTA12560},
	doi = {10.3982/ECTA12560},
	language = {en},
	number = {6},
	urldate = {2022-09-27},
	journal = {Econometrica},
	author = {Su, Liangjun and Shi, Zhentao and Phillips, Peter C. B.},
	year = {2016},
	pages = {2215--2264},
}

@article{qian_shrinkage_2016,
	title = {Shrinkage estimation of common breaks in panel data models via adaptive group fused {Lasso}},
	volume = {191},
	issn = {03044076},
	url = {https://linkinghub.elsevier.com/retrieve/pii/S0304407615002377},
	doi = {10.1016/j.jeconom.2015.09.004},
	language = {en},
	number = {1},
	urldate = {2022-09-27},
	journal = {Journal of Econometrics},
	author = {Qian, Junhui and Su, Liangjun},
	year = {2016},
	pages = {86--109},
}

@article{wang_homogeneity_2018,
	title = {Homogeneity pursuit in panel data models: {Theory} and application},
	volume = {33},
	issn = {1099-1255},
	shorttitle = {Homogeneity pursuit in panel data models},
	url = {http://onlinelibrary.wiley.com/doi/abs/10.1002/jae.2632},
	doi = {10.1002/jae.2632},
	language = {en},
	number = {6},
	urldate = {2022-09-27},
	journal = {Journal of Applied Econometrics},
	author = {Wang, Wuyi and Phillips, Peter C. B. and Su, Liangjun},
	year = {2018},
	pages = {797--815},
}

@article{su_sieve_2019,
	title = {Sieve {Estimation} of {Time}-{Varying} {Panel} {Data} {Models} {With} {Latent} {Structures}},
	volume = {37},
	issn = {0735-0015, 1537-2707},
	url = {https://www.tandfonline.com/doi/full/10.1080/07350015.2017.1340299},
	doi = {10.1080/07350015.2017.1340299},
	language = {en},
	number = {2},
	urldate = {2022-09-27},
	journal = {Journal of Business \& Economic Statistics},
	author = {Su, Liangjun and Wang, Xia and Jin, Sainan},
	year = {2019},
	pages = {334--349},
}

@article{dempster_maximum_1977,
	title = {Maximum {Likelihood} from {Incomplete} {Data} via the {EM} {Algorithm}},
	volume = {39},
	issn = {0035-9246},
	url = {http://www.jstor.org/stable/2984875},
	number = {1},
	urldate = {2022-09-27},
	journal = {Journal of the Royal Statistical Society. Series B (Methodological)},
	author = {Dempster, A. P. and Laird, N. M. and Rubin, D. B.},
	year = {1977},
	pages = {1--38},
}

@article{bonhomme_distributional_2019,
	title = {A {Distributional} {Framework} for {Matched} {Employer} {Employee} {Data}},
	volume = {87},
	issn = {1468-0262},
	url = {http://onlinelibrary.wiley.com/doi/abs/10.3982/ECTA15722},
	doi = {10.3982/ECTA15722},
	language = {en},
	number = {3},
	urldate = {2022-09-27},
	journal = {Econometrica},
	author = {Bonhomme, Stéphane and Lamadon, Thibaut and Manresa, Elena},
	year = {2019},
	pages = {699--739},
}

@article{tanaka_strong_2009,
	title = {Strong {Consistency} of the {Maximum} {Likelihood} {Estimator} for {Finite} {Mixtures} of {Location}-{Scale} {Distributions} {When} {Penalty} is {Imposed} on the {Ratios} of the {Scale} {Parameters}},
	volume = {36},
	issn = {0303-6898},
	url = {http://www.jstor.org/stable/41000314},
	number = {1},
	urldate = {2022-09-27},
	journal = {Scandinavian Journal of Statistics},
	author = {Tanaka, Kentaro},
	year = {2009},
	pages = {171--184},
}

@article{redner_mixture_1984,
	title = {Mixture {Densities}, {Maximum} {Likelihood} and the {Em} {Algorithm}},
	volume = {26},
	issn = {0036-1445},
	url = {http://www.jstor.org/stable/2030064},
	number = {2},
	urldate = {2022-09-27},
	journal = {SIAM Review},
	author = {Redner, Richard A. and Walker, Homer F.},
	year = {1984},
	pages = {195--239},
}

@article{redner_note_1981,
	title = {Note on the {Consistency} of the {Maximum} {Likelihood} {Estimate} for {Nonidentifiable} {Distributions}},
	volume = {9},
	issn = {0090-5364},
	url = {http://www.jstor.org/stable/2240890},
	number = {1},
	urldate = {2022-09-27},
	journal = {The Annals of Statistics},
	author = {Redner, Richard},
	year = {1981},
	pages = {225--228},
}

@article{compiani_using_2016,
	title = {Using mixtures in econometric models: a brief review and some new results},
	volume = {19},
	issn = {1368-423X},
	shorttitle = {Using mixtures in econometric models},
	url = {http://onlinelibrary.wiley.com/doi/abs/10.1111/ectj.12068},
	doi = {10.1111/ectj.12068},
	language = {en},
	number = {3},
	urldate = {2022-09-27},
	journal = {The Econometrics Journal},
	author = {Compiani, Giovanni and Kitamura, Yuichi},
	year = {2016},
	pages = {C95--C127},
}

@article{celeux_classification_1992,
	title = {A classification {EM} algorithm for clustering and two stochastic versions},
	volume = {14},
	number = {3},
	urldate = {2022-09-27},
	journal = {Computational Statistics \& Data Analysis},
	author = {Celeux, Gilles and Govaert, Gérard},
	year = {1992},
	pages = {315--332},
}

@article{keane_career_1997,
	title = {The {Career} {Decisions} of {Young} {Men}},
	volume = {105},
	issn = {0022-3808, 1537-534X},
	url = {https://www.journals.uchicago.edu/doi/10.1086/262080},
	doi = {10.1086/262080},
	language = {en},
	number = {3},
	urldate = {2022-09-27},
	journal = {Journal of Political Economy},
	author = {Keane, Michael P. and Wolpin, Kenneth I.},
	year = {1997},
	pages = {473--522},
}

@article{bryant_large-sample_1991,
	title = {Large-sample results for optimization-based clustering methods},
	volume = {8},
	issn = {0176-4268, 1432-1343},
	url = {https://link.springer.com/10.1007/BF02616246},
	doi = {10.1007/BF02616246},
	language = {en},
	number = {1},
	urldate = {2022-09-27},
	journal = {Journal of Classification},
	author = {Bryant, Peter G.},
	year = {1991},
	pages = {31--44},
}

@article{bryant_asymptotic_1978,
	title = {Asymptotic {Behaviour} of {Classification} {Maximum} {Likelihood} {Estimates}},
	volume = {65},
	issn = {0006-3444},
	url = {http://www.jstor.org/stable/2335205},
	doi = {10.2307/2335205},
	number = {2},
	urldate = {2022-09-27},
	journal = {Biometrika},
	author = {Bryant, Peter and Williamson, John A.},
	year = {1978},
	pages = {273--281},
}

@article{same_online_2007,
	title = {An online classification {EM} algorithm based on the mixture model},
	volume = {17},
	issn = {0960-3174, 1573-1375},
	url = {http://link.springer.com/10.1007/s11222-007-9017-z},
	doi = {10.1007/s11222-007-9017-z},
	language = {en},
	number = {3},
	urldate = {2022-09-27},
	journal = {Statistics and Computing},
	author = {Samé, Allou and Ambroise, Christophe and Govaert, Gérard},
	year = {2007},
	pages = {209--218},
}

@article{chamberlain_feedback_2022,
	title = {Feedback in panel data models},
	volume = {226},
	issn = {03044076},
	url = {https://linkinghub.elsevier.com/retrieve/pii/S0304407621001937},
	doi = {10.1016/j.jeconom.2019.08.018},
	language = {en},
	number = {1},
	urldate = {2022-09-29},
	journal = {Journal of Econometrics},
	author = {Chamberlain, Gary},
	year = {2022},
	pages = {4--20},
}

@article{wooldridge_two-way_2021,
	title = {Two-{Way} {Fixed} {Effects}, the {Two}-{Way} {Mundlak} {Regression}, and {Difference}-in-{Differences} {Estimators}},
	issn = {1556-5068},
	url = {https://www.ssrn.com/abstract=3906345},
	doi = {10.2139/ssrn.3906345},
	language = {en},
	urldate = {2022-09-29},
	journal = {SSRN Electronic Journal},
	author = {Wooldridge, Jeffrey M.},
	year = {2021},
}

@inproceedings{komariah_health_2019,
	title = {Health {State} {Modeling} and {Prediction} based on {Hidden} {Markov} {Models}},
	doi = {10.1109/ICUFN.2019.8806096},
	booktitle = {2019 {Eleventh} {International} {Conference} on {Ubiquitous} and {Future} {Networks} ({ICUFN})},
	author = {Komariah, Kokoy Siti and Sin, Bong-Kee},
	year = {2019},
	pages = {245--250},
}

@article{jones_healthcare_2015,
	title = {Healthcare {Cost} {Regressions}: {Going} {Beyond} the {Mean} to {Estimate} the {Full} {Distribution}},
	volume = {24},
	issn = {1099-1050},
	shorttitle = {Healthcare {Cost} {Regressions}},
	url = {http://onlinelibrary.wiley.com/doi/abs/10.1002/hec.3178},
	doi = {10.1002/hec.3178},
	language = {en},
	number = {9},
	urldate = {2022-09-29},
	journal = {Health Economics},
	author = {Jones, Andrew M. and Lomas, James and Rice, Nigel},
	year = {2015},
	pages = {1192--1212},
}

@article{manning_estimating_2001,
	title = {Estimating log models: to transform or not to transform?},
	language = {en},
	volume = {20},
	journal = {Journal of Health Economics},
	author = {Manning, Willard G and Mullahy, John},
	year = {2001},
	pages = {461--494},
}

@article{luo_bayesian_2021,
	title = {Bayesian latent multi-state modeling for nonequidistant longitudinal electronic health records},
	volume = {77},
	issn = {1541-0420},
	url = {http://onlinelibrary.wiley.com/doi/abs/10.1111/biom.13261},
	doi = {10.1111/biom.13261},
	language = {en},
	number = {1},
	urldate = {2022-09-29},
	journal = {Biometrics},
	author = {Luo, Yu and Stephens, David A. and Verma, Aman and Buckeridge, David L.},
	year = {2021},
	pages = {78--90},
}

@article{norton_specification_2008,
	title = {Specification tests for the sample selection and two-part models},
	volume = {8},
	issn = {1387-3741, 1572-9400},
	url = {http://link.springer.com/10.1007/s10742-008-0037-8},
	doi = {10.1007/s10742-008-0037-8},
	language = {en},
	number = {4},
	urldate = {2022-09-29},
	journal = {Health Services and Outcomes Research Methodology},
	author = {Norton, Edward C. and Dow, William H. and Do, Young Kyung},
	year = {2008},
	pages = {201--208},
}

@article{higgins_bootstrap_2024,
	title = {Bootstrap {Inference} for {Fixed}-{Effect} {Models}},
	volume = {92},
	issn = {1468-0262},
	url = {https://onlinelibrary.wiley.com/doi/abs/10.3982/ECTA20712},
	doi = {10.3982/ECTA20712},
	language = {en},
	number = {2},
	urldate = {2026-01-23},
	journal = {Econometrica},
	author = {Higgins, Ayden and Jochmans, Koen},
	year = {2024},
	pages = {411--427},
}

@article{kasahara_testing_2015,
	title = {Testing the {Number} of {Components} in {Normal} {Mixture} {Regression} {Models}},
	volume = {110},
	issn = {0162-1459, 1537-274X},
	url = {https://www.tandfonline.com/doi/full/10.1080/01621459.2014.986272},
	doi = {10.1080/01621459.2014.986272},
	language = {en},
	number = {512},
	urldate = {2022-09-30},
	journal = {Journal of the American Statistical Association},
	author = {Kasahara, Hiroyuki and Shimotsu, Katsumi},
	year = {2015},
	pages = {1632--1645},
}

@book{hastie_elements_2009,
	address = {New York},
	edition = {2},
	series = {Springer series in statistics},
	title = {The {Elements} of {Statistical} {Learning}},
	language = {en},
	publisher = {Springer International Publishing},
	author = {Hastie, Trevor and Tibshirani, Robert and Friedman, Jerome},
	year = {2009},
}

@book{bishop_pattern_2006,
	address = {New York},
	series = {Information science and statistics},
	title = {Pattern {Recognition} and {Machine} {Learning}},
	isbn = {978-0-387-31073-2},
	language = {en},
	pages = {778},
	publisher = {Springer},
	author = {Bishop, Christopher M.},
	year = {2006},
}

@article{chen_likelihood_2014,
	title = {Likelihood inference in some finite mixture models},
	volume = {182},
	issn = {03044076},
	url = {https://linkinghub.elsevier.com/retrieve/pii/S0304407614000694},
	doi = {10.1016/j.jeconom.2014.04.010},
	language = {en},
	number = {1},
	urldate = {2022-10-08},
	journal = {Journal of Econometrics},
	author = {Chen, Xiaohong and Ponomareva, Maria and Tamer, Elie},
	year = {2014},
	pages = {87--99},
}

@incollection{fruhwirth-schnatter_em_2019,
	address = {Boca Raton, Florida : CRC Press, [2019]},
	edition = {1},
	title = {{EM} {Methods} for {Finite} {Mixtures}},
	isbn = {978-0-429-05591-1},
	url = {https://www.taylorfrancis.com/books/9780429508240/chapters/10.1201/9780429055911-2},
	language = {en},
	urldate = {2022-10-11},
	booktitle = {Handbook of {Mixture} {Analysis}},
	publisher = {Chapman and Hall/CRC},
	author = {Celeux, Gilles},
	editor = {Frühwirth-Schnatter, Sylvia and Celeux, Gilles and Robert, Christian P.},
	year = {2019},
	doi = {10.1201/9780429055911-2},
	pages = {21--39},
}

@book{fruhwirth-schnatter_finite_2006,
	address = {New York},
	series = {Springer series in statistics},
	title = {Finite mixture and {Markov} switching models},
	isbn = {978-0-387-32909-3},
	language = {en},
	publisher = {Springer},
	author = {Frühwirth-Schnatter, Sylvia},
	year = {2006},
}

@article{yakowitz_identifiability_1968,
	title = {On the {Identifiability} of {Finite} {Mixtures}},
	volume = {39},
	issn = {0003-4851},
	url = {http://projecteuclid.org/euclid.aoms/1177698520},
	doi = {10.1214/aoms/1177698520},
	language = {en},
	number = {1},
	urldate = {2022-10-11},
	journal = {The Annals of Mathematical Statistics},
	author = {Yakowitz, Sidney J. and Spragins, John D.},
	year = {1968},
	pages = {209--214},
}

@book{mclachlan_finite_2000,
	edition = {1},
	title = {Finite {Mixture} {Models}},
	isbn = {978-0-471-72118-5},
	url = {http://onlinelibrary.wiley.com/doi/10.1002/0471721182},
	urldate = {2022-10-11},
	publisher = {John Wiley \& Sons, Ltd},
	author = {McLachlan, Geoffrey J. and Peel, David},
	year = {2000},
	doi = {10.1002/0471721182},
}

@article{bai_panel_2009,
	title = {Panel {Data} {Models} {With} {Interactive} {Fixed} {Effects}},
	volume = {77},
	issn = {1468-0262},
	url = {https://onlinelibrary.wiley.com/doi/abs/10.3982/ECTA6135},
	doi = {10.3982/ECTA6135},
	language = {en},
	number = {4},
	urldate = {2022-10-14},
	journal = {Econometrica},
	author = {Bai, Jushan},
	year = {2009},
	pages = {1229--1279},
}

@article{teicher_identifiability_1961,
	title = {Identifiability of {Mixtures}},
	volume = {32},
	issn = {0003-4851},
	url = {http://projecteuclid.org/euclid.aoms/1177705155},
	doi = {10.1214/aoms/1177705155},
	language = {en},
	number = {1},
	urldate = {2022-10-14},
	journal = {The Annals of Mathematical Statistics},
	author = {Teicher, Henry},
	year = {1961},
	pages = {244--248},
}

@article{pollard_strong_1981,
	title = {Strong {Consistency} of {K}-{Means} {Clustering}},
	volume = {9},
	issn = {0090-5364},
	url = {https://projecteuclid.org/journals/annals-of-statistics/volume-9/issue-1/Strong-Consistency-of-K-Means-Clustering/10.1214/aos/1176345339.full},
	doi = {10.1214/aos/1176345339},
	number = {1},
	pages = {135--140},
	urldate = {2022-10-26},
	journal = {The Annals of Statistics},
	author = {Pollard, David},
	year = {1981},
}

@article{dzemski_convergence_2021,
	title = {Convergence rate of estimators of clustered panel models with misclassification},
	volume = {203},
	issn = {01651765},
	url = {https://linkinghub.elsevier.com/retrieve/pii/S016517652100121X},
	doi = {10.1016/j.econlet.2021.109844},
	language = {en},
	urldate = {2022-10-30},
	journal = {Economics Letters},
	author = {Dzemski, Andreas and Okui, Ryo},
	year = {2021},
	pages = {109844},
}

@article{wang_identifying_2021,
	title = {Identifying latent group structures in nonlinear panels},
	volume = {220},
	issn = {03044076},
	url = {https://linkinghub.elsevier.com/retrieve/pii/S0304407620301214},
	doi = {10.1016/j.jeconom.2020.04.003},
	language = {en},
	number = {2},
	urldate = {2022-10-30},
	journal = {Journal of Econometrics},
	author = {Wang, Wuyi and Su, Liangjun},
	year = {2021},
	pages = {272--295},
}

@article{liu_identification_2020,
	title = {Identification and estimation in panel models with overspecified number of groups},
	volume = {215},
	issn = {03044076},
	url = {https://linkinghub.elsevier.com/retrieve/pii/S0304407619302118},
	doi = {10.1016/j.jeconom.2019.09.008},
	language = {en},
	number = {2},
	urldate = {2022-10-30},
	journal = {Journal of Econometrics},
	author = {Liu, Ruiqi and Shang, Zuofeng and Zhang, Yonghui and Zhou, Qiankun},
	month = apr,
	year = {2020},
	pages = {574--590},
}

@article{douc_asymptotics_2001,
	title = {Asymptotics of the {Maximum} {Likelihood} {Estimator} for {General} {Hidden} {Markov} {Models}},
	volume = {7},
	issn = {13507265},
	url = {https://www.jstor.org/stable/3318493?origin=crossref},
	doi = {10.2307/3318493},
	language = {en},
	number = {3},
	urldate = {2023-01-25},
	journal = {Bernoulli},
	author = {Douc, Randal and Matias, Catherine},
	year = {2001},
	pages = {381--420},
}

@book{wooldridge_econometric_2010,
	address = {Cambridge, Massachusetts},
	edition = {2nd},
	title = {Econometric {Analysis} of {Cross} {Section} and {Panel} {Data}},
	isbn = {978-0-262-23258-6},
	publisher = {The MIT Press},
	author = {Wooldridge, Jeffrey M.},
	year = {2010},
}

@article{su_bias_2009,
	title = {Bias in 2-part mixed models for longitudinal semicontinuous data},
	volume = {10},
	issn = {1468-4357, 1465-4644},
	url = {https://academic.oup.com/biostatistics/article-lookup/doi/10.1093/biostatistics/kxn044},
	doi = {10.1093/biostatistics/kxn044},
	language = {en},
	number = {2},
	urldate = {2023-01-28},
	journal = {Biostatistics},
	author = {Su, Li and Tom, Brian D. M. and Farewell, Vernon T.},
	year = {2009},
	pages = {374--389},
}

@article{deb_demand_1997,
	title = {Demand for {Medical} {Care} by the {Elderly}: {A} {Finite} {Mixture} {Approach}},
	volume = {12},
	url = {http://www.jstor.org/stable/2285252},
	language = {en},
	number = {3,},
	journal = {Journal of Applied Econometrics},
	author = {Deb, Partha and Trivedi, Pravin K.},
	year = {1997},
	pages = {313--336},
}

@article{mclachlan_iterative_1975,
	title = {Iterative {Reclassification} {Procedure} for {Constructing} an {Asymptotically} {Optimal} {Rule} of {Allocation} in {Discriminant} {Analysis}},
	volume = {70},
	issn = {0162-1459, 1537-274X},
	url = {http://www.tandfonline.com/doi/abs/10.1080/01621459.1975.10479874},
	doi = {10.1080/01621459.1975.10479874},
	language = {en},
	number = {350},
	urldate = {2023-01-31},
	journal = {Journal of the American Statistical Association},
	author = {McLachlan, G. J.},
	year = {1975},
	pages = {365--369},
}

@article{crane_use_2010,
	title = {Use of an electronic administrative database to identify older community dwelling adults at high-risk for hospitalization or emergency department visits: {The} elders risk assessment index},
	volume = {10},
	number = {338},
	issn = {1472-6963},
	shorttitle = {Use of an electronic administrative database to identify older community dwelling adults at high-risk for hospitalization or emergency department visits},
	url = {https://bmchealthservres.biomedcentral.com/articles/10.1186/1472-6963-10-338},
	doi = {10.1186/1472-6963-10-338},
	language = {en},
	urldate = {2023-01-31},
	journal = {BMC Health Services Research},
	author = {Crane, Sarah J and Tung, Ericka E and Hanson, Gregory J and Cha, Stephen and Chaudhry, Rajeev and Takahashi, Paul Y},
	year = {2010},
}

@article{charlson_charlson_2008,
	title = {The {Charlson} comorbidity index is adapted to predict costs of chronic disease in primary care patients},
	volume = {61},
	issn = {08954356},
	url = {https://linkinghub.elsevier.com/retrieve/pii/S0895435608000309},
	doi = {10.1016/j.jclinepi.2008.01.006},
	language = {en},
	number = {12},
	urldate = {2023-01-31},
	journal = {Journal of Clinical Epidemiology},
	author = {Charlson, Mary E. and Charlson, Robert E. and Peterson, Janey C. and Marinopoulos, Spyridon S. and Briggs, William M. and Hollenberg, James P.},
	year = {2008},
	pages = {1234--1240},
}

@article{heckman_sample_1979,
	title = {Sample {Selection} {Bias} as a {Specification} {Error}},
	volume = {47},
	issn = {0012-9682},
	url = {https://www.jstor.org/stable/1912352},
	doi = {10.2307/1912352},
	number = {1},
	urldate = {2023-01-31},
	journal = {Econometrica},
	author = {Heckman, James J.},
	year = {1979},
	pages = {153--161},
}

@article{chen_consistency_2017,
	title = {Consistency of the {MLE} under {Mixture} {Models}},
	volume = {32},
	issn = {0883-4237},
	url = {https://projecteuclid.org/journals/statistical-science/volume-32/issue-1/Consistency-of-the-MLE-under-Mixture-Models/10.1214/16-STS578.full},
	doi = {10.1214/16-STS578},
	language = {en},
	number = {1},
	urldate = {2023-02-04},
	pages = {47--63},
	journal = {Statistical Science},
	author = {Chen, Jiahua},
	year = {2017},
}

@book{hsiao_analysis_2014,
	address = {New York},
	edition = {3rd},
	title = {Analysis of {Panel} {Data}},
	language = {en},
	publisher = {Cambridge University Press},
	author = {Hsiao, Cheng},
	year = {2014},
}

@article{heckman_method_1984,
	title = {A {Method} for {Minimizing} the {Impact} of {Distributional} {Assumptions} in {Econometric} {Models} for {Duration} {Data}},
	volume = {52},
	issn = {0012-9682},
	url = {https://www.jstor.org/stable/1911491},
	doi = {10.2307/1911491},
	number = {2},
	urldate = {2023-02-13},
	journal = {Econometrica},
	author = {Heckman, J. and Singer, B.},
	year = {1984},
	pages = {271--320},
}

@article{lumsdaine_estimation_2023,
	title = {Estimation of panel group structure models with structural breaks in group memberships and coefficients},
	volume = {233},
	issn = {0304-4076},
	url = {https://www.sciencedirect.com/science/article/pii/S0304407622000033},
	doi = {https://doi.org/10.1016/j.jeconom.2022.01.001},
	number = {1},
	journal = {Journal of Econometrics},
	author = {Lumsdaine, Robin L. and Okui, Ryo and Wang, Wendun},
	year = {2023},
	pages = {45--65},
}

@article{balakrishnan_statistical_2017,
	title = {Statistical guarantees for the {EM} algorithm: {From} population to sample-based analysis},
	volume = {45},
	issn = {0090-5364},
	shorttitle = {Statistical guarantees for the {EM} algorithm},
	url = {https://projecteuclid.org/journals/annals-of-statistics/volume-45/issue-1/Statistical-guarantees-for-the-EM-algorithm--From-population-to/10.1214/16-AOS1435.full},
	doi = {10.1214/16-AOS1435},
	language = {en},
	number = {1},
	pages = {77--120},
	urldate = {2023-05-12},
	journal = {The Annals of Statistics},
	author = {Balakrishnan, Sivaraman and Wainwright, Martin J. and Yu, Bin},
	year = {2017},
}

@article{neelon_bayesian_2011,
	title = {A {Bayesian} {Two}-{Part} {Latent} {Class} {Model} for {Longitudinal} {Medical} {Expenditure} {Data}: {Assessing} the {Impact} of {Mental} {Health} and {Substance} {Abuse} {Parity}},
	volume = {67},
	issn = {1541-0420},
	shorttitle = {A {Bayesian} {Two}-{Part} {Latent} {Class} {Model} for {Longitudinal} {Medical} {Expenditure} {Data}},
	url = {https://onlinelibrary.wiley.com/doi/abs/10.1111/j.1541-0420.2010.01439.x},
	doi = {10.1111/j.1541-0420.2010.01439.x},
	language = {en},
	number = {1},
	urldate = {2023-05-12},
	journal = {Biometrics},
	author = {Neelon, Brian and O'Malley, A. James and Normand, Sharon-Lise T.},
	year = {2011},
	pages = {280--289},
}

@article{provencher_decline_2015,
	title = {Decline in {Activities} of {Daily} {Living} {After} a {Visit} to a {Canadian} {Emergency} {Department} for {Minor} {Injuries} in {Independent} {Older} {Adults}: {Are} {Frail} {Older} {Adults} with {Cognitive} {Impairment} at {Greater} {Risk}?},
	volume = {63},
	issn = {1532-5415},
	shorttitle = {Decline in {Activities} of {Daily} {Living} {After} a {Visit} to a {Canadian} {Emergency} {Department} for {Minor} {Injuries} in {Independent} {Older} {Adults}},
	url = {https://onlinelibrary.wiley.com/doi/abs/10.1111/jgs.13389},
	doi = {10.1111/jgs.13389},
	language = {en},
	number = {5},
	urldate = {2023-05-13},
	journal = {Journal of the American Geriatrics Society},
	author = {Provencher, Véronique and Sirois, Marie-Josée and Ouellet, Marie-Christine and Camden, Stéphanie and Neveu, Xavier and Allain-Boulé, Nadine and Emond, Marcel and CETI on Mobility in Aging},
	year = {2015},
	pages = {860--868},
}

@article{sirven_cost_2017,
	title = {The cost of frailty in {France}},
	volume = {18},
	issn = {1618-7598, 1618-7601},
	url = {http://link.springer.com/10.1007/s10198-016-0772-7},
	doi = {10.1007/s10198-016-0772-7},
	language = {en},
	number = {2},
	urldate = {2023-05-13},
	journal = {The European Journal of Health Economics},
	author = {Sirven, Nicolas and Rapp, Thomas},
	year = {2017},
	pages = {243--253},
}

@article{hussey_continuity_2014,
	title = {Continuity and the {Costs} of {Care} for {Chronic} {Disease}},
	volume = {174},
	issn = {2168-6106},
	url = {http://archinte.jamanetwork.com/article.aspx?doi=10.1001/jamainternmed.2014.245},
	doi = {10.1001/jamainternmed.2014.245},
	language = {en},
	number = {5},
	urldate = {2023-05-13},
	journal = {JAMA Internal Medicine},
	author = {Hussey, Peter S. and Schneider, Eric C. and Rudin, Robert S. and Fox, D. Steven and Lai, Julie and Pollack, Craig Evan},
	year = {2014},
	pages = {742--748},
}

@article{chu_continuity_2012,
	title = {Continuity of {Care}, {Potentially} {Inappropriate} {Medication}, and {Health} {Care} {Outcomes} {Among} the {Elderly}: {Evidence} {From} a {Longitudinal} {Analysis} in {Taiwan}},
	volume = {50},
	issn = {0025-7079},
	shorttitle = {Continuity of {Care}, {Potentially} {Inappropriate} {Medication}, and {Health} {Care} {Outcomes} {Among} the {Elderly}},
	url = {https://journals.lww.com/00005650-201211000-00015},
	doi = {10.1097/MLR.0b013e31826c870f},
	language = {en},
	number = {11},
	urldate = {2023-05-13},
	journal = {Medical Care},
	author = {Chu, Hsuan-Yin and Chen, Chi-Chen and Cheng, Shou-Hsia},
	year = {2012},
	pages = {1002--1009},
}

@article{bice_quantitative_1977,
	title = {A {Quantitative} {Measure} of {Continuity} of {Care}},
	volume = {15},
	issn = {0025-7079},
	url = {https://www.jstor.org/stable/3763789},
	number = {4},
	urldate = {2023-05-13},
	journal = {Medical Care},
	author = {Bice, Thomas W. and Boxerman, Stuart B.},
	year = {1977},
	pages = {347--349},
}

@article{zhang_cross-validation_2015,
	title = {Cross-validation for selecting a model selection procedure},
	volume = {187},
	issn = {03044076},
	url = {https://linkinghub.elsevier.com/retrieve/pii/S0304407615000305},
	doi = {10.1016/j.jeconom.2015.02.006},
	language = {en},
	number = {1},
	urldate = {2023-07-11},
	journal = {Journal of Econometrics},
	author = {Zhang, Yongli and Yang, Yuhong},
	year = {2015},
	pages = {95--112},
}

@article{gourieroux_pseudo_1984,
	title = {Pseudo {Maximum} {Likelihood} {Methods}: {Theory}},
	volume = {52},
	issn = {0012-9682},
	shorttitle = {Pseudo {Maximum} {Likelihood} {Methods}},
	url = {https://www.jstor.org/stable/1913471},
	doi = {10.2307/1913471},
	number = {3},
	urldate = {2023-11-29},
	journal = {Econometrica},
	author = {Gourieroux, C. and Monfort, A. and Trognon, A.},
	year = {1984},
	pages = {681--700},
}

@inproceedings{bottou_convergence_1994,
	title = {Convergence {Properties} of the {K}-{Means} {Algorithms}},
	volume = {7},
	url = {https://proceedings.neurips.cc/paper/1994/hash/a1140a3d0df1c81e24ae954d935e8926-Abstract.html},
	urldate = {2023-12-08},
	booktitle = {Advances in {Neural} {Information} {Processing} {Systems}},
	author = {Bottou, Léon and Bengio, Yoshua},
	year = {1994},
}

@article{cattaneo_inference_2018,
	title = {Inference in {Linear} {Regression} {Models} with {Many} {Covariates} and {Heteroscedasticity}},
	volume = {113},
	issn = {0162-1459, 1537-274X},
	url = {https://www.tandfonline.com/doi/full/10.1080/01621459.2017.1328360},
	doi = {10.1080/01621459.2017.1328360},
	language = {en},
	number = {523},
	urldate = {2024-02-01},
	journal = {Journal of the American Statistical Association},
	author = {Cattaneo, Matias D. and Jansson, Michael and Newey, Whitney K.},
	year = {2018},
	pages = {1350--1361},
}

@article{wang_panel_2024,
	title = {Panel data models with time-varying latent group structures},
	volume = {240},
	issn = {03044076},
	url = {https://linkinghub.elsevier.com/retrieve/pii/S0304407624000319},
	doi = {10.1016/j.jeconom.2024.105685},
	language = {en},
	number = {1},
	urldate = {2024-05-01},
	journal = {Journal of Econometrics},
	author = {Wang, Yiren and Phillips, Peter C.B. and Su, Liangjun},
	year = {2024},
	pages = {105685},
}

@article{aragam_uniform_2023,
	title = {Uniform consistency in nonparametric mixture models},
	volume = {51},
	issn = {0090-5364},
	url = {https://projecteuclid.org/journals/annals-of-statistics/volume-51/issue-1/Uniform-consistency-in-nonparametric-mixture-models/10.1214/22-AOS2255.full},
	doi = {10.1214/22-AOS2255},
	language = {en},
	number = {1},
	pages = {362--390},
	urldate = {2024-05-12},
	journal = {The Annals of Statistics},
	author = {Aragam, Bryon and Yang, Ruiyi},
	year = {2023},
}

@misc{kitamura_nonparametric_2018,
	title = {Nonparametric {Analysis} of {Finite} {Mixtures}},
	url = {http://arxiv.org/abs/1811.02727},
	language = {en},
	urldate = {2024-05-12},
	publisher = {arXiv},
	author = {Kitamura, Yuichi and Laage, Louise},
	year = {2018},
	note = {arXiv preprint:1811.02727 [econ]},
}

@article{ahn_dierence_2024,
	title = {Diﬀerence in {Diﬀerences} with {Latent} {Group} {Structures}},
	language = {en},
	journal = {Working Paper, University of Pennsylvania},
	author = {Ahn, Young and Kasahara, Hiroyuki},
	year = {2024},
}

@article{dzemski_confidence_2024,
	title = {Confidence {Set} for {Group} {Membership}},
	volume = {15},
	copyright = {https://creativecommons.org/licenses/by-nc/4.0/legalcode},
	issn = {1759-7323},
	url = {https://www.econometricsociety.org/doi/10.3982/QE1975},
	doi = {10.3982/QE1975},
	language = {en},
	number = {2},
	urldate = {2024-07-22},
	journal = {Quantitative Economics},
	author = {Dzemski, Andreas and Okui, Ryo},
	year = {2024},
	pages = {245--277},
}

@article{mclachlan_finite_2019,
	title = {Finite {Mixture} {Models}},
	volume = {6},
	issn = {2326-8298, 2326-831X},
	url = {https://www.annualreviews.org/doi/10.1146/annurev-statistics-031017-100325},
	doi = {10.1146/annurev-statistics-031017-100325},
	language = {en},
	number = {1},
	urldate = {2024-07-22},
	journal = {Annual Review of Statistics and Its Application},
	author = {McLachlan, Geoffrey J. and Lee, Sharon X. and Rathnayake, Suren I.},
	year = {2019},
	pages = {355--378},
}

@article{kolle_reversing_2023,
	title = {Reversing frailty in older adults: a scoping review},
	volume = {23},
	issn = {1471-2318},
	shorttitle = {Reversing frailty in older adults},
	url = {https://bmcgeriatr.biomedcentral.com/articles/10.1186/s12877-023-04309-y},
	doi = {10.1186/s12877-023-04309-y},
	language = {en},
	number = {751},
	urldate = {2024-07-22},
	journal = {BMC Geriatrics},
	author = {Kolle, Aurélie Tonjock and Lewis, Krystina B. and Lalonde, Michelle and Backman, Chantal},
	year = {2023},
}

@article{boot_optimal_2018,
	title = {Optimal {Forecasts} from {Markov} {Switching} {Models}},
	volume = {36},
	issn = {0735-0015, 1537-2707},
	url = {https://www.tandfonline.com/doi/full/10.1080/07350015.2016.1219264},
	doi = {10.1080/07350015.2016.1219264},
	language = {en},
	number = {4},
	urldate = {2024-07-23},
	journal = {Journal of Business \& Economic Statistics},
	author = {Boot, Tom and Pick, Andreas},
	year = {2018},
	pages = {628--642},
}

@book{chen_statistical_2023,
	address = {Singapore},
	series = {{ICSA} {Book} {Series} in {Statistics}},
	title = {Statistical {Inference} {Under} {Mixture} {Models}},
	copyright = {https://www.springernature.com/gp/researchers/text-and-data-mining},
	isbn = {978-981-9961-39-9 978-981-9961-41-2},
	url = {https://link.springer.com/10.1007/978-981-99-6141-2},
	language = {en},
	urldate = {2024-08-31},
	publisher = {Springer Nature Singapore},
	author = {Chen, Jiahua},
	year = {2023},
	doi = {10.1007/978-981-99-6141-2},
}

@article{iacus_causal_2012,
	title = {Causal {Inference} without {Balance} {Checking}: {Coarsened} {Exact} {Matching}},
	volume = {20},
	copyright = {https://www.cambridge.org/core/terms},
	issn = {1047-1987, 1476-4989},
	shorttitle = {Causal {Inference} without {Balance} {Checking}},
	url = {https://www.cambridge.org/core/product/identifier/S1047198700012985/type/journal_article},
	doi = {10.1093/pan/mpr013},
	language = {en},
	number = {1},
	urldate = {2024-09-09},
	journal = {Political Analysis},
	author = {Iacus, Stefano M. and King, Gary and Porro, Giuseppe},
	year = {2012},
	pages = {1--24},
}

@article{haggerty_experienced_2013,
	title = {Experienced {Continuity} of {Care} {When} {Patients} {See} {Multiple} {Clinicians}: {A} {Qualitative} {Metasummary}},
	volume = {11},
	issn = {1544-1709, 1544-1717},
	shorttitle = {Experienced {Continuity} of {Care} {When} {Patients} {See} {Multiple} {Clinicians}},
	url = {http://www.annfammed.org/cgi/doi/10.1370/afm.1499},
	doi = {10.1370/afm.1499},
	language = {en},
	number = {3},
	urldate = {2024-10-18},
	journal = {The Annals of Family Medicine},
	author = {Haggerty, J. L. and Roberge, D. and Freeman, G. K. and Beaulieu, C.},
	year = {2013},
	pages = {262--271},
}

@article{budanova_penalized_2025,
	title = {Penalized estimation of finite mixture models},
	volume = {249},
	issn = {03044076},
	url = {https://linkinghub.elsevier.com/retrieve/pii/S0304407625000120},
	doi = {10.1016/j.jeconom.2025.105958},
	language = {en},
	urldate = {2025-07-05},
	journal = {Journal of Econometrics},
	author = {Budanova, Sofya},
	year = {2025},
	pages = {105958},
}

@article{nguyen_approximations_2019,
	title = {On approximations via convolution-defined mixture models},
	volume = {48},
	issn = {0361-0926},
	url = {https://doi.org/10.1080/03610926.2018.1487069},
	doi = {10.1080/03610926.2018.1487069},
	number = {16},
	urldate = {2025-08-11},
	journal = {Communications in Statistics - Theory and Methods},
	author = {Nguyen, Hien D. and McLachlan, Geoffrey},
	year = {2019},
	pages = {3945--3955},
}

@article{hathaway_constrained_1985,
	title = {A {Constrained} {Formulation} of {Maximum}-{Likelihood} {Estimation} for {Normal} {Mixture} {Distributions}},
	volume = {13},
	issn = {0090-5364},
	url = {https://projecteuclid.org/journals/annals-of-statistics/volume-13/issue-2/A-Constrained-Formulation-of-Maximum-Likelihood-Estimation-for-Normal-Mixture/10.1214/aos/1176349557.full},
	doi = {10.1214/aos/1176349557},
	number = {2},
	pages = {795--800},
	urldate = {2025-08-13},
	journal = {The Annals of Statistics},
	author = {Hathaway, Richard J.},
	year = {1985},
}

@article{li_non-finite_2009,
	title = {Non-finite {Fisher} information and homogeneity: an {EM} approach},
	volume = {96},
	issn = {0006-3444},
	shorttitle = {Non-finite {Fisher} information and homogeneity},
	url = {https://www.jstor.org/stable/27798833},
	number = {2},
	urldate = {2025-08-14},
	journal = {Biometrika},
	author = {Li, P. and Chen, J. and Marriott, P.},
	year = {2009},
	pages = {411--426},
}
\newpage
\appendix
\setcounter{table}{0}
\setcounter{figure}{0}
\renewcommand{\thetable}{C.\arabic{table}}
\renewcommand{\thefigure}{C.\arabic{figure}}
\section{Appendix - Proofs}\label{sec1:A}
\subsection{Lemma \ref{lem1:34}}
\begin{proof}
	Unbiasedness of all three classifiers are treated separately. For notational convenience, I drop the ``0'' subscript in the expected value $\mathbb{E}_0$ for all remaining proofs. All expectations are taken with respect to the true generating density $f_{z^{0}_{i}}(y_{i}|\theta^0_{z^{0}_{i}})$ for the $i^{th}$ observation unless stated otherwise.
	\begin{enumerate}
		\item \textbf{Joint density classifier} From Definition \ref{def1:2} and Definition \ref{def1:5}$(i)$, we can write that
		\begin{align*}
			\mathbb{E}[f_{z^{0}_{i}}(y_{i}|\theta^0_{z^{0}_{i}})] > \mathbb{E}[f_{j}(y_{i}|\theta^0_{j})] \Rightarrow \arg\underset{g\in \mathbb{G}}\max \ \mathbb{E}[f_g(y_{i}|\theta^0_{g})] = z^0_{i},
		\end{align*}
		for any value $j \ne z^0_{i}$ if $\theta^0_j = \theta^0_g \Leftrightarrow j = g$, which is true if $\boldsymbol{\mu}^0_j = \boldsymbol{\mu}^0_g \Leftrightarrow j = g$. Furthermore, we can write that
		\begin{align*}
			\arg\underset{g\in \mathbb{G}}\max \ \mathbb{E}[f_g(y_{i}|\theta^0_{g})] \Leftrightarrow  \arg\underset{g\in \mathbb{G}}\max \ \mathbb{E}\left[\log \left( \frac{f_g(y_{i}|\theta^0_{g})}{f_{z^{0}_{i}}(y_{i}|\theta^0_{z^{0}_{i}})} \right)\right],
		\end{align*}
		given that $f_{z^{0}_{i}}(y_{i}|\theta^0_{z^{0}_{i}}) >0$ for any $y_{i} \in \mathcal{Y}$ and any $z^{0}_{i} \in \mathbb{G}$ by Assumption \ref{ass1:1}(ii). Therefore, we can apply Jensen's inequality on the LHS and obtain the following
		\begin{align*}
			\mathbb{E}\left[\log \left( \frac{f_g(y_{i}|\theta^0_{g})}{f_{z^{0}_{i}}(y_{i}|\theta^0_{z^{0}_{i}})} \right)\right] &\le \log \left( \mathbb{E}\left[\frac{f_g(y_{i}|\theta^0_{g})}{f_{z^{0}_{i}}(y_{i}|\theta^0_{z^{0}_{i}})}\right]\right),\\
			&\le \log \left( \int_{\mathcal{Y}}  \frac{f_g(y_{i}|\theta^0_{g})}{f_{z^{0}_{i}}(y_{i}|\theta^0_{z^{0}_{i}})} f_{z^{0}_{i}}(y_{i}|\theta^0_{z^{0}_{i}})\upsilon(d y_{i}) \right),\\
			&\le \log \left( \int_{\mathcal{Y}}  f_g(y_{i}|\theta^0_{g})\upsilon(d y_{i}) \right),\\
			&\le \log \left( 1 \right) = 0.
		\end{align*}
		By Assumption \ref{ass1:1}$(ii)-(iii)$ and Assumption \ref{ass1:2}$(i)$, the upper bound of Jensen's inequality will be reached if and only if $j = z^{0}_{i}$, thus implying that $\arg\underset{g\in \mathbb{G}}\max \ \mathbb{E}[f_g(y_{i}|\theta^0_{g})] = z^0_{i}$.
		\item \textbf{Euclidean distance classifier} Following the same logic as above, we have that
		\begin{align*}
			\mathbb{E}[||y_{i} - \boldsymbol{\mu}^0_{z^0_{i}}||^2] < \mathbb{E}[||y_{i} - \boldsymbol{\mu}^0_{j}||^2] \Rightarrow \arg\underset{g\in \mathbb{G}}\max \ \mathbb{E}[-||y_{i} - \boldsymbol{\mu}_{g}||^2] = z^0_{i},
		\end{align*}
		for any value $j \ne z^0_{i}$. From Appendix \ref{lem1:S1}, we also have that
		\begin{align*}
			\mathbb{E}[||y_{i} - \boldsymbol{\mu}_{j}||^2] &= \sum_{l=1}^p \sigma^2_{z^0_{i},ll} + \sum_{l=1}^p (a_{jz^0_{i},l})^2 > \sum_{l=1}^p \sigma^2_{z^0_{i},ll} = \mathbb{E}[||y_{i} - \boldsymbol{\mu}_{z^0_{i}}||^2],
		\end{align*}
		given that at least one element in $\mathbf{a}_{jz^0_{i}} = \boldsymbol{\mu}^0_{j} - \boldsymbol{\mu}^0_{z^0_{i}}$ is non-zero if $j \ne z^0_{i}$.
		\item \textbf{Mahalanobis distance classifier} Following the same logic as above, we have that
		\begin{align*}
			\mathbb{E}[d^2(y_{i}, \boldsymbol{\mu}^0_{z^0_{i}},\Sigma^0_{z^0_{i}})] < \mathbb{E}[d^2(y_{i}, \boldsymbol{\mu}^0_{j},\Sigma^0_{j})] \Rightarrow \arg\underset{g\in \mathbb{G}}\max \ \mathbb{E}[-d^2(y_{i}, \boldsymbol{\mu}^0_{g},\Sigma^0_{g})] = z^0_{i},
		\end{align*}
		for any value $j \ne z^0_{i}$. From Appendices \ref{lem1:S2} and \ref{lem1:S3}, we also have that
		\begin{align*}
			\mathbb{E}[d^2(y_{i}, \boldsymbol{\mu}^0_{j},\Sigma^0_{j})] = p - 2\sum_{l=1}^p v_{jz^0_{i},ll}  + \sum_{l=1}^p\sum_{m\le l} (v_{jz^0_{i},lm})^2 +  \sum_{l=1}^p (b_{jz^0_{i},l})^2 >  p = \mathbb{E}[d^2(y_{i}, \boldsymbol{\mu}^0_{z^0_{i}},\Sigma^0_{z^0_{i}})]
		\end{align*}
		if $\Sigma_j = \Sigma_{z^0_{i}}$ since this implies that all terms $v_{jz^0_{i},lm}$ will be equal to zero for any pair $(l,m) \in \{1,2...,p\}^2$.
	\end{enumerate}
\end{proof}
\subsection{Lemma \ref{lem1:35}}
\begin{proof}
	Similar to the proof of Lemma \ref{lem1:34}, we can write that
	\begin{align*}
		\mathbb{E}[d^2(y_{i}, \boldsymbol{\mu}^0_{z^0_{i}}, \Sigma^0_{z^0_{i}})] < \mathbb{E}[d^2(y_{i}, \boldsymbol{\mu}^0_{j}, \Sigma^0_{j})] \Rightarrow \arg\underset{g\in \mathbb{G}}\max \ \mathbb{E}[-d^2(y_{i}, \boldsymbol{\mu}_g, \Sigma_g)] = z^0_{i},
	\end{align*}
	for any $j \ne z^0_{i}$. Using the results from Appendix \ref{lem1:S2} and Appendix \ref{lem1:S3}, we have that
	\begin{align*}
		\mathbb{E}[d^2(y_{i}, \boldsymbol{\mu}^0_{z^0_{i}}, \Sigma^0_{z^0_{i}})] &= p,
	\end{align*}
	and that
	\begin{align*}
		\mathbb{E}[d^2(y_{i}, \boldsymbol{\mu}^0_{j}, \Sigma^0_{j})] &= p - 2\sum_{l=1}^p v_{jz^0_{i},ll}  + \sum_{l=1}^p\sum_{m\le l} (v_{jz^0_{i},lm})^2 +  \sum_{l=1}^p (b_{jz^0_{i},l})^2,\\
		&= p - 2\sum_{l=1}^p v_{jz^0_{i},ll}  + \sum_{l=1}^p\sum_{m\le l} (v_{jz^0_{i},lm})^2,
	\end{align*}
	if $\boldsymbol{\mu}^0_j = \boldsymbol{\mu}^0_{z^0_{i}}$. This implies that the Mahalanobis distance classifier will be unbiased if
	\begin{align*}
		2\sum_{l=1}^p v_{jz^0_{i},ll} &<  \sum_{l=1}^p\sum_{m\le l} (v_{jz^0_{i},lm})^2,
	\end{align*}
	for any $j \ne z^0_{i}$. This inequality will be satisfied if $p$ is sufficiently large given that the expression on the RHS is a sum of $p(p+1)/2$ positive terms under Assumption \ref{ass1:2}. On the other hand, the expression on the LHS is a sum of $p$ terms that can be either positive, negative, or null depending on the sign of the diagonal elements of $A_{jz^0_{i}} = W_{z^0_{i}} - W_j$.
\end{proof}
\subsection{Theorem \ref{th1:1}}
\begin{proof}
	From Definition \ref{def1:5}$(iii)$, we have that
	\begin{align*}
		\mathbb{P}[\cup_{g=1}^G (z^{M}_{i(g)}(\boldsymbol{\mu}^0,\Sigma^0) \ne z^0_{ig})] &\Leftrightarrow \mathbb{P}[d^2(y_{i}, \boldsymbol{\mu}^0_{z^0_{i}}, \Sigma^0_{z^0_{i}}) \ge d^2(y_{i}, \boldsymbol{\mu}^0_{j}, \Sigma^0_{j})]
	\end{align*}
		for at least one value $j \ne z^0_{i}$. Hence, we can also write
	\begin{align*}
		\mathbb{P}[\cup_{g=1}^G (z^{M}_{i(g)}(\boldsymbol{\mu}^0,\Sigma^0) \ne z^0_{ig})] &=
		1 -  \mathbb{P}[\cap_{j\ne z^0_{i}} (d^2(y_{i}, \boldsymbol{\mu}^0_{z^0_{i}}, \Sigma^0_{z^0_{i}}) < d^2(y_{i}, \boldsymbol{\mu}^0_{j}, \Sigma^0_{j}))],\\
		&= 1 - \prod_{j\ne z^0_{i}} \mathbb{P}[ d^2(y_{i}, \boldsymbol{\mu}^0_{z^0_{i}}, \Sigma^0_{z^0_{i}}) < d^2(y_{i}, \boldsymbol{\mu}^0_{j}, \Sigma^0_{j})],\\
		&= 1 - \prod_{j\ne z^0_{i}} (1 - \mathbb{P}[ d^2(y_{i}, \boldsymbol{\mu}^0_{z^0_{i}}, \Sigma^0_{z^0_{i}}) \ge d^2(y_{i}, \boldsymbol{\mu}^0_{j}, \Sigma^0_{j})]),
	\end{align*}
	where the second equality comes from the fact that $\boldsymbol{\mu}^0_{j}$ and $\Sigma^0_{j}$ are constant for any $j \in \mathbb{G}$. 
	\par
	Using the result from Appendix \ref{lem1:S4}, we have that
	\begin{align*}
		\mathbb{P}[d^2(y_{i}, z^0_{i}) \ge d^2(y_{i}, j)] &\le \frac{\mathbb{E}[ d^2(y_{i},z^0_{i})] + \frac{1}{2}\left|\sqrt{\text{Var}[d^2(y_{i}, j)]}\right|} {\mathbb{E}[ d^2(y_{i}, j)]},
	\end{align*}
	where $d^2(y_{i}, j) = d^2(y_{i}, \boldsymbol{\mu}^0_{j}, \Sigma^0_{j})$ for any $j \in \mathbb{G}$. If we define $W_{j}$ as a $p \times p$ lower triangular matrix such that $\{\Sigma^0_{j}\}^{-1} = W_jW_j^{\top}$ for any $j \in \mathbb{G}$, and $\check{y}_{ij} = W_j^{\top}(y_{i} - \boldsymbol{\mu}^0_j)$, then
	\begin{align*}
		\text{Var}[d^2(y_{i}, j)] &= \text{Var}[(\check{y}_{ij})^{\top} \check{y}_{ij}] = \text{Var}[\sum_{l=1}^p (\check{y}_{ij,l})^2],\\
		&= \sum_{l=1}^p \sum_{m=1}^p \text{Cov}[(\check{y}_{ij,l})^2,(\check{y}_{ij,m})^2],
	\end{align*}
	where $\check{y}_{ij,l}$ is the $l^{th}$ element in the vector $\check{y}_{ij}$. This shows that $\text{Var}[d^2(y_{i}, j)]$ is a sum of $p^2$ finite constants (not necessarily all positive) under Assumption \ref{ass1:2}$(iv)$. To see this, let's apply the Cauchy-Schwarz inequality to the above covariance term
	\begin{align*}
		|\text{Cov}[(\check{y}_{ij,l})^2,(\check{y}_{ij,m})^2]| \le (\text{Var}[(\check{y}_{ij,l})^2]\text{Var}[(\check{y}_{ij,m})^2])^{1/2},
	\end{align*}
	where each variance term is bounded from above if $\mathbb{E}[(y_{il})^4] < \infty$ (given that all other lower-order moments will be bounded as well). This comes from the fact that
	\begin{align*}
		\mathbb{E}[(\check{y}_{ij,l})^4] = \mathbb{E}[(\sum_{m=1}^p w_{j,ml}(y_{im}-\mu^0_{j,m}))^4],
	\end{align*}
	where $w_{j,ml}$ is the element located at the $m^{th}$ row and the $j^{th}$ column of the matrix $W_j$. This implies that there exists a finite constant $M>0$ such that $\text{Var}[d^2(y_{i}, j)] \le p^2M$. Note that $p^2M$ is a sharp bound if all covariance terms are strictly positive. Combining the former inequality and the result of Appendix \ref{lem1:S3} with the first inequality leads to
	\begin{align*}
		\mathbb{P}[d^2(y_{i}, z^0_{i}) \ge d^2(y_{i}, j)] &\le \frac{p(1+\frac{1}{2}\sqrt{M})}{2 \sum_{l=1}^pd_{jz^0_{i},ll}  + \sum_{l=1}^p\sum_{m\le l} (v_{jz^0_{i},lm})^2 - p + \sum_{l=1}^p(b_{jz^0_{i},l})^2},
	\end{align*}
	where $d_{jz^0_{i},ll} > 0$. Because all terms in the denominators are positive except $-p$, there exists a constant $0 < K < \infty$ such that
	\begin{align*}
		2 \sum_{l=1}^p d_{jz^0_{i},ll}  +  \sum_{l=1}^p \sum_{m\le l} (v_{jz^0_{i},lm})^2 +  \sum_{l=1}^p (b_{jz^0_{i},l})^2 &\ge 3Kp  +  \frac{Kp(p+1)}{2} = p\left(\frac{K(p+1)}{2} +3K\right).
	\end{align*}
	Note that $K$ also accommodates for the eventuality that a nonvanishing proportion of values in $V_{jz^0_{i}} = W_{z^0_{i}}^{-1} A_{jz^0_{i}}$ might be exactly equal to zero under Assumption \ref{ass1:2}$(v)$ due to null elements in $A_{jz^0_{i}}$, in which case the term $\sum_{l=1}^p \sum_{m\le l} (v_{jz^0_{i},lm})^2$ still goes to infinity but at a slower rate. That said, we can write
	\begin{align*}
		\frac{p(1+\frac{1}{2}\sqrt{M})}{2 \sum_{l=1}^pd_{jz^0_{i},ll}  + \sum_{l=1}^p\sum_{m\le l} (v_{jz^0_{i},lm})^2 - p + \sum_{l=1}^p(b_{jz^0_{i},l})^2} &\le \frac{p(1+\frac{1}{2}\sqrt{M})}{p\left(\frac{K(p+1)}{2} +3K-1\right)},\\
		& = O(p^{-1}),
	\end{align*}
	which leads to
	\begin{align*}
		\mathbb{P}[d^2(y_{i}, z^0_{i}) \ge d^2(y_{i}, j)] & = O(p^{-1}),
	\end{align*}
	for any $j \ne z^0_{i}$ since $K>0$ is fixed as $p \to \infty$. Note also that if Assumption \ref{ass1:2}$(v)$ is violated and $W_j \ne W_g$ for a \textit{vanishing} proportion of the elements in each matrix, then the term $\sum_{l=1}^p\sum_{m\le l} (v_{jz^0_{i},lm})^2$ cannot be lower bounded by $Kp(p+1)/2$ for all values of $p > 0$ and a given value of $K$. If Assumption \ref{ass1:2}$(v)$ is not violated, then there exists a constant $0 < \tilde{K} \le 1$ such that
	\begin{align*}
		\mathbb{P}[\cup_{g=1}^G (z^{M}_{i(g)}(\boldsymbol{\mu}^0,\Sigma^0) \ne z^0_{ig})] 
		&\le 1 - (1 - \tilde{K}p^{-1})^{(G-1)},
	\end{align*}
	for any discrete $p>0$. If $G$ is a finite constant, then $\mathbb{P}[\cup_{g=1}^G (z^{M}_{i(g)}(\boldsymbol{\mu}^0,\Sigma^0) \ne z^0_{ig})] = O(p^{-1})$ since 
	\begin{align*}
		p(1 - (1 - \tilde{K}p^{-1})^{(G-1)}) = p - p\left(1-\frac{(G-1)\tilde{K}}{p} + O(p^{-2})\right) \to 	(G-1)\tilde{K} \ \text{as $p \to \infty$}.
	\end{align*}
\end{proof}
\subsection{Theorem \ref{th1:2}}
\begin{proof}
	From Definition \ref{def1:3} and the proof of Theorem \ref{th1:1}, we can write that
	\begin{align*}
		\mathbb{P}[\cup_{i=1}^N \cup_{g=1}^G (z^{M}_{i(g)}(\boldsymbol{\mu}^0,\Sigma^0) \ne z^0_{ig})] &=  1- \mathbb{P}[\cap_{i=1}^N  \cap_{g=1}^G (z^{M}_{i(g)}(\boldsymbol{\mu}^0,\Sigma^0) = z^0_{ig})],\\
		&=1- \prod_{i=1}^N  \mathbb{P}[\cap_{g=1}^G (z^{M}_{i(g)}(\boldsymbol{\mu}^0,\Sigma^0) = z^0_{ig})],\\
		&= 1- \prod_{i=1}^N  (1 - \mathbb{P}[\cup_{g=1}^G (z^{M}_{i(g)}(\boldsymbol{\mu}^0,\Sigma^0) \ne z^0_{ig})]),\\
		&\le 1- (1 - \tilde{K}p^{-1})^{N(G-1)},
	\end{align*}
	with $0 < \tilde{K} \le 1$ for any discrete $p>0$. The second equality comes from the fact that all $y_{i}$ are independent from each other by Assumption \ref{ass1:2}$(iv)$. Relaxing this assumption to introduce cross-sectional dependence in $y_{i}$ would lead to
	\begin{align*}
		\mathbb{P}[\cap_{i=1}^N  \cap_{g=1}^G (z^{M}_{i(g)}(\boldsymbol{\mu}^0,\Sigma^0) = z^0_{ig})] &= \prod_{i=1}^N  \mathbb{P}[\cap_{g=1}^G (z^{M}_{i(g)}(\boldsymbol{\mu}^0,\Sigma^0) = z^0_{ig})|\cap_{j\ne i}\cap_{g=1}^G (z^{M}_{j(g)}(\boldsymbol{\mu}^0,\Sigma^0) = z^0_{jg})],
	\end{align*}
	which may be different from $\prod_{i=1}^N  \mathbb{P}[\cap_{g=1}^G (z^{M}_{i(g)}(\boldsymbol{\mu}^0,\Sigma^0) = z^0_{ig})]$. However, relaxing this independence assumption does not necessarily invalidate the conclusion of the theorem.
	\par
	If we assume that $p = aN^{b}$ with $(a,b) \in \mathbb{R}^2_{>0}$ and that $G>1$ is constant, then
	\begin{align*}
		\lim_{N \to \infty} 1-(1-\tilde{K}(aN^{b})^{-1})^{N(G-1)}= \begin{cases}
			1  &\text{if $0<b<1$,}\\
			1-\exp^{-\tilde{K}(G-1)/a}  &\text{if $b=1$,} \\
			0 &\text{if $b>1$},
		\end{cases}
	\end{align*}
	which implies that $z^{M}_{ig}(\boldsymbol{\mu}^0,\Sigma^0)$ is a uniformly consistent classifier if the number of covariates $p$ increases at a faster rate than the number of observations $N$ since the upper bound on the probability that at least one observation in the sample is misclassified will be equal to zero if and only if $p/N \to \infty$ as $N,p \to \infty$.
\end{proof}
\subsection{Corollary \ref{cor1:34}}
\begin{proof}
	From Definition \ref{def1:4}, we can write that
	\begin{align*}
		\hat{E}_{Np}(\theta^0) &= \sum_{i=1}^N \sum_{g=1}^G \frac{\mathbbm{1}[z^0_{ig} \ne z^{M}_{i(g)}(\boldsymbol{\mu}^0,\Sigma^0)]}{2N},\\
		&= \sum_{i=1}^N  \frac{\mathbbm{1}[z^0_{i} \ne z^{M}_{i}(\boldsymbol{\mu}^0,\Sigma^0)]}{N}\xrightarrow{p} \mathbb{P}[z^0_{i} \ne z^{M}_{i}(\boldsymbol{\mu}^0,\Sigma^0)] \ \text{as $N \to \infty$},
	\end{align*}
	which is equivalent to $\mathbb{P}[\cup_{g=1}^G (z^{M}_{i(g)}(\boldsymbol{\mu}^0,\Sigma^0) \ne z^0_{ig})]$. By the proof of Theorem \ref{th1:1}, we know that
	\begin{align*}
		\mathbb{P}[\cup_{g=1}^G (z^{M}_{i(g)}(\boldsymbol{\mu}^0,\Sigma^0) \ne z^0_{ig})] 
		&\le 1 - (1 - \tilde{K}p^{-1})^{(G-1)}.
	\end{align*}
	Assuming that $G = ap^b$ with $(a,b) \in \mathbb{R}^2_{>0}$, we obtain that
	\begin{align*}
		\lim_{p \to \infty} 1 - (1 - \tilde{K}p^{-1})^{(ap^b-1)} = \begin{cases}
			0  &\text{if $0<b<1$},\\
			1-\exp^{-\tilde{K}a}  &\text{if $b=1$},\\
			1  &\text{if $b>1$},
		\end{cases} 
	\end{align*}
	which implies that $p$ has to grow at a strictly higher rate than $G$ for the consistency of $z^{M}_{i(g)}(\boldsymbol{\mu}^0,\Sigma^0)$ to be generally verified.
\end{proof}
\subsection{Corollary \ref{cor1:35}}
\begin{proof}
	The first part of the corollary is a direct consequence of Theorems \ref{th1:1} and \ref{th1:2} when $N$ and $G$ are held constant. In this case, the probability that at least one observation in the sample gets misclassified when the classifier is evaluated at the true parameter values converges almost surely to zero as $p \to \infty$. This implies that the estimates provided by the maximization of the M-CL function are exactly the same as those produced by the maximization of the infeasible complete-data log likelihood when $p \to \infty$. If the two sets of estimates are the same, this implies that they share the same bias as long as both objective functions are identical for any sample size, which is verified if $p/N \to \infty$ as $N,p \to \infty$ by Theorem \ref{th1:2}, and which corresponds to the second part of the corollary.
\end{proof}
\subsection{Theorem \ref{the1:33}}
\begin{proof}
	Using the same strategy as in the proof of Theorem \ref{th1:1} and the results from Appendices \ref{lem1:S1} and \ref{lem1:S4}, we can write that
	\begin{align*}
		\mathbb{P}[||y_{i} - \boldsymbol{\mu}^0_{z^0_{i}}||^2 \ge ||y_{i} - \boldsymbol{\mu}^0_{j}||^2] &\le \frac{\mathbb{E}[||y_{i} - \boldsymbol{\mu}^0_{z^0_{i}}||^2] + \frac{1}{2}\left|\sqrt{\text{Var}[||y_{i} - \boldsymbol{\mu}^0_{j}||^2]}\right|}{\mathbb{E}[||y_{i} - \boldsymbol{\mu}^0_{j}||^2]},\\
		&\le \frac{\sum_{l=1}^p \sigma^2_{z^0_{i},ll}+\frac{1}{2}\left|\sqrt{\text{Var}[||y_{i} - \boldsymbol{\mu}^0_{j}||^2]}\right|}{\sum_{l=1}^p \sigma^2_{z^0_{i},ll} + \sum_{l=1}^p (a_{jz^0_{i},l})^2} .
	\end{align*}
	As in the proof of Theorem \ref{th1:1}, we can write the variance term as follows
	\begin{align*}
		\text{Var}[||y_{i} - \boldsymbol{\mu}^0_{j}||^2] &= \text{Var}[\sum_{l=1}^p (x_{il} - \mu^0_{j,l})^2],\\
		&= \sum_{l=1}^p \sum_{m=1}^p\text{Cov}[ (y_{il} - \mu^0_{j,l})^2,(y_{im} - \mu^0_{j,m})^2],
	\end{align*}
	which is a sum of $p^2$ finite terms under Assumption \ref{ass1:2}$(iv)$. This means there exists a finite constant $M>0$ such that $\text{Var}[||y_{i} - \boldsymbol{\mu}^0_{j}||^2] \le p^2M$. This leads to
	\begin{align*}
		&\mathbb{P}[||y_{i} - \boldsymbol{\mu}^0_{z^0_{i}}||^2 \ge ||y_{i} - \boldsymbol{\mu}^0_{j}||^2] \le 	{c}_{jz^0_{i},p},
	\end{align*}
	where
	\begin{align*}
		&{c}_{jz^0_{i},p} = \left(1+\frac{p\sqrt{M}}{2\sum_{l=1}^p \sigma^2_{z^0_{i},ll}}\right) \bigg/ \left(1 + \frac{\sum_{l=1}^p (a_{jz^0_{i},l})^2}{ \sum_{l=1}^p \sigma^2_{z^0_{i},ll}}\right).
	\end{align*}
	By Assumption \ref{ass1:2}$(iv)$, there exists a strictly positive value $\bar{\sigma}^2_{z^0_{i},p} = \frac{1}{p}\sum_{l=1}^p \sigma^2_{z^0_{i},ll}$ that is finite for any $p \in \mathbb{N}_{>0}$. Therefore, we can write that
	\begin{align*}
		{c}_{jz^0_{i},p} &= (1+Q_{i,p})/(1 + \tilde{a}_{jz^0_{i},p}).
	\end{align*}
	where $ 0<Q_{i,p}< \infty$ with $ Q_{i,p}= \frac{\sqrt{M}}{2\bar{\sigma}^2_{z^0_{i},p}}$, and where $\tilde{a}_{jz^0_{i},p} = \frac{\sum_{l=1}^p (a_{jz^0_{i},l})^2}{p\bar{\sigma}^2_{z^0_{i},p}}$. It is then easy to see that ${c}_{jz^0_{i},p} \to 0$ as $p \to \infty$ if and only if $\tilde{a}_{jz^0_{i},p} \to \infty$ as $p \to \infty$. This allows us to write
	\begin{align*}
		\mathbb{P}[\cup_{g=1}^G (z^{E}_{i(g)}(\boldsymbol{\mu}^0) \ne z^0_{ig})] &=\mathbb{P}[||y_{i} - \boldsymbol{\mu}^0_{z^0_{i}}||^2 \ge ||y_{i} - \boldsymbol{\mu}^0_{j}||^2] \ \ \text{for at least one $j\ne z^0_{i}$},\\
		&=1 - \prod_{j \ne z^0_{i}} \mathbb{P}[||y_{i} - \boldsymbol{\mu}^0_{z^0_{i}}||^2 < ||y_{i} - \boldsymbol{\mu}^0_{j}||^2],\\
		&=1 - \prod_{j \ne z^0_{i}} (1- \mathbb{P}[||y_{i} - \boldsymbol{\mu}^0_{z^0_{i}}||^2 \ge ||y_{i} - \boldsymbol{\mu}^0_{j}||^2]),\\
		&\le 1 - \prod_{j \ne z^0_{i}} (1- {c}_{jz^0_{i},p}).
	\end{align*}
	Note that when $\tilde{a}_{jz^0_{i},p} \to 0$, then both $||y_{i} - \boldsymbol{\mu}^0_{z^0_{i}}||^2 \to ||y_{i} - \boldsymbol{\mu}^0_{j}||^2$ and $M$ converges to zero as a consequence of Lemma \ref{lem1:S4}. Therefore, we have that ${c}_{jz^0_{i},p} \to 1$ as $\boldsymbol{\mu}^0_g \to \boldsymbol{\mu}_j^0$ for any pair $(g,j) \in \mathbb{G}^2$ and any $p>0$. Note also that the inequality will become an equality if $||y_{i} - \boldsymbol{\mu}^0_{z^0_{i}}||^2 = ||y_{i} - \boldsymbol{\mu}^0_{j}||^2$ for any $j \ne z^0_{i}$ and any $y_i \in \mathcal{Y}$.
\end{proof}
\subsection{Corollary \ref{cor1:36}}
\begin{proof}
	Akin to the proof of Corollary \ref{cor1:34}, we can write that
	\begin{align*}
		\hat{E}_{Np}(\theta^0) &= \sum_{i=1}^N \frac{\mathbbm{1}[z^0_{i} \ne z^{E}_{i}(\boldsymbol{\mu}^0)]}{N}\xrightarrow{p} \mathbb{P}[z^0_{i} \ne z^{E}_{i}(\boldsymbol{\mu}^0)] \ \text{as $N \to \infty$.}
	\end{align*}
	By Definition \ref{def1:5}$(ii)$ and as shown in the proof of Theorem \ref{the1:33}, we know that
	\begin{align*}
		\mathbb{P}[z^0_{i} \ne z^{E}_{i}(\boldsymbol{\mu}^0)] \le 1 - \prod_{j \ne z^0_{i}} (1-{c}_{jz^0_{i},p}) \to d_{i} \ \text{as $p \to \infty$},
	\end{align*}
	with $d_{i} \ge 0$ and where ${c}_{jz^0_{i},p}$ is defined as in the proof of Theorem \ref{the1:33}. Therefore, we have that $0 \le \mathbb{P}[z^0_{i} \ne z^{E}_{i}(\boldsymbol{\mu}^0)] \le d_{i}$ as $p \to \infty$, where $d_{i}$ will be equal to zero and lead to a consistent classifier if the ratio $\frac{||\boldsymbol{\mu}^0_g - \boldsymbol{\mu}^0_{j}||^2}{\text{tr}(\Sigma^0_{g})} \to \infty$ as $p\to \infty$ for any pair $(g,j) \in \mathbb{G}\times \mathbb{G}\backslash g$.
\end{proof}
\subsection{Lemma \ref{lem1:cem}}
\begin{proof}
	To prove this lemma, we need to show that the C-ML function never decreases when $l^{CM}(\theta^{(k)},\mathbf{z}^{(k)}) \to l^{CM}(\theta^{(k+1)},\mathbf{z}^{(k)})$ and when $l^{CM}(\theta^{(k+1)},\mathbf{z}^{(k)}) \to l^{CM}(\theta^{(k+1)},\mathbf{z}^{(k+1)})$, where $\mathbf{z}^{(k)}$ denotes the assignment values produced by the joint density classifier at the $k^{th}$ iteration of the C-EM algorithm.
	\par
	Based on Definition \ref{def1:0}$(iii)$, we can write that
	\begin{align*}
		l^{CM}(\theta^{(k+1)},\mathbf{z}^{(k)}) &= \sum_{i=1}^N\sum_{g=1}^G z_{ig}^{(k)} \log(f_g(y_i|\theta^{(k+1)}_g)),\\
		&\ge \sum_{i=1}^N\sum_{g=1}^G z_{ig}^{(k)} \log(f_g(y_i|\theta^{(k)}_g)),
	\end{align*}
	since $\theta^{(k+1)} = \arg \max_{\{\theta_g\}_{g=1}^G \in \Theta} \sum_{i=1}^N\sum_{g=1}^G z_{ig}^{(k)} \log(f_g(y_i|\theta_g))$ by definition of $\theta^{(k+1)}$. Based on the same definition, we also have that
	\begin{align*}
		l^{CM}(\theta^{(k+1)},\mathbf{z}^{(k+1)}) &= \sum_{i=1}^N\sum_{g=1}^G z_{ig}^{(k+1)} \log(f_g(y_i|\theta^{(k+1)}_g)),\\
		&\ge \sum_{i=1}^N\sum_{g=1}^G z_{ig}^{(k)} \log(f_g(y_i|\theta^{(k+1)}_g)),
	\end{align*}
	since $z_{ig}^{(k+1)} = 1 \Leftrightarrow \arg\max_{j \in \mathbb{G}} f_j(y_i|\theta^{(k+1)}_j) = g$ and $z_{ig}^{(k+1)}$ selects the component density that maximizes the (log) density value across $\mathbb{G}$. The same logic holds if $f_g(y_i|\theta^{(k)}_g)$ and $f_g(y_i|\theta^{(k+1)}_g)$ are respectively replaced by $f_g(y_i,x_i|\theta^{(k)}_g,\psi^{(k)}_g)$ and $f_g(y_i,x_i|\theta^{(k+1)}_g,\psi^{(k+1)}_g)$ everywhere and the M-step includes the computation of $\psi^{(k+1)}$. However, using the Mahalanobis distance classifier at each classification step may decrease the C-ML function between two consecutive iterations since it may not select the value $g$ that maximizes either $f_g(y_i|\theta^{(k+1)}_g)$ or $f_g(y_i,x_i|\theta^{(k+1)}_g,\psi^{(k+1)}_g)$.
\end{proof}
\subsection{Theorem \ref{the1:35}}
\begin{proof}
	Let's denote any consistent classifier by $z_{ig}(\theta^0)$. The first M-step of the C-EM algorithm can be written as follows~:
	\begin{align*}
		\hat{\theta}^{(1)} &:= \arg \max_{\{\theta_g\}_{g=1}^G \in \Theta} N^{-1}\sum_{i=1}^N \sum_{g=1}^G z_{ig}(\theta^{(0)}) \log( f_g(y_{i}|\theta_g)),
	\end{align*}
	Given that $z_{ig}(\theta)$ is binary and that $h_j(y_i|\theta_j)$ is a continuous function of $\theta_j$ for any $j \in \mathbb{G}$, there exists a ball centered at $\theta^0$, denoted $\mathcal{A}_{\theta^0}$, where all $\theta^* \in \mathcal{A}_{\theta^0}$ are such that $z_{ig}(\theta^*) = z_{ig}(\theta^0)$ for all pairs $(i,g) \in [N]\times \mathbb{G}$ when $N$ is held constant, and where $z_{ig}(\theta^*) = z_{ig}(\theta^0)$ for almost every pair $(i,g) \in [N]\times \mathbb{G}$ when $N \to \infty$. If $\theta^{(0)} \in \mathcal{A}_{\theta^0}$, then we have
	\begin{align*}
		N^{-1}\sum_{i=1}^N \sum_{g=1}^G z_{ig}(\theta^{(0)}) \log( f_g(y_{i}|\theta_g)) \equiv N^{-1} \sum_{i=1}^N \sum_{g=1}^G z_{ig}(\theta^0) \log( f_g(y_{i}|\theta_g)).
	\end{align*}
	As $p\to \infty$, the consistency of $z_{ig}(\theta^0)$ implies that $z_{ig}(\theta^{(0)}) \xrightarrow{p} 1 \Leftrightarrow g= z^0_{i}$ and $z_{ig}(\theta^{(0)}) \xrightarrow{p} 0 \Leftrightarrow g\ne z^0_{i}$  for all pairs $(i,g) \in [N]\times \mathbb{G}$ since $\theta^{(0)} \in \mathcal{A}_{\theta^0}$. Therefore, we can write that
	\begin{align*}
		\hat{\theta}^{(1)} &\xrightarrow{p} \arg \max_{\{\theta_g\}_{g=1}^G \in \Theta} N^{-1} \sum_{i=1}^N \sum_{g=1}^G \mathbbm{1}[z^0_{i}=g] \log(f_g(y_{i}|\theta_{g})) \ \ \text{as $p \to \infty$},
	\end{align*}
	by the continuous mapping theorem. We can then apply the weak law of large numbers (WLLN) to obtain the following
	\begin{align*}
		\hat{\theta}^{(1)}&\xrightarrow{p} \arg \max_{\{\theta_g\}_{g=1}^G \in \Theta} \sum_{g=1}^G\mathbb{E}_0[\mathbbm{1}[z^0_{i}=g] \log(f_g(y_{i}|\theta_{g}))] \ \ \text{as $N,p \to \infty$},\\
		&= \arg \max_{\{\theta_g\}_{g=1}^G \in \Theta} \sum_{g=1}^G \int_{\mathcal{Y}:z^0_{i}=g} \log(f_g(y_{i}|\theta_{g}))f(y_{i}|\xi^0) \upsilon(dy_i),\\
		&= \arg \max_{\{\theta_g\}_{g=1}^G \in \Theta} \sum_{g=1}^G \int_{\mathcal{Y}} \log(f_g(y_{i}|\theta_{g})) f_g(y_{i}|\theta^0_{g})   \upsilon(dy_i),\\
		&= \arg \max_{\{\theta_g\}_{g=1}^G \in \Theta} \sum_{g=1}^G \mathbb{E}_{0,g}[\log(f_g(y_{i}|\theta_{g}))]= \theta^0,
	\end{align*}
	where the last equality comes from the fact that $\arg \max_{\{\theta_g\}_{g=1}^G \in \Theta} \sum_{g=1}^G \mathbb{E}_{0,g}[\log(f_g(y_{i}|\theta_{g}))] = \theta^0$ by consistency of regular single-component MLE under Assumption \ref{ass1:1}$(iii)$, and where $\mathbb{E}_{0,g}[\cdot]$ corresponds to the expected value with respect to the true $g^{th}$ component density. If $p$ and $N$ are not large enough, then it may that $\mathbf{z}(\hat{\theta}^{(1)}) \ne \mathbf{z}(\hat{\theta}^{(0)})$ and the algorithm has not converged. If $\mathbf{z}(\hat{\theta}^{(1)}) = \mathbf{z}(\hat{\theta}^{(2)})$ with $\hat{\theta}^{(2)} \notin \mathcal{A}_{\theta^0}$, then the algorithm has converged after two iterations, but the estimated parameters $\hat{\theta}^{(2)}$ may be trapped in a local maximum and not converge to $\theta^0$ as $N,p \to \infty$.
	However, consistency of $\hat{\theta}^{(1)}$ and continuity of $h_j(y_i|\theta_j)$ guarantees that there exists fixed values $(p^*,N^*)$ such that $\mathbf{z}(\hat{\theta}^{(1)}) = \mathbf{z}(\theta^0)$ with $\mathbf{z}(\hat{\theta}^{(1)}) \in \mathcal{A}_{\theta^0}$ for any pair $(p,N) > (p^*,N^*)$, which implies both the consistency of $\mathbf{z}(\hat{\theta}^{(1)})$ and the convergence of the algorithm.
	\par
	If $\theta^{(0)} \notin \mathcal{A}_{\theta^0}$, then $z_{ig}(\theta^{(0)}) \ne z_{ig}(\theta^0)$ for all pairs $(i,g) \in [N]\times \mathbb{G}$ and additional iterations of the algorithm might be required in order to reach $\theta^{(k)} \in \mathcal{A}_{\theta^0}$ such that $z_{ig}(\hat{\theta}^{(k)}) = z_{ig}(\theta^0)$ for all pairs $(i,g) \in [N]\times \mathbb{G}$. In this case, applying the same logic as above will lead to the convergence of the algorithm, meaning that $\hat{\theta}^{(k)} = \hat{\theta}^{(k-1)}$ such that $\hat{\theta}^{(k)} \xrightarrow{p} \theta^0$ as $N,p \to \infty$.
	\par
	Finally, it might be the case that $\theta^{(0)} \notin \mathcal{A}_{\theta^0}$ with $\hat{\theta}^{(k)} = \hat{\theta}^{(k-1)}$, but without $\hat{\theta}^{(k)} \xrightarrow{p} \theta^0$ as $N,p \to \infty$. Then the algorithm has converged to a local maximum of the C-ML objective function that does not coincide with the global maximum of the C-ML function as $N,p \to \infty$. This implies that other values of $\theta^{(0)}$ have to be used for the C-EM algorithm to reach the global maximum of the C-ML objective.
\end{proof}
\subsection{Theorem \ref{the1:34}}\label{app:A18}
\begin{proof}
	The derivation of the asymptotic distribution of the C-EM algorithm is identical to the derivation of the asymptotic distribution of any ML estimator with the inclusion of the classifier $z_{ig}(\theta)$ in the definition of $s_{ig}(\theta)$. See Section 12.3 of \cite{wooldridge_econometric_2010} for a general proof. The scaling factor $\sqrt{n_g(\hat{\theta}^{(k)})}$ comes from the fact that only a non-vanishing proportion of the observations (by Assumption \ref{ass1:3}$(iii)$) contributes to the estimation of the parameters $\hat{\theta}_{g}^{(k)}$ for each $g \in \mathbb{G}$. 
\end{proof}
\subsection{Lemma S.1}\label{lem1:S1}
\textit{Let Assumptions \ref{ass1:1}(ii)-(iii) and \ref{ass1:2} hold. Then
	\begin{align*}
		\mathbb{E}[||x_{i} - \boldsymbol{\mu}^0_j||^2] =  \sum_{l=1}^p \sigma^2_{z^0_{i},ll} + \sum_{l=1}^p (a_{jz^0_{i},l})^2,
	\end{align*}
	where $a_{jg,l}$ is the $l^{th}$ element of the $p$-sized column-vector $\boldsymbol{a}_{jg} := \boldsymbol{\mu}^0_j-\boldsymbol{\mu}^0_{g}$ for any pair $(g,j) \in \mathbb{G}\times \mathbb{G}\backslash g$.}
\begin{proof}
	Given that $\mathbb{E}[||x_{i} - \boldsymbol{\mu}^0_j||^2]$ is a scalar, we have that
	\begin{align*}
		\mathbb{E}[||x_{i} - \boldsymbol{\mu}^0_j||^2] &= \text{tr}(\mathbb{E}[||x_{i} - \boldsymbol{\mu}^0_j||^2]),\\
		&= \text{tr}(\mathbb{E}[(x_{i} - \boldsymbol{\mu}^0_j)^{\top}(x_{i} - \boldsymbol{\mu}^0_j)]),\\
		&= \text{tr}(\mathbb{E}[(x_{i} - \boldsymbol{\mu}^0_{z^0_{i}} - \boldsymbol{a}_{jz^0_{i}})(x_{i} - \boldsymbol{\mu}^0_{z^0_{i}} - \boldsymbol{a}_{jz^0_{i}})^{\top}]),\\
		&= \text{tr}(\Sigma^0_{z^0_{i}}) - 2\times\text{tr}(\mathbb{E}[(x_{i} - \boldsymbol{\mu}^0_{z^0_{i}})\boldsymbol{a}_{jz^0_{i}}^{\top}]) + \text{tr}(\boldsymbol{a}_{jz^0_{i}}\boldsymbol{a}_{jz^0_{i}}^{\top}),\\
		&= \sum_{l=1}^p \sigma^2_{z^0_{i},ll} + \sum_{l=1}^p (a_{jz^0_{i},l})^2,
	\end{align*}
	given that $\mathbb{E}[x_{i}] = \boldsymbol{\mu}^0_{z^0_{i}}$ and that $\boldsymbol{a}_{jz^0_{i}}$ is non-random, and where $\text{tr}(\cdot)$ denotes the trace operator.
\end{proof}
\subsection{Lemma S.2}\label{lem1:S2}
\textit{Let Assumptions \ref{ass1:1}(ii)-(iii) and \ref{ass1:2} hold. Then
	\begin{align*}
		\text{tr}(\Sigma_j^{-1} \Sigma^0_{z^0_{i}}) = 2\sum_{l=1}^p d_{jz^0_{i},ll}  + \sum_{l=1}^p\sum_{m\le l} (v_{jz^0_{i},lm})^2 - p,
	\end{align*}
	where $v_{jg,lm}$ corresponds to the entry located at the $l^{th}$ row and at the $m^{th}$ column of $V_{jg} = W_{g}^{-1} A_{jg}$, with $W_g$ and $A_{jg}$ both corresponding to $p \times p$ lower triangular matrices for any pair $(g,j) \in \mathbb{G}\times \mathbb{G}\backslash g$.}
\begin{proof}
	Given that $\Sigma^0_{j}$ is a $p \times p$ symmetric, positive-definite matrix, there exists a lower triangular matrix $W_j$ such that $\{\Sigma^0_{j}\}^{-1} = W_jW_j^{\top}$ for each $j \in \mathbb{G}$. If we define the lower triangular matrix $A_{jz^0_{i}} := W_{z^0_{i}} - W_j$ for any $j \ne z^0_{i}$, we have that	
	\begin{align*}
		\text{tr}(\{\Sigma^0_{j}\}^{-1} \Sigma^0_{z^0_{i}}) & = \text{tr}( W_jW_j^{\top} \Sigma^0_{z^0_{i}} ) ,\\
		& = \text{tr}( (W_{z^0_{i}} - A_{jz^0_{i}})(W_{z^0_{i}} - A_{jz^0_{i}})^{\top}\Sigma^0_{z^0_{i}} ),\\	
		& = p + \text{tr}(A_{jz^0_{i}}A_{jz^0_{i}}^{\top}\Sigma^0_{z^0_{i}} - A_{jz^0_{i}}W_{z^0_{i}}^{\top}\Sigma^0_{z^0_{i}}- W_{z^0_{i}}A_{jz^0_{i}}^{\top}\Sigma^0_{z^0_{i}}),\\
		& = p + \text{tr}(A_{jz^0_{i}}A_{jz^0_{i}}^{\top}\{W_{z^0_{i}}^{\top}\}^{-1} W_{z^0_{i}}^{-1} - A_{jz^0_{i}}W_{z^0_{i}}^{\top} \{W_{z^0_{i}}^{\top}\}^{-1} W_{z^0_{i}}^{-1} - W_{z^0_{i}}A_{jz^0_{i}}^{\top}\{W_{z^0_{i}}^{\top}\}^{-1} W_{z^0_{i}}^{-1}),\\
		& = p + \text{tr}(A_{jz^0_{i}}^{\top}\{W_{z^0_{i}}^{\top}\}^{-1} W_{z^0_{i}}^{-1}A_{jz^0_{i}}) -\text{tr}(A_{jz^0_{i}} W_{z^0_{i}}^{-1}) - \text{tr}(A_{jz^0_{i}}^{\top}\{W_{z^0_{i}}^{\top}\}^{-1}),\\
		& = p + \text{tr}(\{W_{z^0_{i}}^{-1} A_{jz^0_{i}}\}^{\top} W_{z^0_{i}}^{-1}A_{jz^0_{i}}) - 2\text{tr}(W_{z^0_{i}}^{-1} A_{jz^0_{i}}),\\
		& = p + \text{tr}(V_{jz^0_{i}}^{\top} V_{jz^0_{i}}) - 2\text{tr}(V_{jz^0_{i}}),\\
		& = p - 2\text{tr}(V_{jz^0_{i}}) + \sum_{l=1}^p\sum_{m\le l} (v_{jz^0_{i},lm})^2,
	\end{align*}
	given that each element of $V_{jz^0_{i}} = W_{z^0_{i}}^{-1} A_{jz^0_{i}}$ above the main diagonal is necessarily equal to zero.  Then, we have that
	\begin{align*}
		2\text{tr}(V_{jz^0_{i}}) &= 2\text{tr}(W_{z^0_{i}}^{-1} A_{jz^0_{i}}),\\
		&= 2\text{tr}(I_p - W_{z^0_{i}}^{-1}W_j),\\
		&= 2p -  2\sum_{l=1}^p {\tilde{w}_{z^0_{i},ll}w_{j,ll}},
	\end{align*}
	where $\tilde{w}_{z^0_{i},ll}$ and $w_{j,ll}$ are the $l^{th}$ diagonal element of $W_{z^0_{i}}^{-1}$ and $W_{j}$ respectively, and both are positive as a result of the Cholesky decomposition of a positive-definite matrix. Hence, we can write that
	\begin{align*}
		\text{tr}(\Sigma^{-1}_j \Sigma^0_{z^0_{i}}) & = 2\sum_{l=1}^p d_{jz^0_{i},ll}  + \sum_{l=1}^p\sum_{m\le l} (v_{jz^0_{i},lm})^2 - p,
	\end{align*}
	where $d_{jz^0_{i},ll} = \tilde{w}_{z^0_{i},ll}w_{j,ll} \ge 0$ for any value of $l \in \{1,2...,p\}$.
\end{proof}
\subsection{Lemma S.3}\label{lem1:S3}
\textit{Let Assumptions \ref{ass1:1}(ii)-(iii) and \ref{ass1:2} hold. Then
	\begin{align*}
		\mathbb{E}[(x_{i} - \boldsymbol{\mu}^0_j)\{\Sigma^0_{j}\}^{-1}(x_{i} - \boldsymbol{\mu}^0_j)^{\top}] = 2\sum_{l=1}^p d_{jz^0_{i},ll}  + \sum_{l=1}^p\sum_{m\le l} (v_{jz^0_{i},lm})^2 - p +  \sum_{l=1}^p (b_{jz^0_{i},l})^2,    
	\end{align*}
	where $v_{jg,lm}$ is defined as in Appendix \ref{lem1:S2}, and where $b_{jg,l}$ corresponds to the $l^{th}$ element of the $p$-sized row-vector $\mathbf{b}_{jg} = \mathbf{a}_{jg}^{\top}W_{j}$ for any pair $(g,j) \in \mathbb{G}\times \mathbb{G}\backslash g$, with $\mathbf{a}_{jg}$ defined as in Appendix \ref{lem1:S1} and where $W_{j}$ is a $p \times p$ lower triangular matrix such that $\{\Sigma^0_{j}\}^{-1} = W_jW_j^{\top}$ for any $j \in \mathbb{G}$.}
\begin{proof}
	Using the property of the trace operator, we can write that
	\begin{align*}
		\mathbb{E}[(x_{i} - \boldsymbol{\mu}^0_j)\{\Sigma^0_{j}\}^{-1}(x_{i} - \boldsymbol{\mu}^0_j)^{\top}]
		&=\text{tr}(\mathbb{E}[(x_{i} - \boldsymbol{\mu}^0_j)^{\top}\{\Sigma^0_{j}\}^{-1}(x_{i} - \boldsymbol{\mu}^0_j)]),\\
		&= \text{tr}(\{\Sigma^0_{j}\}^{-1} \mathbb{E}[(x_{i} - \boldsymbol{\mu}^0_{z^0_{i}} - \boldsymbol{a}_{jz^0_{i}})(x_{i} - \boldsymbol{\mu}^0_{z^0_{i}} - \boldsymbol{a}_{jz^0_{i}})^{\top}]),\\
		&=  \text{tr}(\{\Sigma^0_{j}\}^{-1}\Sigma^0_{z^0_{i}}) + \text{tr} (\{\Sigma^0_{j}\}^{-1} \mathbb{E}[2 (\boldsymbol{\mu}^0_{z^0_{i}} -x_{i}  )\boldsymbol{a}_{jz^0_{i}}^{\top} + \boldsymbol{a}_{jz^0_{i}}\boldsymbol{a}_{jz^0_{i}}^{\top}]),\\
		&= \text{tr}(\{\Sigma^0_{j}\}^{-1}\Sigma^0_{z^0_{i}}) + \text{tr}( W_jW_j^{\top} \boldsymbol{a}_{jz^0_{i}}\boldsymbol{a}_{jz^0_{i}}^{\top}),\\
		&= \text{tr}(\{\Sigma^0_{j}\}^{-1}\Sigma^0_{z^0_{i}}) + \sum_{l=1}^p (b_{jz^0_{i},l})^2,
	\end{align*}
	The fourth equality comes from the fact that $\mathbb{E}[2(x_{i} -\boldsymbol{\mu}^0_{z^0_{i}} )\mathbf{a}_{jz^0_{i}}^{\top}] = 2\mathbb{E}[x_{i}-\boldsymbol{\mu}^0_{z^0_{i}} ]\mathbf{a}_{jz^0_{i}}^{\top} = 0$. From the proof in Appendix \ref{lem1:S2}, we also know that
	\begin{align*}
		\text{tr}(\Sigma^{-1}_j\Sigma^0_{z^0_{i}}) &= 2\sum_{l=1}^p d_{jz^0_{i},ll}  + \sum_{l=1}^p\sum_{m\le l} (v_{jz^0_{i},lm})^2 - p,
	\end{align*}
	where $v_{jz^0_{i},lm}$ and $d_{jz^0_{i},ll}$ are defined as in Appendix \ref{lem1:S2}. Combining the two results completes the proof.
\end{proof}
\subsection{Lemma S.4}\label{lem1:S4}
\textit{Let $f(x)$ and $g(x)$ be two distinct, univariate and positive functions of the $p$-variate random variable $X$, where $f: \mathbb{R}^{p} \to \mathbb{R}_{\ge0}$ and $g: \mathbb{R}^{p} \to \mathbb{R}_{\ge0}$ such that the two functions share the same support and the same image. Let also $f(x) \ne g(x)$ for at least one $x \in \mathcal{X}$, with $E[f(x)] < \infty$, $E[f(x)^2] < \infty$, $E[g(x)] < \infty$, and $E[g(x)^2] < \infty$ for any $|x|_1 < \infty$, where $|\cdot|_1$ denotes the $\textit{l}_1$ norm. Then
	\begin{align*}
		\mathbb{P}[f(x) \ge g(x)] \le \frac{\mathbb{E}[f(x)] +\frac{1}{2}\left|\sqrt{\text{Var}[g(x)]}\right|}{\mathbb{E}[g(x)]},
	\end{align*}
	provided that $\mathbb{E}[g(x)] \ne 0$.}
\begin{proof}
	This lemma is a generalization of Markov's inequality to univariate functions of $X$ on both sides of the inequality that is contained within the probability. The standard Markov's inequality for functions of $X$ states that
	\begin{align*}
		\mathbb{P}[f(x) \ge a] \le \frac{\mathbb{E}[f(x)]}{a},
	\end{align*}
	for any constant $a > 0$ and any function $f(x)$, which is a special case of the generalized inequality when $g(x) = a$. The argument for the generalized Markov's inequality goes as follows. Let the indicator function $\mathbbm{1}{[f(x)\ge g(x)]}$ be equal to one if and only if $f(x)\ge g(x)$ and zero otherwise. Therefore, we have that
	\begin{align*}
		f(x) \ge g(x)\mathbbm{1}{[f(x) \ge g(x)]},
	\end{align*}
	which is always satisfied. Hence, we can take expectation on both sides of the inequality and it will never be reversed since 
	\begin{align*}
		\mathbb{E}[g(x)\mathbbm{1}{[f(x)\ge g(x)]}] &= \int_{\mathcal{X}} g(x)\mathbbm{1}{[f(x)\ge g(x)]} p_X(x) dx,\\
		&= \int_{f(x)\ge g(x)} g(x) p_X(x) dx,\\
		& \le \int_{f(x)\ge g(x)} f(x) p_X(x) dx \le \int_{\mathcal{X}} f(x) p_X(x) dx = \mathbb{E}[f(x)],
	\end{align*}
		where $p_X(x)$ corresponds to the true generating density of $X$. The last inequality comes from the fact that $f(x) \ge 0$ for any $x \in \mathcal{X}$. Note also that equality occurs when $g(x) = f(x)$ for all $x \in \mathcal{X}\backslash_{x^*}$, where $x^*$ is a set with measure zero in $\mathcal{X}$. Therefore, we can write that
	\begin{align*}
		\mathbb{E}[f(x)] &\ge  \text{Cov}[g(x),\mathbbm{1}{[f(x)\ge g(x)]}] + \mathbb{E}[g(x)]\mathbb{P}[f(x)\ge g(x)],
	\end{align*}
	given that both $g(x)$ and $\mathbbm{1}{[f(x)\ge g(x)]}$ have finite second moments. Subtracting the term $\text{Cov}[g(x),\mathbbm{1}{[f(x)\ge g(x)]}]$ and dividing by $\mathbb{E}[g(x)]$ on each side of the inequality leads to the following
	\begin{align*}
		\mathbb{P}[f(x)\ge g(x)] \le \frac{\mathbb{E}[f(x)] - \text{Cov}[g(x),\mathbbm{1}{[f(x)\ge g(x)]}] }{\mathbb{E}[g(x)]}.
	\end{align*}
	By the Cauchy-Schwarz inequality, we know that
	\begin{align*}
		\left|\text{Cov}[g(x),\mathbbm{1}{[f(x)\ge g(x)]}]\right| \le (\text{Var}[g(x)]\text{Var}[\mathbbm{1}{[f(x)\ge g(x)]}])^{1/2},
	\end{align*}
	and we also know that the variance of any binary variable can be at most equal to 0.25. This allows us to write the following
	\begin{align*}
		 -\frac{1}{2}\left|\sqrt{\text{Var}[g(x)]}\right| \le \text{Cov}[g(x),\mathbbm{1}{[f(x)\ge g(x)]}]  \le \frac{1}{2}\left|\sqrt{\text{Var}[g(x)]}\right|.
	\end{align*}
	Substituting $\text{Cov}[g(x),\mathbbm{1}{[f(x)\ge g(x)]}]$ by its lower bound in the above inequality completes the proof.
\end{proof}
\newpage
\section{Details of the Estimation Procedure for the Second Simulation Exercise and the Empirical Analysis}\label{sec1:B}
In practice, both the EM and the C-EM algorithms that are employed in the second simulation exercise correspond to an iterative weighted generalized least squares (IWGLS) procedure where the weights depend on the chosen algorithm. It is a form of generalized least squares (GLS) procedure given that the variance of the unit-random effects $\sigma^2_{\alpha,j}$ is taken into account through the use of the following equation
\begin{align}\tag{A.1} 
	\hat{\tilde{\beta}}^{(k+1)}_{g} &= (\sum_{i=1}^N \{\tilde{\mathbf{X}}^{(k)}_{ig}\}^{\top} \{\hat{\Omega}_g^{(k)}\}^{-1} \tilde{\mathbf{X}}^{(k)}_{ig})^{-1} (\sum_{i=1}^N \{\tilde{\mathbf{X}}^{(k)}_{ig}\}^{\top} \{\hat{\Omega}_g^{(k)}\}^{-1} \tilde{\mathbf{y}}^{(k)}_{ig}),
\end{align}
where $\tilde{\mathbf{X}}^{(k)}_{ig} = (w^{(k)}_{i1g}X_{i1,1}^{\top},...,w^{(k)}_{iTg}X_{iT,1}^{\top})^{\top}$, $\tilde{\mathbf{y}}^{(k)}_{ig} = (w^{(k)}_{i1g}y_{i1},...,w^{(k)}_{iTg}y_{iT})^{\top}$ with $X_{it,1}$ defined as in Section \ref{subsec1:412}, and with  
\begin{align}\tag{A.2} 
	w^{(k)}_{itg} =
	\begin{cases}
		z^{D}_{itg}(\hat{\tilde{\beta}}^{(k)},\hat{\Omega}^{(k-1)},\hat{\psi}^{(k-1)}) \ \text{if the C-EM algorithm is used},\\
		\tau_{itg}(\hat{\tilde{\beta}}^{(k)},\hat{\Omega}^{(k-1)},\hat{\pi}^{(k-1)}) \ \text{if the EM algorithm is used}.
	\end{cases}
\end{align}
Each weight is defined as follows
\begin{align*}
	z^{D}_{itg}(\hat{\tilde{\beta}}^{(k)},\hat{\Omega}^{(k-1)},\hat{\psi}^{(k-1)}) &:= \mathbbm{1}[\arg\max_{j\in \mathbb{G}} f_j(y_{it},x_{it}|\hat{\tilde{\beta}}^{(k)}_{j},\hat{\Omega}^{(k-1)}_j,\hat{\psi}^{(k-1)}_j) = g],\\
	\tau_{itg}(\hat{\tilde{\beta}}^{(k)},\hat{\Omega}^{(k-1)},\hat{\pi}^{(k-1)}) &:= \frac{\hat{\pi}^{(k-1)}_g f_N(y_{it}|x_{it}; \hat{\tilde{\beta}}^{(k)}_{g},\hat{\Omega}^{(k-1)}_j)}{\sum_{j=1}^G \hat{\pi}^{(k-1)}_jf_N(y_{it}|x_{it}; \hat{\tilde{\beta}}^{(k)}_{j},\hat{\Omega}^{(k-1)}_j)},
\end{align*}
where $f_j(y_{it},x_{it}|\hat{\tilde{\beta}}_{j},\hat{\Omega}_j,\hat{\psi}_j) = f_N(y_{it}|x_{it};\hat{\tilde{\beta}}_{j},\hat{\Omega}_j)f_N(x_{it}|\hat{\psi}_j)$ refers to the joint density of the outcome and covariates with $f_N(\cdot|\theta)$ representing the normal univariate (for $y_{it}$) or multivariate (for $x_{it}$) pdf evaluated at $x$ and conditional on parameters $\theta$, $\hat{\Omega}^{(k)} = (\hat{\Omega}^{(k)}_1,...,\hat{\Omega}^{(k)}_G)$ refers to the set of variance-covariance matrices of the outcome for each group, and where $\hat{\psi}^{(k-1)} = (\hat{\psi}^{(k-1)}_1,...,\hat{\psi}^{(k-1)}_G)$ refers to the estimated parameters of the covariates' density. The mixing weight $\hat{\pi}^{(k)} = (\hat{\pi}^{(k)}_1,...,\hat{\pi}^{(k)}_G)$ is obtained using 
\begin{align*}
	\hat{\pi}^{(k)}_g = \frac{1}{NT} \sum_{i=1}^N\sum_{t=1}^T \tau_{itg}(\hat{\tilde{\beta}}^{(k)},\hat{\Omega}^{(k-1)},\hat{\pi}^{(k-1)}),
\end{align*}
for every $g \in \mathbb{G}$. The elements on the main diagonal of $\hat{\Omega}_g^{(k)}$ are estimated using
\begin{align*}
	\hat{\sigma}^{2(k)}_{\alpha+\epsilon,g} = \frac{\sum_{i=1}^N \sum_{t=1}^T w^{(k)}_{itg}(\hat{\epsilon}_{itg}^{(k)})^2}{\sum_{i=1}^N\sum_{t=1}^T w^{(k)}_{itg}-2-T},
\end{align*}
were the denominator accounts for the number of mean parameters to estimate, while the off-diagonal elements are estimated using
\begin{align*}
	\hat{\sigma}^{2(k)}_{\alpha,g} = \sum_{i=1}^N \alpha_{ig}^{(k)} \left(\frac{\sum_{t=1}^T w^{(k)}_{itg}\hat{\epsilon}_{itg}^{(k)}}{\sum_{t=1}^T w^{(k)}_{itg}} - \sum_{j=1}^N\alpha_{jg}^{(k)}\left(\frac{\sum_{t=1}^T w^{(k)}_{jtg}\hat{\epsilon}_{jtg}^{(k)}}{\sum_{t=1}^T w^{(k)}_{jtg}}\right) \right)^2,
\end{align*}
where $\hat{\epsilon}_{itg}^{(k)} = y_{it} - X_{it}\hat{\tilde{\beta}}^{(k)}_{g}$, $\alpha_{ig}^{(k)} = \frac{\sum_{t=1}^T w^{(k)}_{itg}}{\sum_{i=1}^N\sum_{t=1}^T w^{(k)}_{itg}}$, and where $w^{(k)}_{itg}$ is defined as above. It is then possible to compute $\hat{\sigma}^{2(k)}_{\epsilon,g} = \hat{\sigma}^{2(k)}_{\alpha+\epsilon,g}- \hat{\sigma}^{2(k)}_{\alpha,g}$ for each $g \in \mathbb{G}$. To avoid undefined cases and singleton mean values, I remove all units $i$ such that $\sum_{t=1}^T w^{(k)}_{itg} = \{0,1\}$ from the computation of $\hat{\sigma}^{2(k)}_{\alpha,g}$ for each value of $k$ when using the C-EM algorithm. The use of the weights $\alpha_{ig}^{(k)}$ leads to improved finite-sample results since it weights the contribution of each unit to the variance term $\hat{\sigma}^{2(k)}_{\alpha,g}$ based on their aggregate probability of belonging to group $g$ over time. Finally, $\hat{\psi}^{(k)} = (\boldsymbol{\hat{\mu}}^{(k)}, \hat{\Sigma}^{(k)})$ is computed using the following equations
\begin{align*}
	&\hat{\boldsymbol{\mu}}^{(k)}_g = \sum_{i=1}^N \sum_{t=1}^T \frac{w^{(k)}_{itg} x_{it}}{\sum_{j=1}^N \sum_{l=1}^T w^{(k)}_{jlg}}, &\hat{\Sigma}^{(k)}_g = \frac{\sum_{i=1}^N w^{(k-1)}_{itg}(x_{it} - \hat{\boldsymbol{\mu}}^{(k)}_g)(x_{it} - \hat{\boldsymbol{\mu}}^{(k)}_g)^{\top}}{(\sum_{j=1}^N \sum_{t=1}^T w^{(k)}_{jtg})-p}.
\end{align*}
\par
Estimation of the parameters for the empirical analysis follows an approach that is similar to the ones shown above, the main difference being the estimation of a random effects Probit model using a Newton-Raphson iterative procedure. The details of the Newton-Raphson procedure are available upon request from the author. To perform inference, a cluster-robust variance estimator has been used for both parts of the model. For the continuous part, such an estimator is described as follows~:
\begin{align*}
	&\widehat{\text{Var}[\hat{\tilde{\beta}}^{k}_g]} = \{\mathbf{\tilde{Q}}^{(k)}_g\}^{-1} \sum_{i=1}^N \left[\{\tilde{\mathbf{X}}^{(k)}_{ig}\}^{\top} \{\hat{\Omega}_g^{(k)}\}^{-1} \hat{\epsilon}_{itg}^{(k)} \{\hat{\epsilon}_{itg}^{(k)}\}^{\top} \{\hat{\Omega}_g^{(k)}\}^{-1} \tilde{\mathbf{X}}^{(k)}_{ig} \right] \{\mathbf{\tilde{Q}}^{(k)}_g\}^{-1},
\end{align*}
where $\mathbf{\tilde{Q}}^{(k)}_g = \sum_{i=1}^N \{\tilde{\mathbf{X}}^{(k)}_{ig}\}^{\top} \{\hat{\Omega}_g^{(k)}\}^{-1} \tilde{\mathbf{X}}^{(k)}_{ig}$ and where all other elements have the same meaning as above. For the binary part, Eq. (\ref{eqn1:11}) has been used to obtain cluster-robust standard errors on the estimated parameters $\hat{\tilde{\beta}}^{(k),b}_g$, where the $b$ superscript stands for ``binary''. The formal definitions of the ``component-wise'' score function and its derivative for the random effects Probit model are standard and can be found in Chapter 15 of \cite{wooldridge_econometric_2010}.
\newpage
\section{Additional Simulation Results}\label{sec1:C}
\subsection{First Simulation Exercise}\label{subsec1:C1}
\begin{figure}[h!]
	\centering
	\begin{subfigure}[Estimation biases for different sample sizes]{
			\centering
			\includegraphics[width=14.5cm]{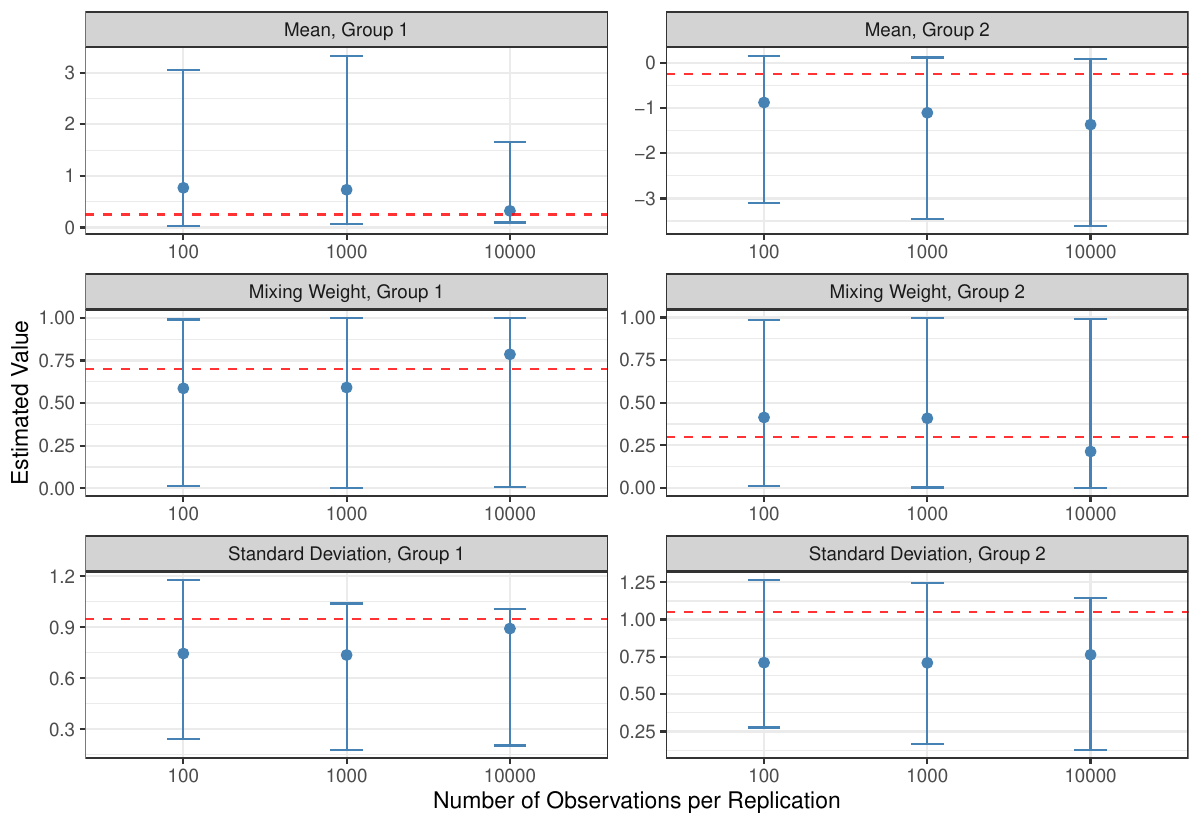}
	}\end{subfigure}
	\begin{subfigure}[Estimated parameters for N=10,000]{
			\centering
			\includegraphics[width=14.5cm]{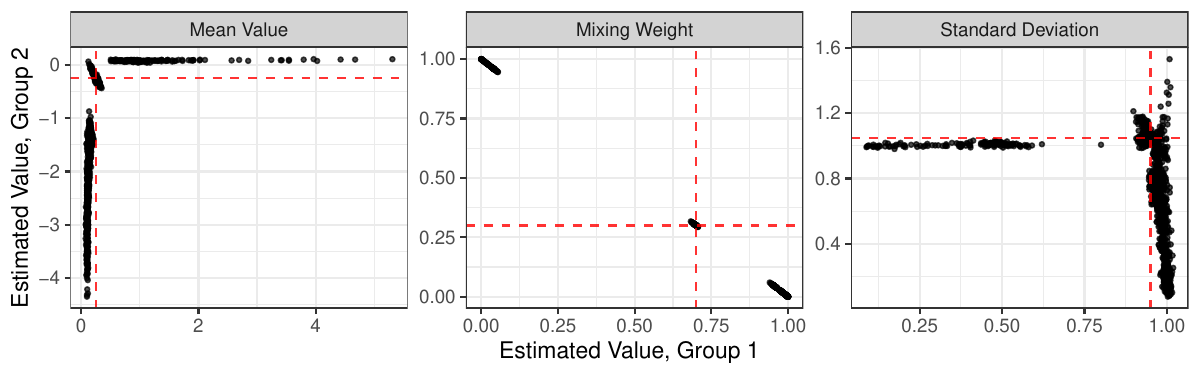}
	}\end{subfigure}
	\caption{Simulation results for the mixture of normal distributions with true values $\boldsymbol{\mu}^0 = (0.25,-0.25)$, $\sigma^0 = (0.95,1.05)$, and $\pi^0_1 = 0.7$. The blue dots correspond to the estimated value averaged across 1,000 replications, while the error bars represent the $2.5^{th}$ and $97.5^{th}$ percentiles of the estimated parameter's empirical distribution. The red dashed lines correspond to the true parameter values.\label{fig1:C1}}
\end{figure}
\newpage
\begin{figure}[h!]
	\centering
	\begin{subfigure}[Estimation biases for different sample sizes]{
			\centering
			\includegraphics[width=14.5cm]{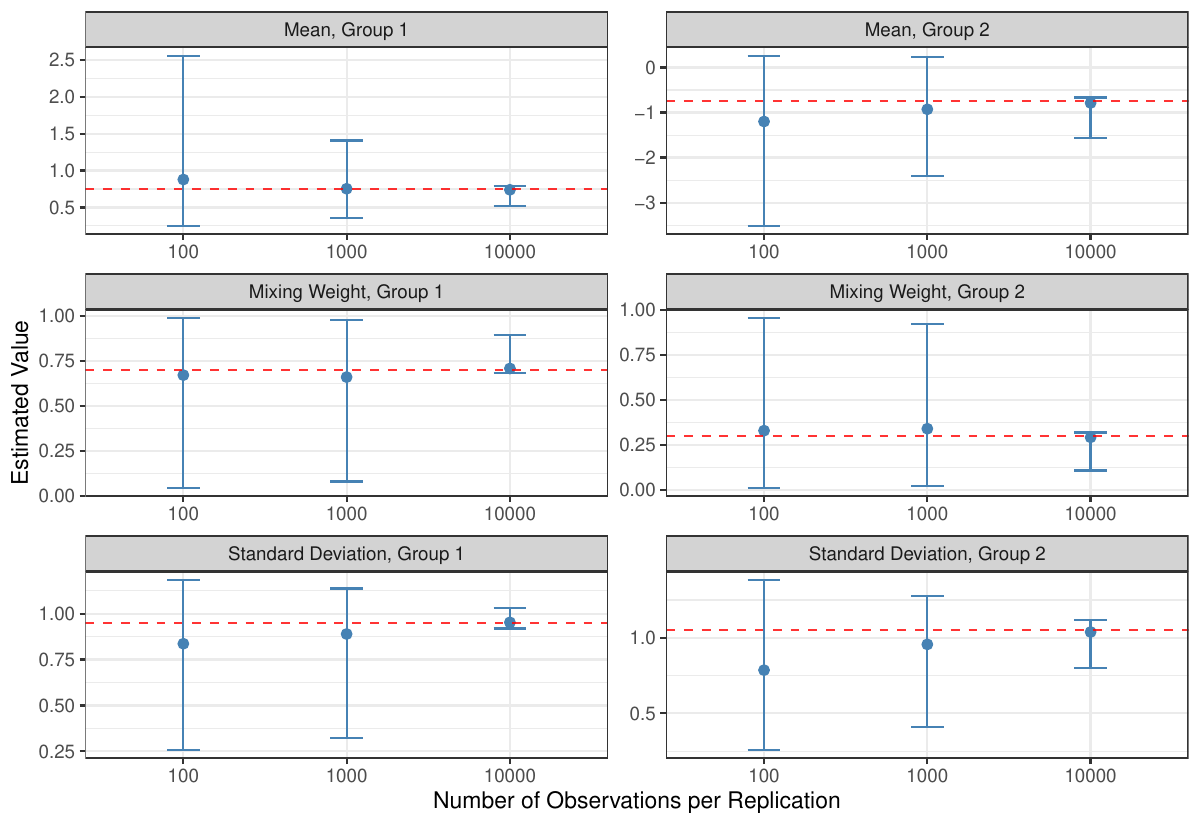}
	}\end{subfigure}
	\begin{subfigure}[Estimated parameters for N=10,000]{
			\centering
			\includegraphics[width=14.5cm]{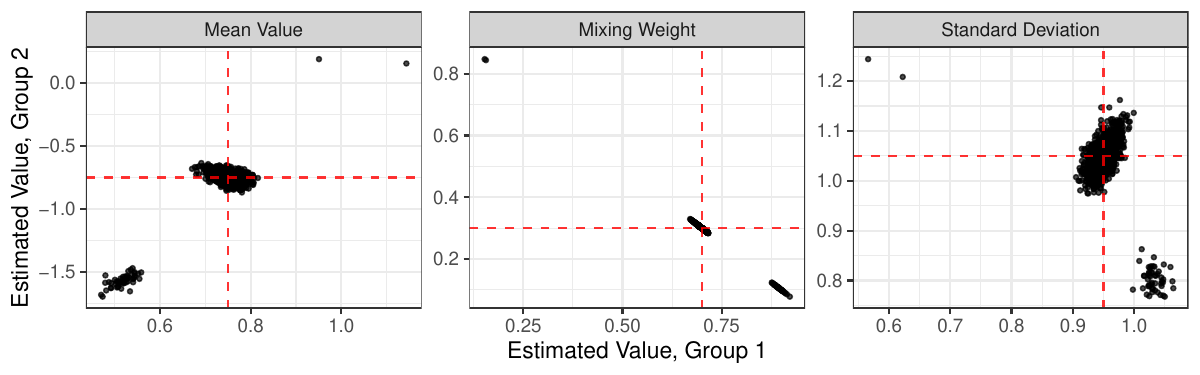}
	}\end{subfigure}
	\caption{Simulation results for the mixture of normal distributions with true values $\boldsymbol{\mu}^0 = (0.75,-0.75)$, $\sigma^0 = (0.95,1.05)$, and $\pi^0_1 = 0.7$. The blue dots correspond to the estimated value averaged across 1,000 replications, while the error bars represent the $2.5^{th}$ and $97.5^{th}$ percentiles of the estimated parameter's empirical distribution. The red dashed lines correspond to the true parameter values.\label{fig1:C2}}
\end{figure}
\newpage
\begin{figure}[h!]
	\centering
	\begin{subfigure}[Estimation biases for different sample sizes]{
			\centering
			\includegraphics[width=14.5cm]{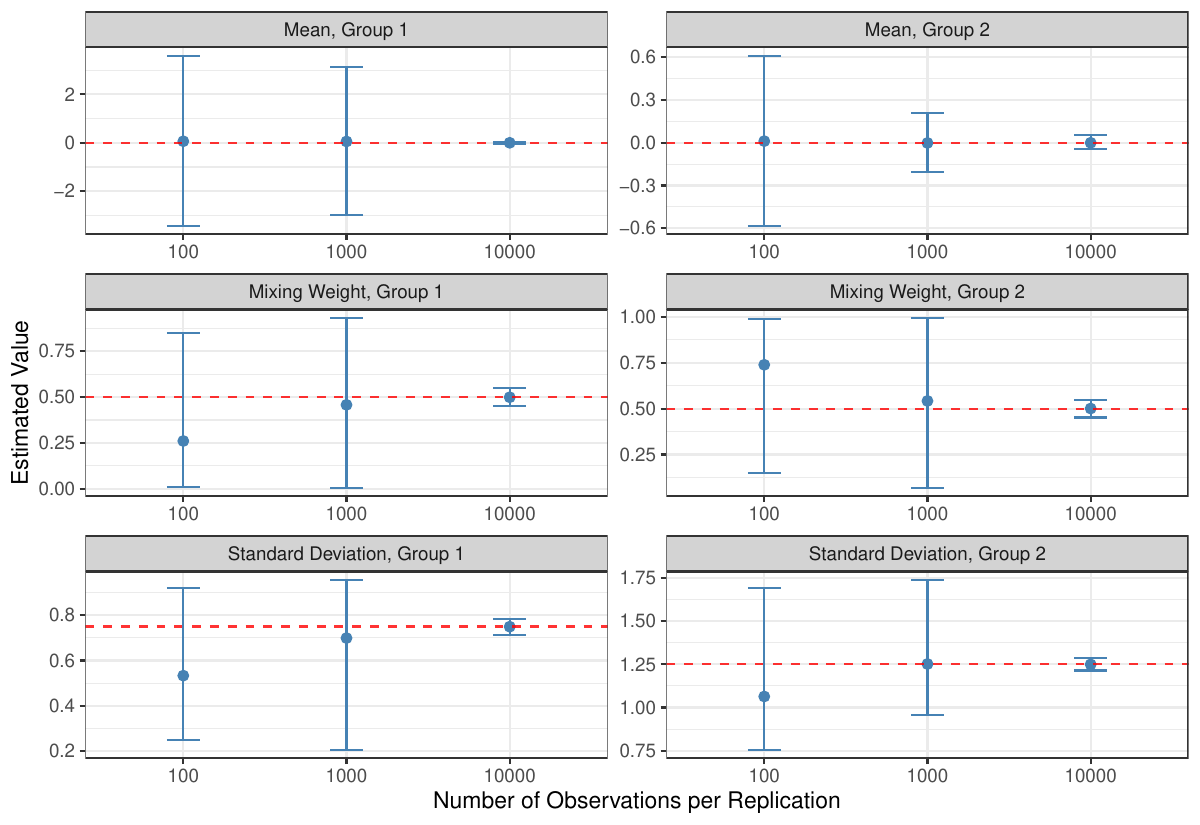}
	}\end{subfigure}
	\begin{subfigure}[Estimated parameters for N=10,000]{
			\centering
			\includegraphics[width=14.5cm]{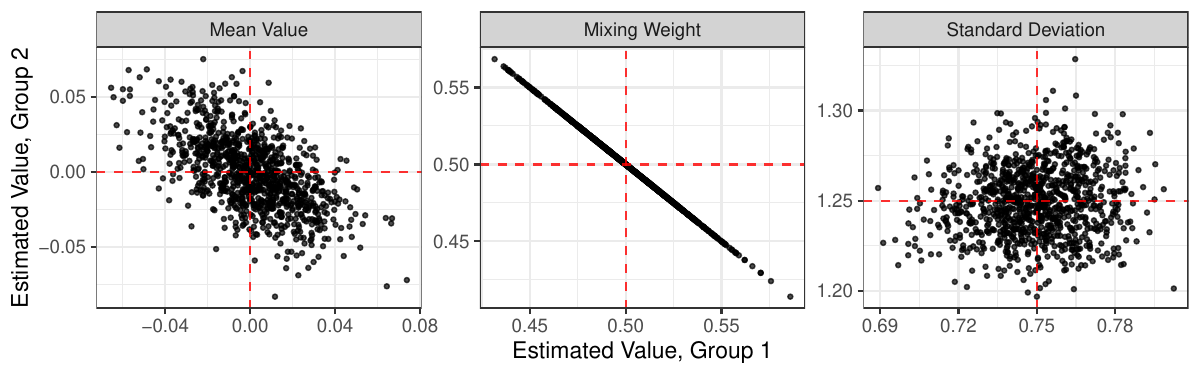}
	}\end{subfigure}
	\caption{Simulation results for the mixture of normal distributions with true values $\boldsymbol{\mu}^0 = (0,0)$, $\sigma^0 = (0.75,1.25)$, and $\pi^0_1 = 0.5$. The blue dots correspond to the estimated value averaged across 1,000 replications, while the error bars represent the $2.5^{th}$ and $97.5^{th}$ percentiles of the estimated parameter's empirical distribution. The red dashed lines correspond to the true parameter values.\label{fig1:C3}}
\end{figure}
\newpage
\begin{figure}[h!]
	\centering
	\begin{subfigure}[Estimation biases for different sample sizes]{
			\centering
			\includegraphics[width=14.5cm]{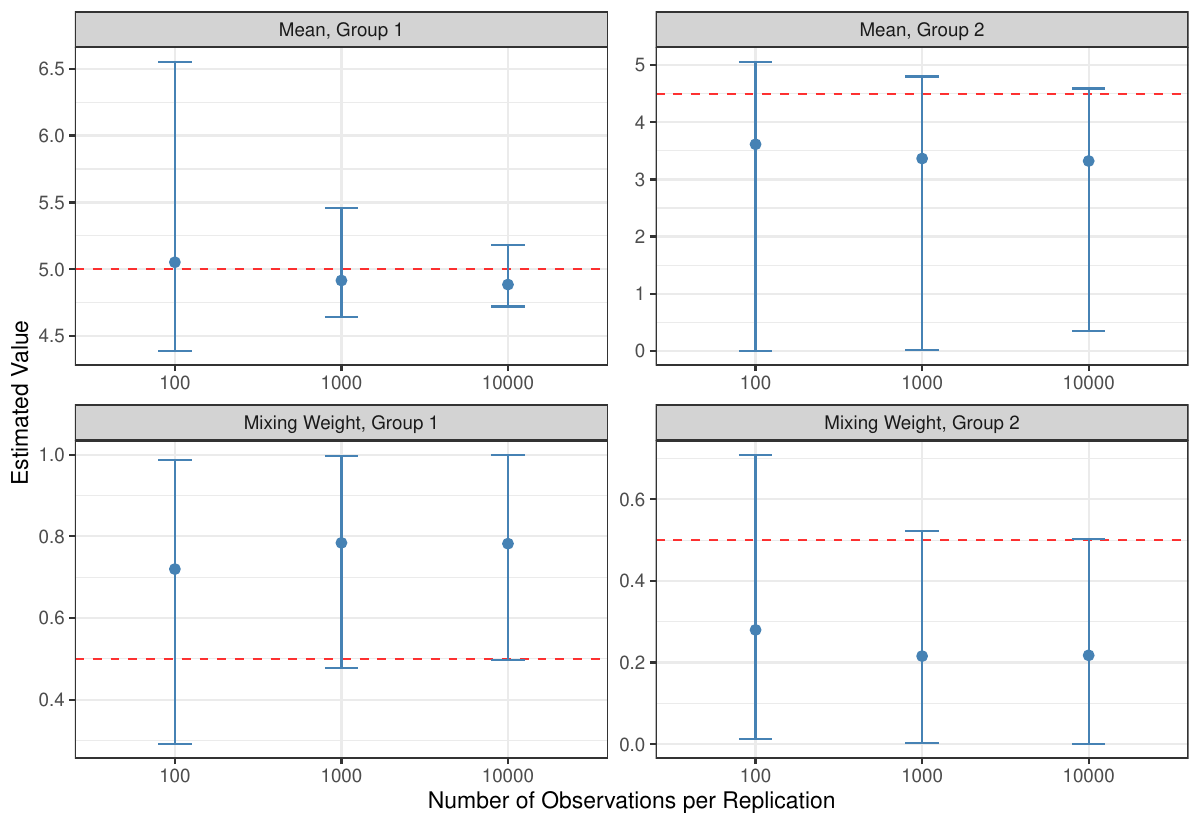}
	}\end{subfigure}
	\begin{subfigure}[Estimated parameters for N=10,000]{
			\centering
			\includegraphics[width=14.5cm]{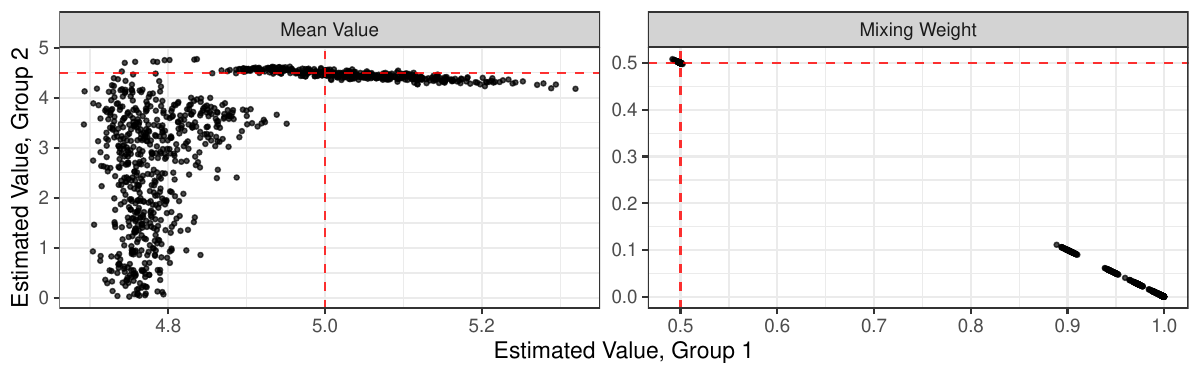}
	}\end{subfigure}
	\caption{Simulation results for the mixture of Poisson distributions with true values $\boldsymbol{\mu}^0 = (5,4.5)$ and $\pi^0_1 = 0.5$. The blue dots correspond to the estimated value averaged across 1,000 replications, while the error bars represent the $2.5^{th}$ and $97.5^{th}$ percentiles of the estimated parameter's empirical distribution. The red dashed lines correspond to the true parameter values.\label{fig1:C4}}
\end{figure}
\newpage
\begin{figure}[h!]
	\centering
	\begin{subfigure}[Estimation biases for different sample sizes]{
			\centering
			\includegraphics[width=14.5cm]{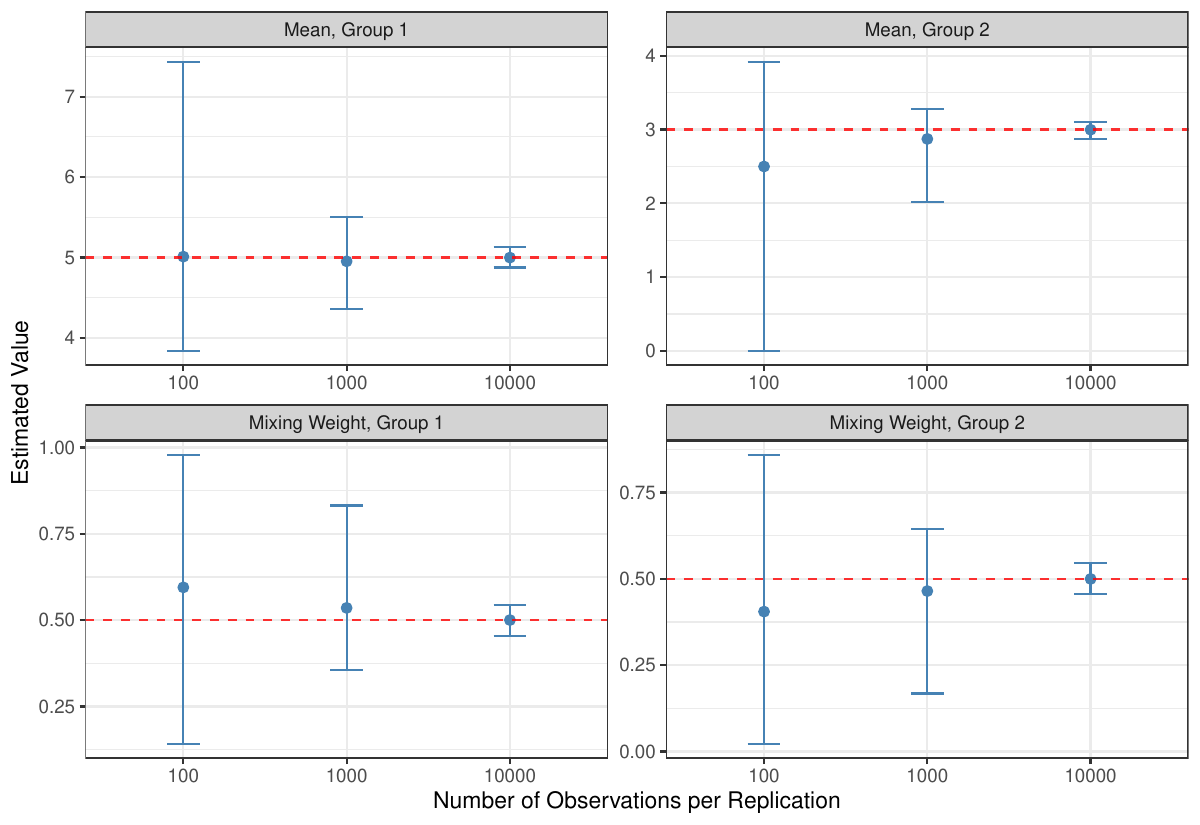}
	}\end{subfigure}
	\begin{subfigure}[Estimated parameters for N=10,000]{
			\centering
			\includegraphics[width=14.5cm]{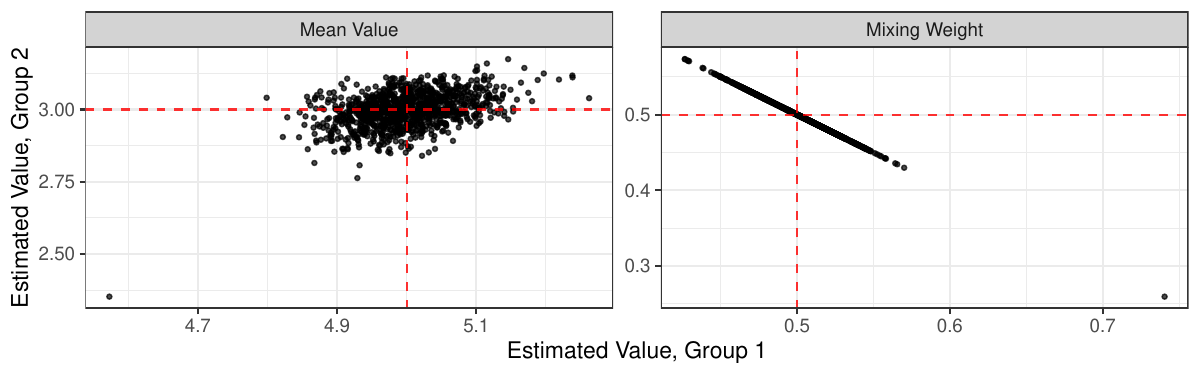}
	}\end{subfigure}
	\caption{Simulation results for the mixture of Poisson distributions with true values $\boldsymbol{\mu}^0 = (5,3)$ and $\pi^0_1 = 0.5$. The blue dots correspond to the estimated value averaged across 1,000 replications, while the error bars represent the $2.5^{th}$ and $97.5^{th}$ percentiles of the estimated parameter's empirical distribution. The red dashed lines correspond to the true parameter values.\label{fig1:C5}}
\end{figure}
\newpage
\begin{figure}[h!]
	\centering
	\begin{subfigure}[Estimation biases for different sample sizes]{
			\centering
			\includegraphics[width=14.5cm]{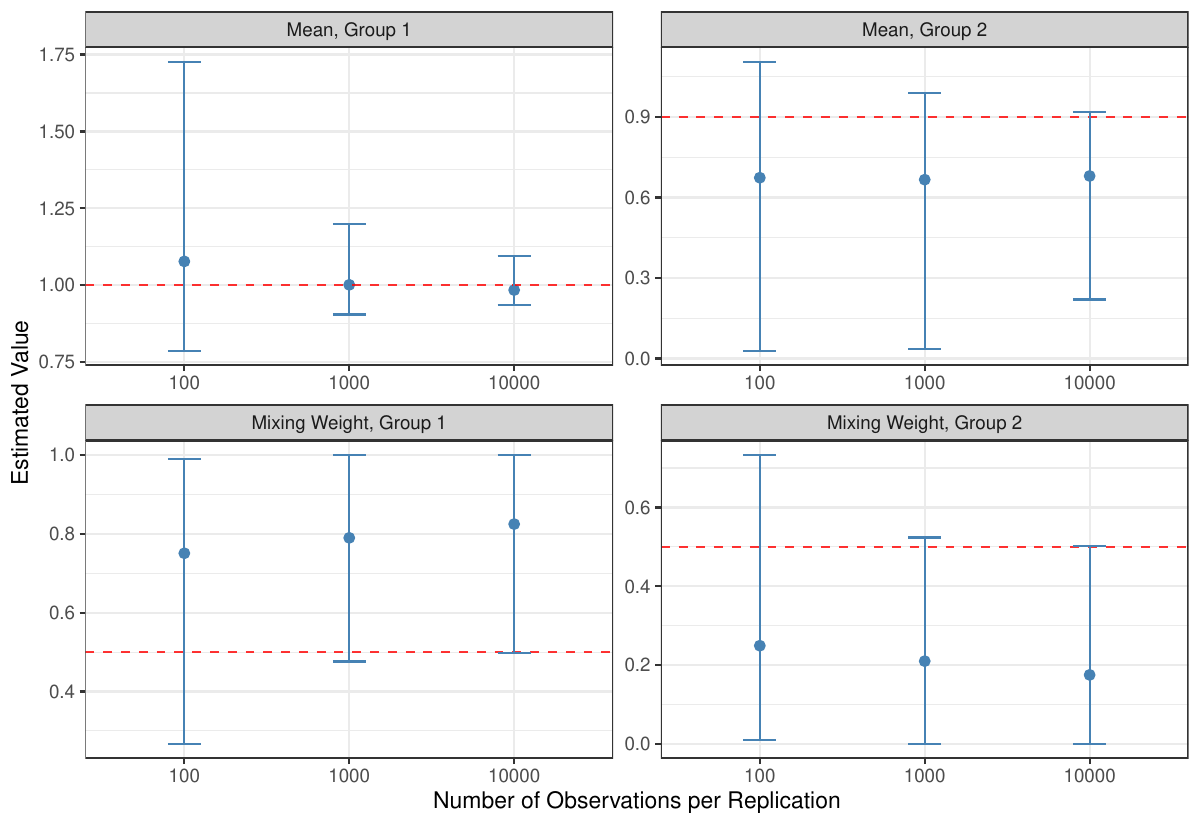}
	}\end{subfigure}
	\begin{subfigure}[Estimated parameters for N=10,000]{
			\centering
			\includegraphics[width=14.5cm]{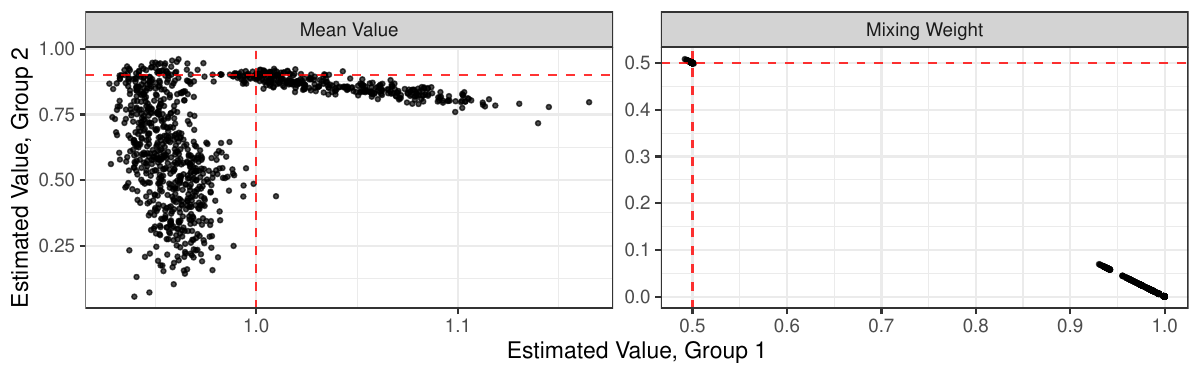}
	}\end{subfigure}
	\caption{Simulation results for the mixture of exponential distributions with true values $\boldsymbol{\mu}^0 = (1,0.9)$ and $\pi^0_1 = 0.5$. The blue dots correspond to the estimated value averaged across 1,000 replications, while the error bars represent the $2.5^{th}$ and $97.5^{th}$ percentiles of the estimated parameter's empirical distribution. The red dashed lines correspond to the true parameter values.\label{fig1:C6}}
\end{figure}
\newpage
\begin{figure}[h!]
	\centering
	\begin{subfigure}[Estimation biases for different sample sizes]{
			\centering
			\includegraphics[width=14.5cm]{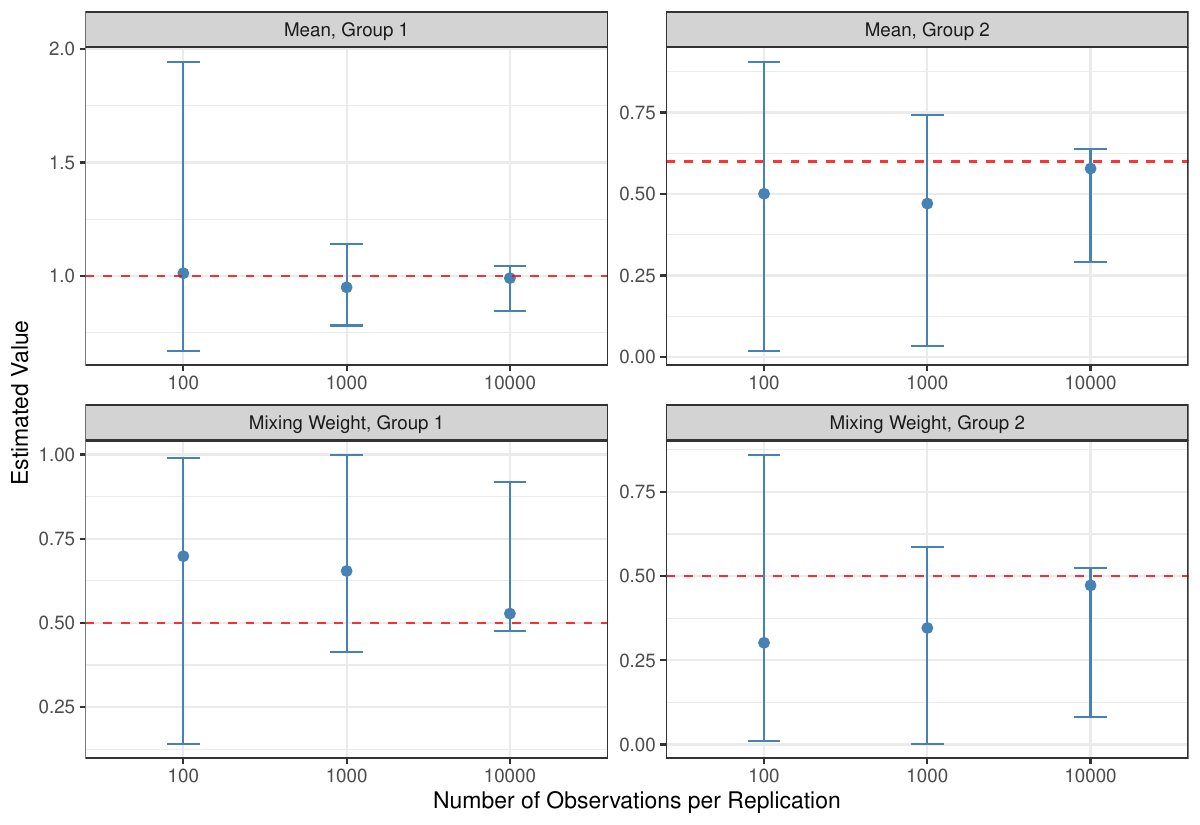}
	}\end{subfigure}
	\begin{subfigure}[Estimated parameters for N=10,000]{
			\centering
			\includegraphics[width=14.5cm]{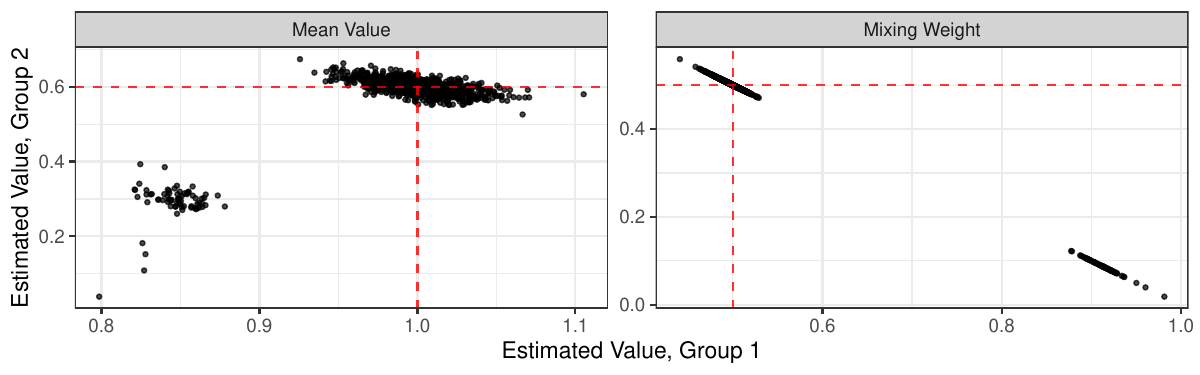}
	}\end{subfigure}
	\caption{Simulation results for the mixture of exponential distributions with true values $\boldsymbol{\mu}^0 = (1,0.6)$ and $\pi^0_1 = 0.5$. The blue dots correspond to the estimated value averaged across 1,000 replications, while the error bars represent the $2.5^{th}$ and $97.5^{th}$ percentiles of the estimated parameter's empirical distribution. The red dashed lines correspond to the true parameter values.\label{fig1:C7}}
\end{figure}
\newpage
\subsection{Second Simulation Exercise}\label{subsec1:C2}
\begin{figure}[h!]
	\setstretch{1.15}
	\centering
	\subfigure{
		\includegraphics[width=14.5cm]{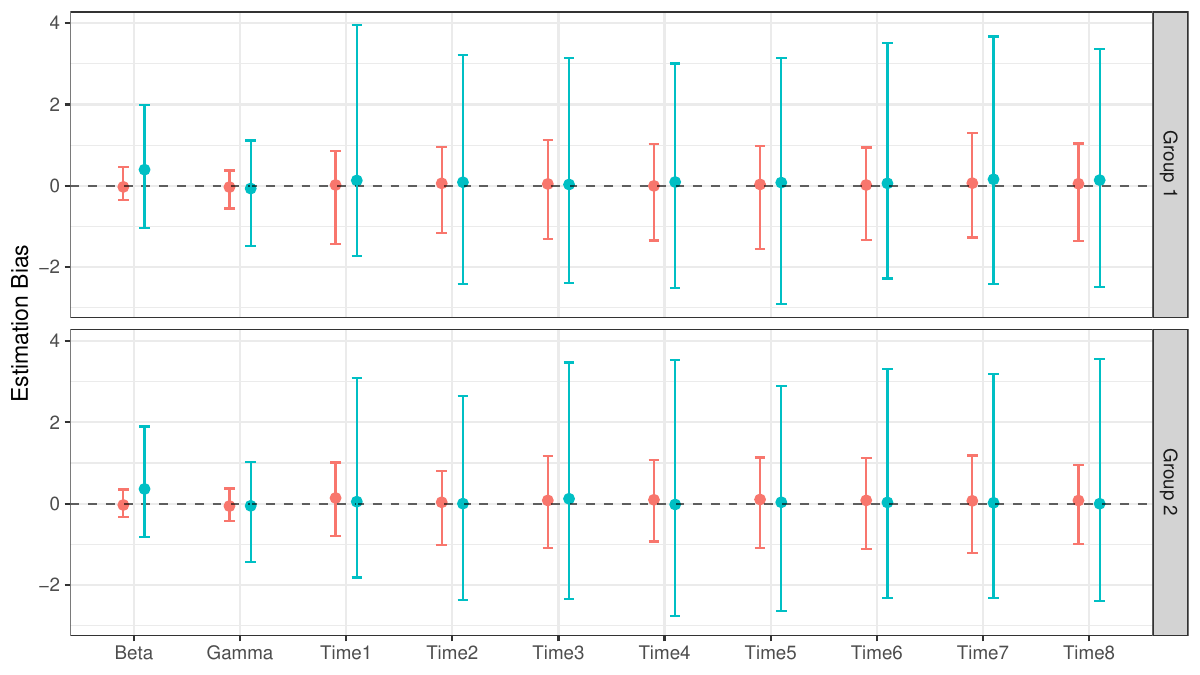}
	}
	\subfigure{
		\includegraphics[width=14.5cm]{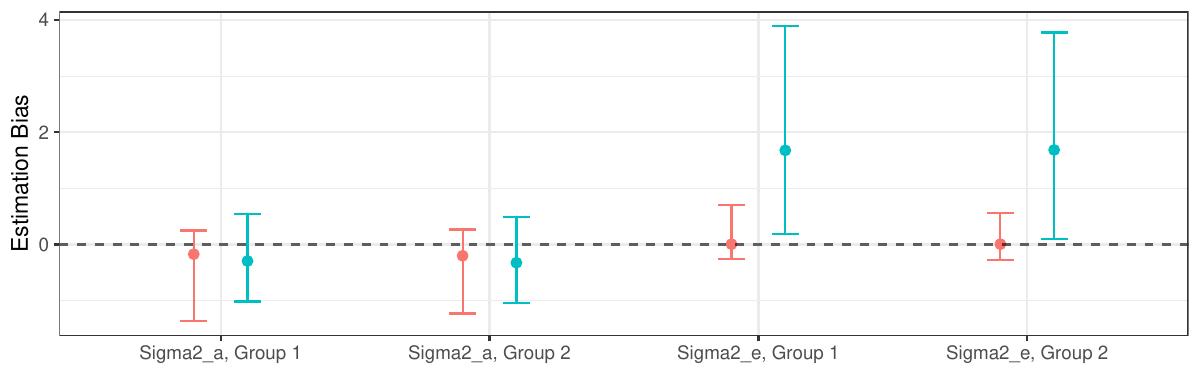}
	}
	\subfigure{
		\includegraphics[width=14.5cm]{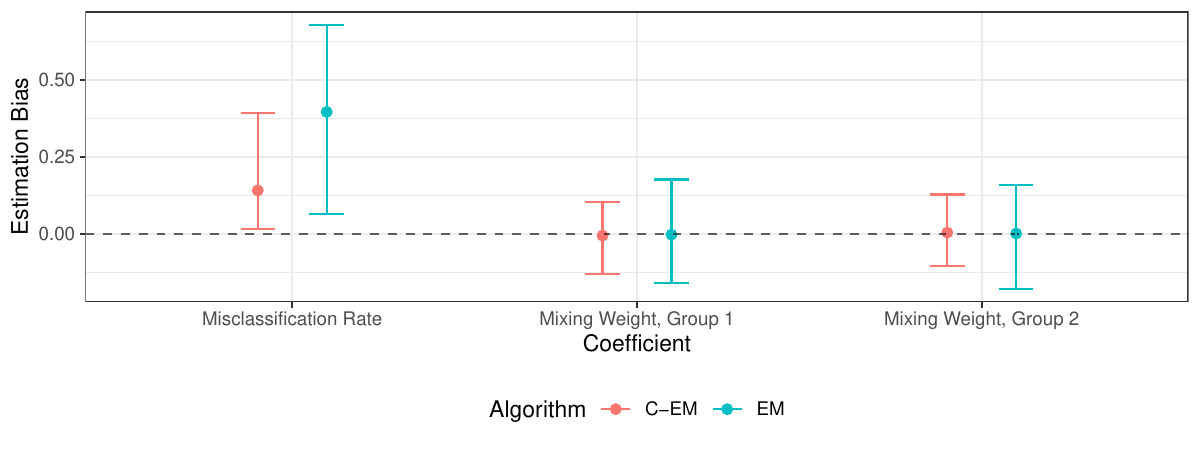}
	}
	\caption{Estimation bias for all parameters in the mixture when $N = 750$, $T=8$, $G=2$, and $p=1$. The dots correspond to the estimated bias over 250 replications, while the error bars represent the $2.5^{th}$ and $97.5^{th}$ percentiles of the empirical distribution of the bias. The coefficients $Time1,....,Time8$ represent the time-fixed effects, with $Sigma2\_a = \sigma^2_{\alpha,g}$ and $Sigma2\_e = \sigma^2_{\epsilon}$.\label{fig1:C8}}
\end{figure}
\newpage
\begin{figure}[h!]
	\setstretch{1.15}
	\centering
	\subfigure{
		\includegraphics[width=14.5cm]{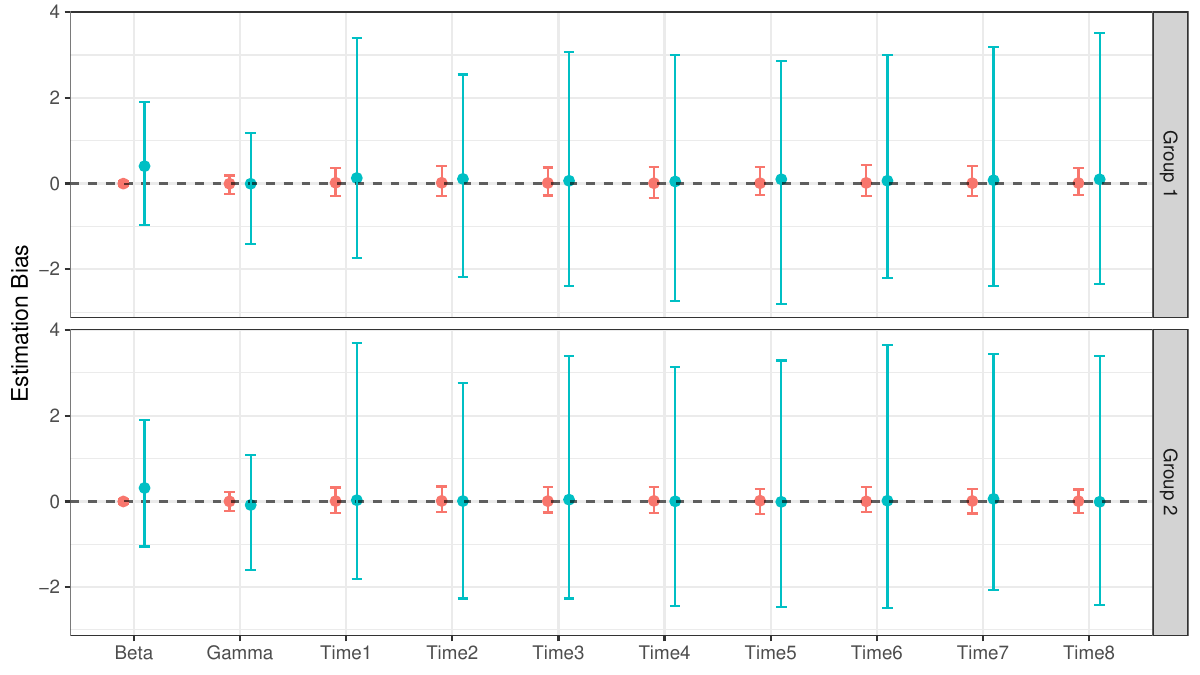}
	}
	\subfigure{
		\includegraphics[width=14.5cm]{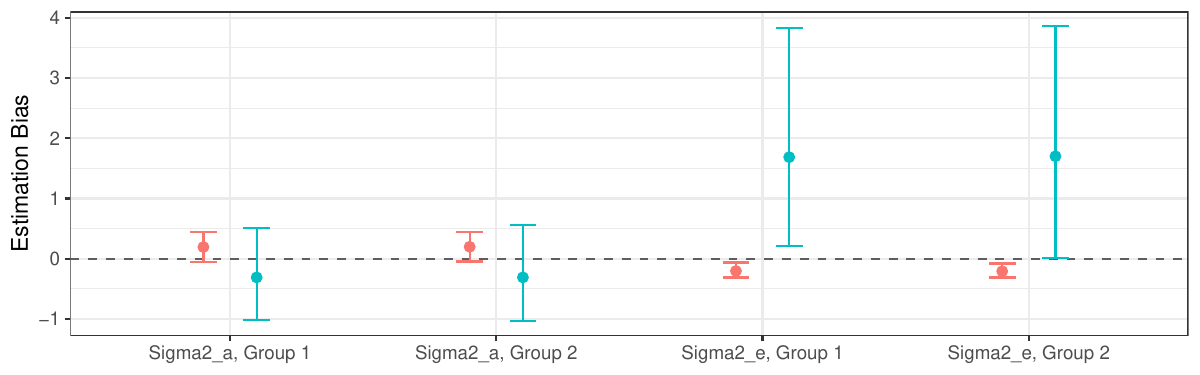}
	}
	\subfigure{
		\includegraphics[width=14.5cm]{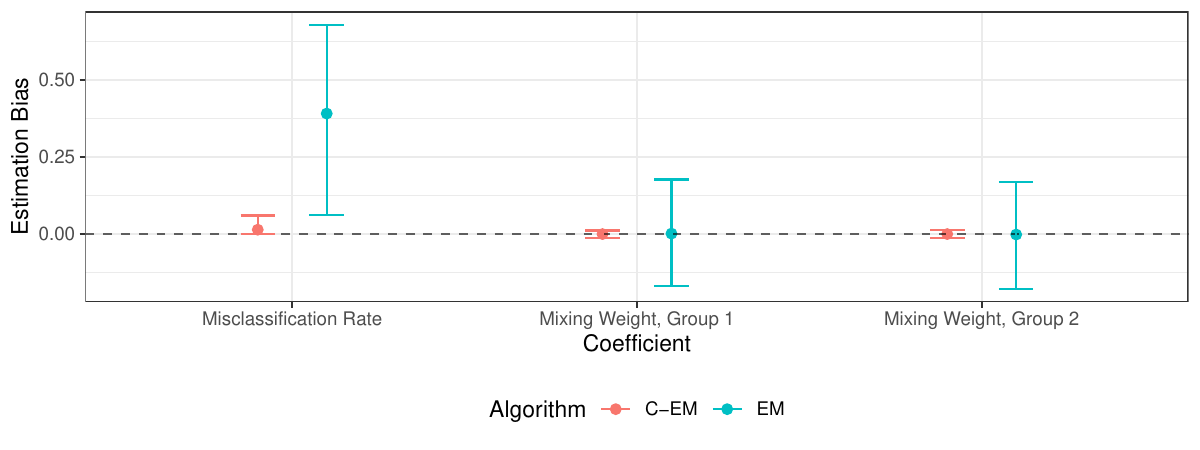}
	}
	\caption{Estimation bias for all parameters in the mixture when $N = 750$, $T=8$, $G=2$, and $p=5$. The dots correspond to the estimated bias over 250 replications, while the error bars represent the $2.5^{th}$ and $97.5^{th}$ percentiles of the empirical distribution of the bias. The coefficients $Time1,....,Time8$ represent the time-fixed effects, with $Sigma2\_a = \sigma^2_{\alpha,g}$ and $Sigma2\_e = \sigma^2_{\epsilon}$.\label{fig1:C9}}
\end{figure}
\newpage
\begin{figure}[h!]
	\setstretch{1.15}
	\centering
	\subfigure{
		\includegraphics[width=14.5cm]{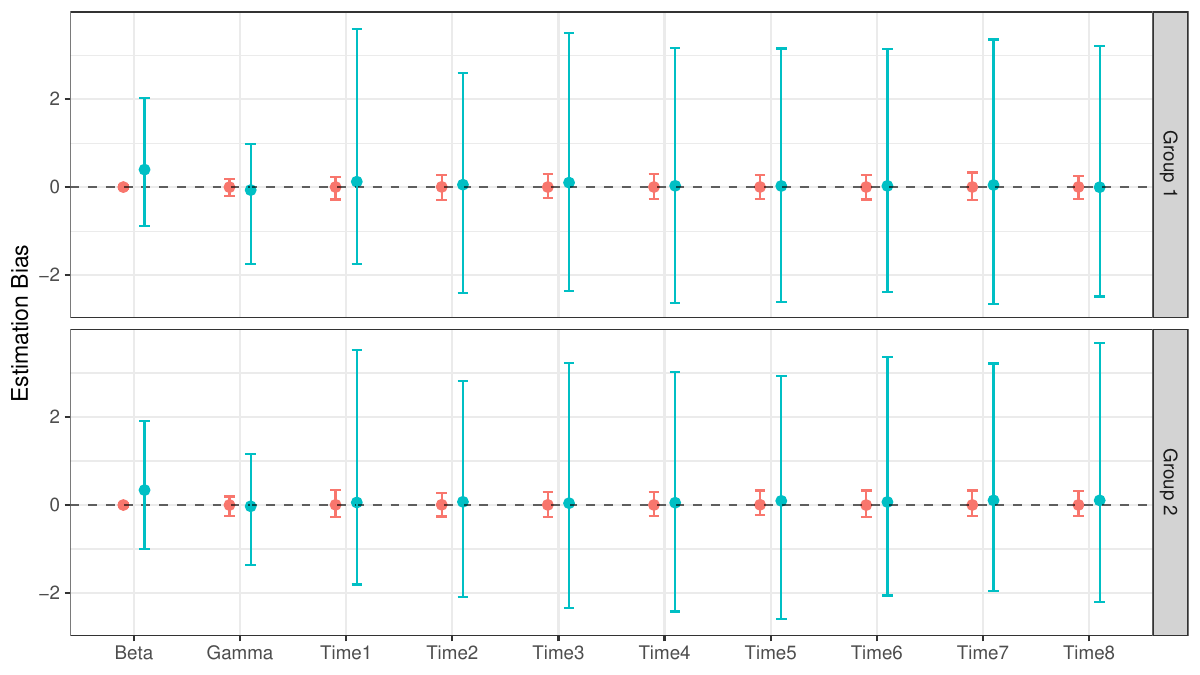}
	}
	\subfigure{
		\includegraphics[width=14.5cm]{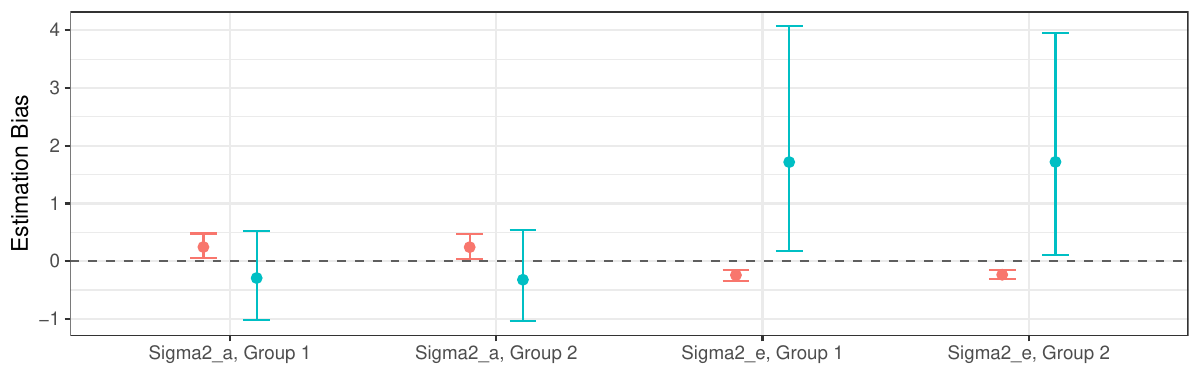}
	}
	\subfigure{
		\includegraphics[width=14.5cm]{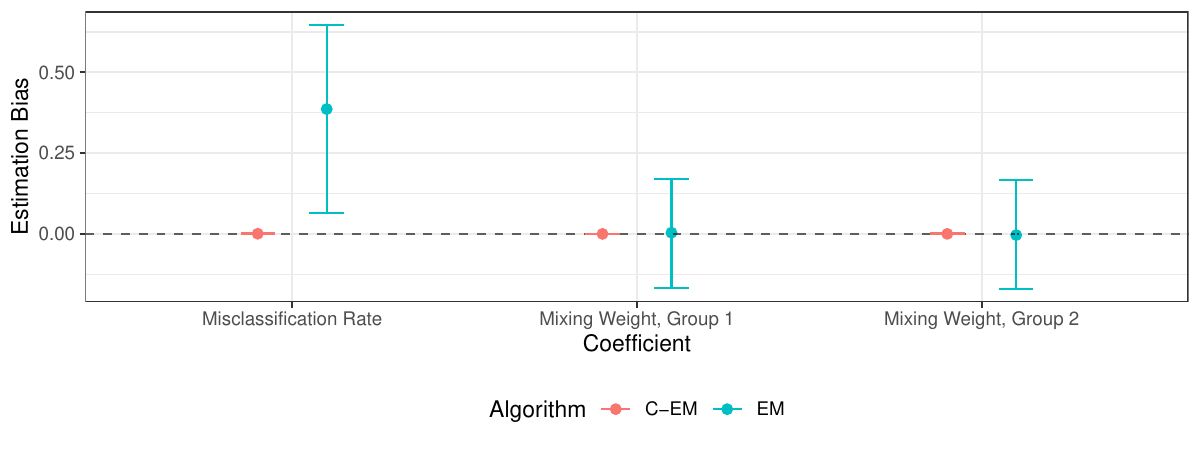}
	}
	\caption{Estimation bias for all parameters in the mixture when $N = 750$, $T=8$, $G=2$, and $p=10$. The dots correspond to the estimated bias over 250 replications, while the error bars represent the $2.5^{th}$ and $97.5^{th}$ percentiles of the empirical distribution of the bias. The coefficients $Time1,....,Time8$ represent the time-fixed effects, with $Sigma2\_a = \sigma^2_{\alpha,g}$ and $Sigma2\_e = \sigma^2_{\epsilon}$.\label{fig1:C10}}
\end{figure}
\newpage
\begin{figure}[h!]
	\setstretch{1.15}
	\centering
	\subfigure{
		\includegraphics[width=14.5cm]{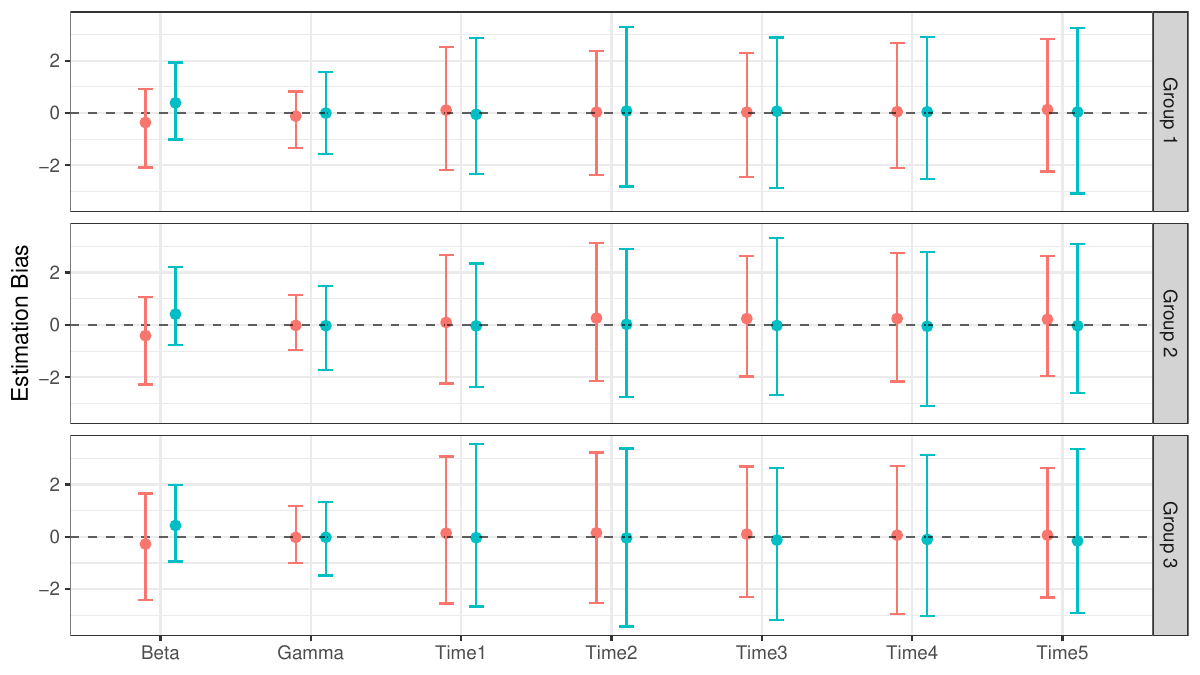}
	}
	\subfigure{
		\includegraphics[width=14.5cm]{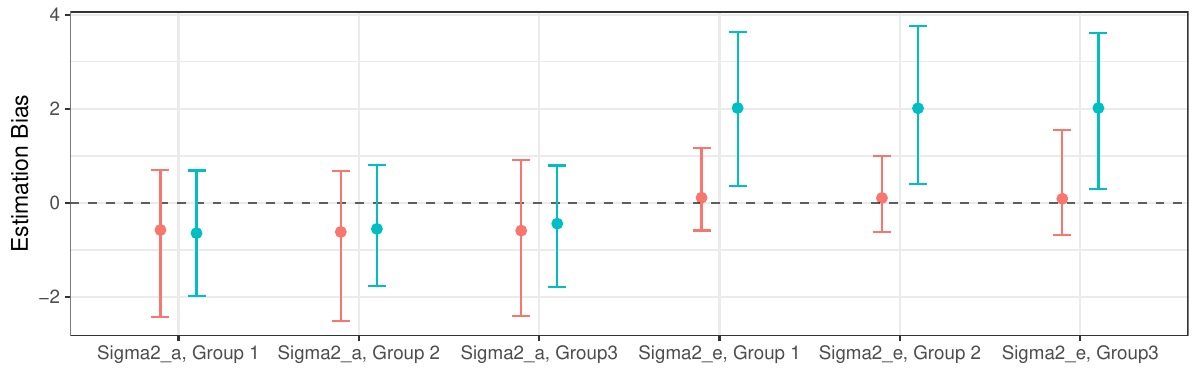}
	}
	\subfigure{
		\includegraphics[width=14.5cm]{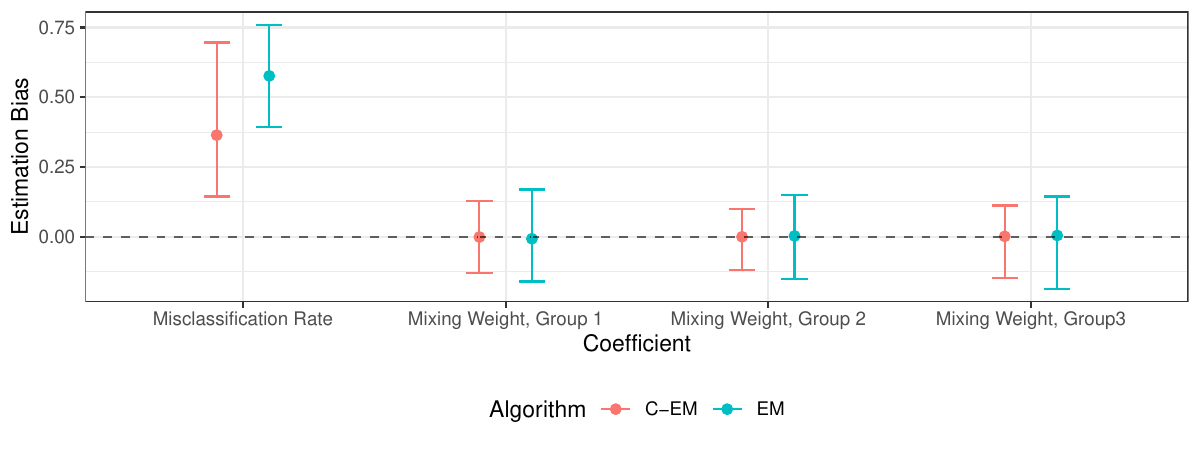}
	}
	\caption{Estimation bias for all parameters in the mixture when $N = 500$, $T=5$, $G=3$, and $p=1$. The dots correspond to the estimated bias over 250 replications, while the error bars represent the $2.5^{th}$ and $97.5^{th}$ percentiles of the empirical distribution of the bias. The coefficients $Time1,....,Time5$ represent the time-fixed effects, with $Sigma2\_a = \sigma^2_{\alpha,g}$ and $Sigma2\_e = \sigma^2_{\epsilon}$.\label{fig1:C11}}
\end{figure}
\newpage
\begin{figure}[h!]
	\setstretch{1.15}
	\centering
	\subfigure{
		\includegraphics[width=14.5cm]{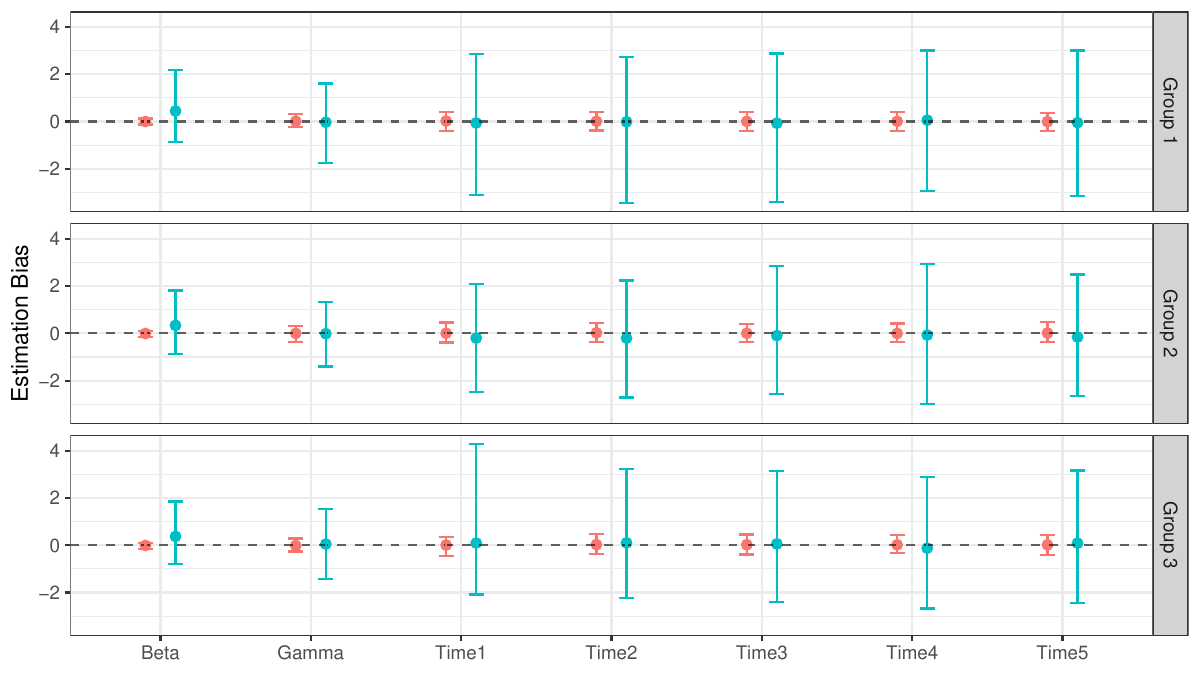}
	}
	\subfigure{
		\includegraphics[width=14.5cm]{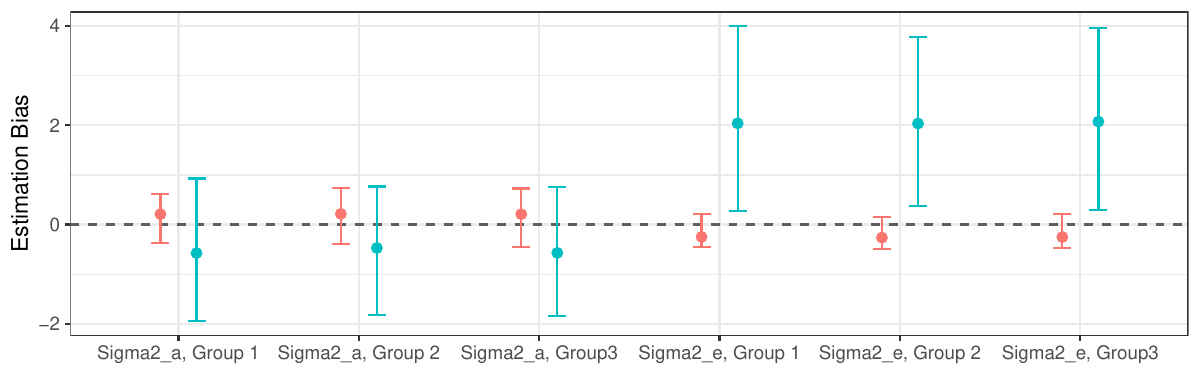}
	}
	\subfigure{
		\includegraphics[width=14.5cm]{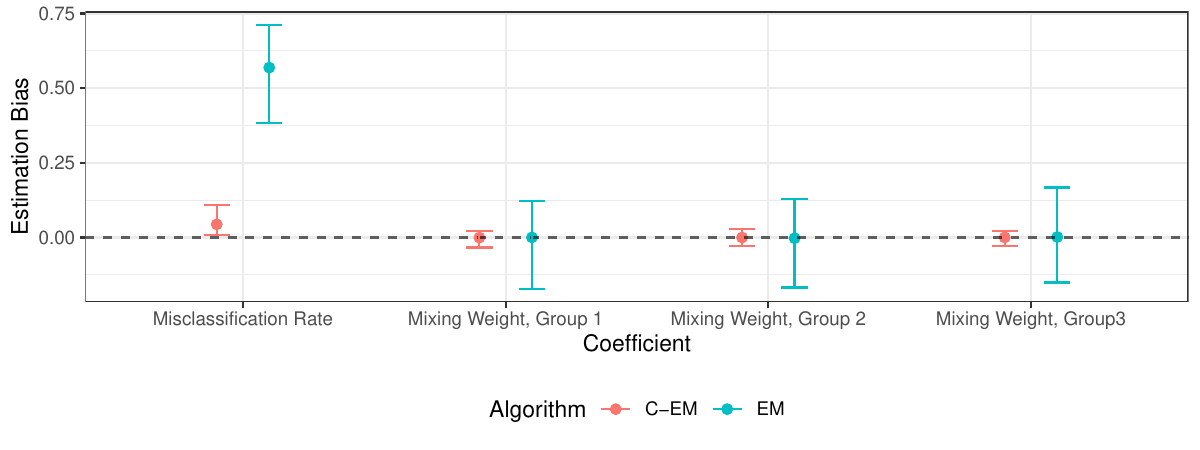}
	}
	\caption{Estimation bias for all parameters in the mixture when $N = 500$, $T=5$, $G=3$, and $p=5$. The dots correspond to the estimated bias over 250 replications, while the error bars represent the $2.5^{th}$ and $97.5^{th}$ percentiles of the empirical distribution of the bias. The coefficients $Time1,....,Time5$ represent the time-fixed effects, with $Sigma2\_a = \sigma^2_{\alpha,g}$ and $Sigma2\_e = \sigma^2_{\epsilon}$.\label{fig1:C12}}
\end{figure}
\newpage
\begin{figure}[h!]
	\setstretch{1.15}
	\centering
	\subfigure{
		\includegraphics[width=14.5cm]{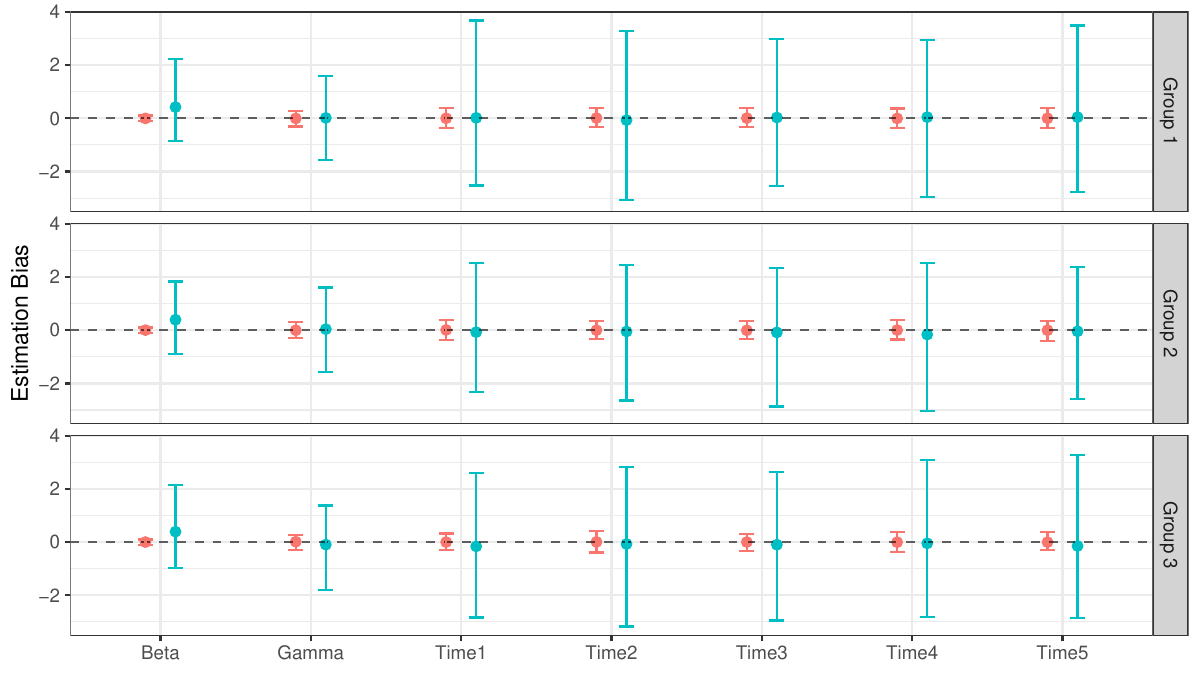}
	}
	\subfigure{
		\includegraphics[width=14.5cm]{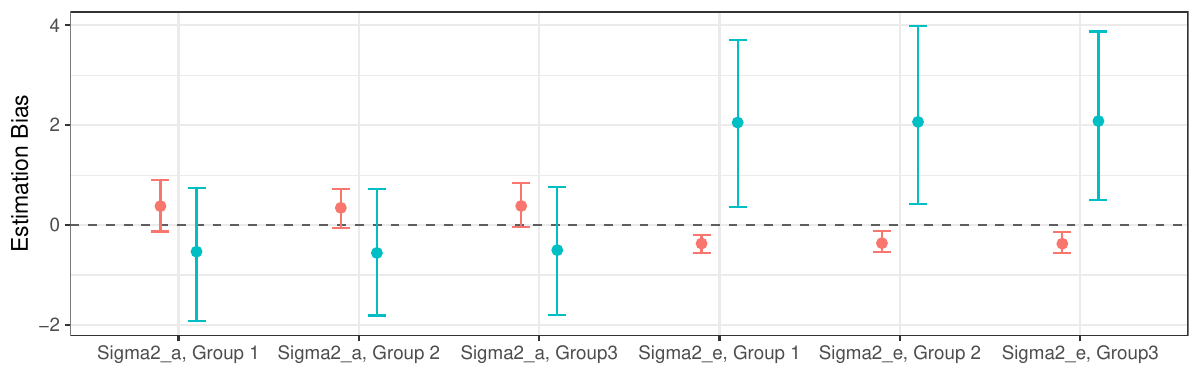}
	}
	\subfigure{
		\includegraphics[width=14.5cm]{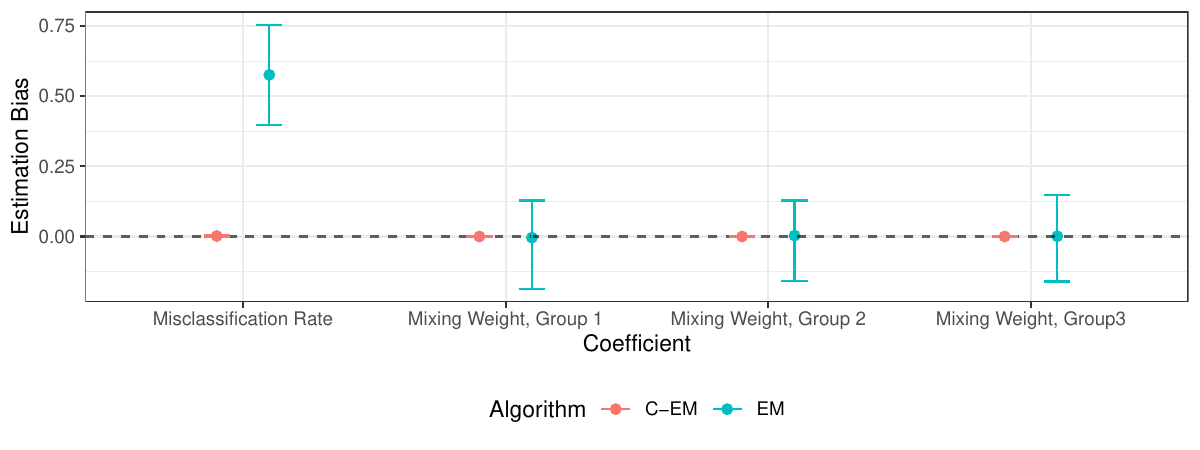}
	}
	\caption{Estimation bias for all parameters in the mixture when $N = 500$, $T=5$, $G=3$, and $p=10$. The dots correspond to the estimated bias over 250 replications, while the error bars represent the $2.5^{th}$ and $97.5^{th}$ percentiles of the empirical distribution of the bias. The coefficients $Time1,....,Time5$ represent the time-fixed effects, with $Sigma2\_a = \sigma^2_{\alpha,g}$ and $Sigma2\_e = \sigma^2_{\epsilon}$.\label{fig1:C13}}
\end{figure}
\newpage
\setcounter{table}{0}
\setcounter{figure}{0}
\renewcommand{\thetable}{D.\arabic{table}}
\renewcommand{\thefigure}{D.\arabic{figure}}
\section{Covariates Included in the Empirical Analysis}\label{sec1:D}
\begin{table}[h]
	\begin{center}
		\caption{\centering Covariates included in all component densities for each part of the empirical model \label{tab1:4}}
		\begin{tabular}{p{4.2cm} c c}\toprule\toprule
			{Covariate} & {Binary Part} & {Continuous Part} \\
			Time-varying ERA& X &  X \\ [1mm]
			Time-varying COCI&  &X \\[1mm]
			Time-averaged ERA& X & X\\[1mm]
			Time-averaged Charlson& X &X \\[1mm]
			Time-averaged COCI& X  & X\\[1mm]
			Gender & X  &X \\[1mm]
			Time-fixed effects & X &X \\[1mm]
			Unit-random effects & X& X\\[1mm]
			Intercept & X & X\\
			\bottomrule\bottomrule
		\end{tabular}\\
		\vspace{1mm}
		\small \textbf{Note}~: COCI = Continuity of care indicator, ERA = Elder's risk assessment.
	\end{center}
\end{table}
\newpage
\setcounter{table}{0}
\setcounter{figure}{0}
\renewcommand{\thetable}{E.\arabic{table}}
\renewcommand{\thefigure}{E.\arabic{figure}}
\section{Definition of the ERA and the Charlson Indices}\label{sec1:E}

\begin{table}[h]
	\begin{center}
		\caption{\centering The scoring systems of the modified ERA index and the modified Charlson index}
		\begin{tabular}{p{5.5cm} c c p{5.5cm} c}\toprule\toprule
			\multicolumn{2}{c}{\textbf{Charlson Index}} &&\multicolumn{2}{c}{\textbf{ERA Index}}\\
			\cmidrule{1-2} \cmidrule{4-5}
			{Condition} & {Weight} && Parameter& Weight\\
			\midrule	
			Peripheral vascular disease & 1  && \textbf{Social support index}&\\ [1mm]
			Chronic pulmonary disease &  1 &&\ \ \ 1st and 2nd quintiles &-1\\[1mm]
			Connective tissue disease & 1 &&\ \ \ 3rd quintile and null value&0\\[1mm]
			Ulcer disease & 1 &&\ \ \ 4th and 5th quintiles &1 \\[1mm]
			Mild liver disease & 1  &&\textbf{Age}&\\[1mm]
			Depression  & 1   &&\ \ \ 65-69&0\\[1mm]
			Use of warfarin & 1  &&\ \ \ 70-79&1\\[1mm]
			Hypertension & 1 &&\ \ \ 80-89&3\\[1mm]
			Hemiplegia & 2 &&\ \ \ $\ge$ 90&7\\
			Moderate or severe renal disease & 2&&\multirow{2}{7cm}{\textbf{Number of days in hospital during the last 2 years}} & \\
			Skin ulcers/cellulitis & 2&&&\\
			Moderate or severe liver disease & 3 &&\ \ \ 1 to 5 days & 5\\
			AIDS & 6 &&\ \ \ $\ge$ 6 days & 11 \\
			&&& \textbf{History of comorbidities} &\\
			&&& \ \ \ Diabetes & 2\\
			&&& \ \ \ CAD/MI/CHF & 3\\
			&&& \ \ \ Stroke & 2\\
			&&& \ \ \ COPD & 5\\
			&&& \ \ \ Cancer & 1\\
			&&& \ \ \ Dementia & 3\\
			\bottomrule\bottomrule
		\end{tabular}\\
		\vspace{1mm}
		\small \textbf{Note}~: CAD = Coronary artery disease; MI = Myocardial infarction; CHF = Chronic heart failure; COPD = Chronic obstructive pulmonary disorder; AIDS = Acquired immune deficiency syndrome. Overlapping items between the two indices were removed from the Charlson index to reduce collinearity.
	\end{center}
\end{table}
\newpage

\section{Additional Empirical Results}\label{sec1:F}
\subsection{Estimates Obtained from the Best Model, C-EM Algorithm}\label{sec1:F1}
\setcounter{table}{0}
\setcounter{figure}{0}
\renewcommand{\thetable}{F.\arabic{table}}
\renewcommand{\thefigure}{F.\arabic{figure}}
\begin{table}[h!]
	\begin{center}
		\caption{\centering Additional estimates associated with the optimal set of estimates generated by the C-EM Algorithm \label{tab1:16}}		
		\begin{tabular}{p{4.5cm} ccccc } \toprule\toprule
			\multirow{2}{2cm}{Coefficients} &\multicolumn{5}{c}{\textbf{Group/Component}}   \\ 
			&1&2&3&4&5 \\ \midrule
			\multicolumn{6}{c}{\textbf{Binary Part}}     \\ [2mm]
			\multirow{2}{*}{ERA} &-5.413***&19.666***&-14.607***&4.341*&174.416**\\
			&(0.904)&(5.046)&(1.755)&(1.785)&(55.578)\\[1mm]
			\multirow{2}{*}{Time-averaged Charlson} &-46.186***&31.002***&52.267***&35.537***&-758.674***\\
			&(2.099)&(4.890)&(2.355)&(4.457)&(65.068)\\[1mm]
			\multirow{2}{*}{Time-averaged COCI} &-2.129***&329.400***&-6.345***&-2.077*&1789.412***\\
			&(0.279)&(29.245)&(0.600)&(0.867)&(122.124)\\[1mm]
			\multirow{2}{*}{Time-averaged ERA}&1.179&47.679***&8.680***&0.449&659.342***\\
			&(0.923)&(5.181)&(1.553)&(2.298)&(110.359)\\[1mm]
			\multirow{2}{*}{Male} &1.157&277.050***&-9.908***&5.760&381.782***\\
			&(1.355)&(27.497)&(1.693)&(3.222)&(22.582)\\ 
			\multirow{2}{*}{Number of observations}&\multirow{2}{*}{2142}&\multirow{2}{*}{1392}&\multirow{2}{*}{2075}&\multirow{2}{*}{1774}&\multirow{2}{*}{1863}\\
			&&&&&\\ \midrule
			\multicolumn{6}{c}{\textbf{Continuous Part}}     \\ [2mm]
			\multirow{2}{*}{ERA} &-0.340***&-0.003&-0.669***&0.085*&-0.119***\\
			&(0.050)&(0.028)&(0.066)&(0.037)&(0.034)\\[1mm]
			\multirow{2}{*}{COCI} &0.608***&0.391&0.349***&-0.103***&-0.085***\\
			&(0.043)&(2.4E06)&(0.039)&(0.018)&(0.013)\\[1mm]
			\multirow{2}{*}{Time-averaged Charlson} &0.102&0.005&0.014&0.043*&-0.116***\\
			&(0.070)&(0.021)&(0.081)&(0.021)&(0.028)\\[1mm]
			\multirow{2}{*}{Time-averaged COCI} &-0.171***&0.013&-0.128***&-0.032&-0.005\\
			&(0.017)&(0.012)&(0.027)&(0.020)&(0.011)\\[1mm]
			\multirow{2}{*}{Time-averaged ERA}&0.549***&0.043&0.809***&-0.062&0.170***\\
			&(0.052)&(0.031)&(0.063)&(0.043)&(0.035)\\[1mm]
			\multirow{2}{*}{Male} &0.251***&-0.012&0.078&-0.051&-0.015\\
			&(0.066)&(0.043)&(0.079)&(0.061)&(0.038)\\ 
			\multirow{2}{*}{Number of observations}&\multirow{2}{*}{1801}&\multirow{2}{*}{1330}&\multirow{2}{*}{1097}&\multirow{2}{*}{1720}&\multirow{2}{*}{1828}\\
			&&&&&\\ \bottomrule\bottomrule
		\end{tabular}
	\end{center}
	\small \textbf{Note}~: Cluster-robust standard deviations are shown in parentheses. The number of observations refers to the sum of the estimated group memberships within each group. * = p-value$<0.05$, ** = p-value$<0.01$, *** = p-value$<0.001$.
\end{table}
\pagebreak
\subsection{Estimates Obtained from the Best Model, EM Algorithm}\label{sec1:F2}
\begin{table}[h!]
	\begin{center}
		\caption{\centering Additional estimates associated with the optimal set of estimates generated by the EM Algorithm \label{tab1:17}}		
		\begin{tabular}{p{4.5cm} cccc } \toprule\toprule
			\multirow{2}{2cm}{Coefficients} &\multicolumn{4}{c}{\textbf{Group/Component}}     \\ 
			&1&2&3&4 \\ \midrule
			\multicolumn{5}{c}{\textbf{Binary Part}}     \\ 
			\multirow{2}{*}{ERA} &-110.673***&-922.474&-1.195&54144.590***\\
			&(29.091)&(476.582)&(0.705)&(1830.669)\\[1mm]
			\multirow{2}{*}{Time-averaged Charlson} &-420.397***&205.818&-42.752***&487254.501***\\
			&(14.880)&(360.249)&(2.808)&(20966.975)\\[1mm]
			\multirow{2}{*}{Time-averaged COCI} &-2329.817***&-1363.916***&-1.363***&22241.703***\\
			&(78.736)&(118.584)&(0.291)&(751.892)\\[1mm]
			\multirow{2}{*}{Time-averaged ERA}&705.402***&3824.340***&2.434**&-63320.044***\\
			&(34.534)&(582.868)&(0.815)&(2140.879)\\[1mm]
			\multirow{2}{*}{Male}&-232.246***&2166.655***&-1.726&-44511.269***\\
			&(28.839)&(644.333)&(1.056)&(1504.569)\\ 
			\multirow{2}{*}{Number of observations}&\multirow{2}{*}{2452.8}&\multirow{2}{*}{2784.1}&\multirow{2}{*}{1956.5}&\multirow{2}{*}{2052.5}\\
			&&&& \\ \midrule
			\multicolumn{5}{c}{\textbf{Continuous Part}}     \\ 
			\multirow{2}{*}{ERA} &-0.124***&-0.011&-0.469***&-0.272***\\
			&(0.030)&(0.031)&(0.076)&(0.043)\\[1mm]
			\multirow{2}{*}{COCI} &-0.292***&-0.006&0.153***&0.098***\\
			&(0.027)&(0.006)&(0.017)&(0.018)\\[1mm]
			\multirow{2}{*}{Time-averaged Charlson} &-0.132***&0.060**&-15.542***&-11.622\\
			&(0.025)&(0.021)&(1.573)&(51.432)\\[1mm]
			\multirow{2}{*}{Time-averaged COCI} &-0.073***&-0.026*&-0.021&0.089***\\
			&(0.017)&(0.012)&(0.025)&(0.023)\\[1mm]
			\multirow{2}{*}{Time-averaged ERA}&0.230***&0.039&0.541***&0.372***\\
			&(0.033)&(0.034)&(0.085)&(0.055)\\[1mm]
			\multirow{2}{*}{Male} &0.150**&0.000&0.014&0.186*\\
			&(0.049)&(0.041)&(0.090)&(0.092)\\  
			\multirow{2}{*}{Number of observations}&\multirow{2}{*}{2402.0}&\multirow{2}{*}{2782.9}&\multirow{2}{*}{540.6}&\multirow{2}{*}{2050.6}\\
			&&&& \\ \bottomrule\bottomrule
		\end{tabular}
	\end{center}
	\small \textbf{Note}~: Cluster-robust standard deviations are shown in parentheses. The number of observations refers to the sum of the estimated posterior probabilities within each group. * = p-value$<0.05$, ** = p-value$<0.01$, *** = p-value$<0.001$.
\end{table}
\pagebreak
\subsection{Estimated Time-Fixed Effects, EM Algorithm}\label{sec1:F3}
\begin{figure}[h!]
	\begin{center}
		\includegraphics[width=14.5cm]{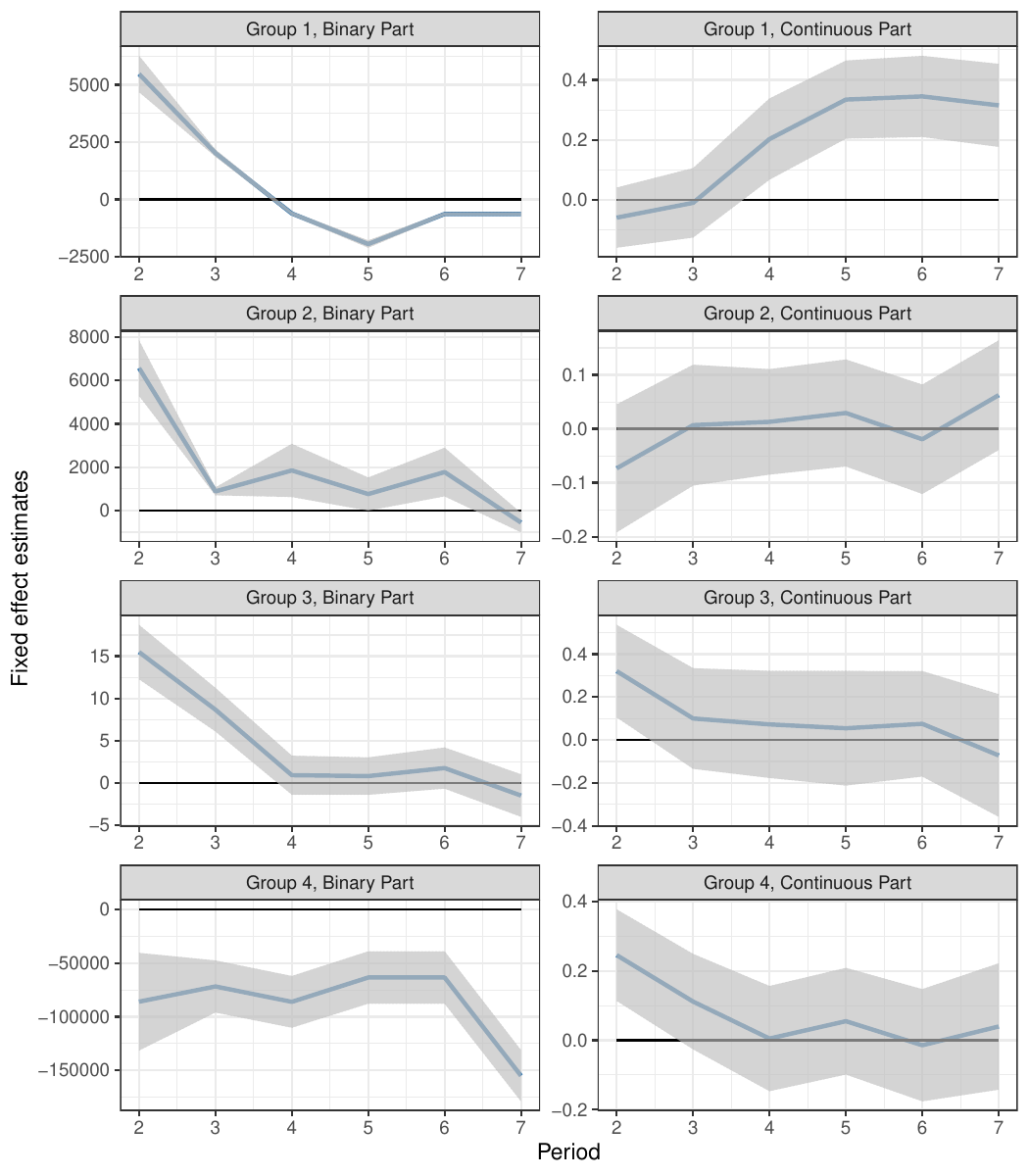}
	\end{center}
	\caption{Time-fixed effects associated of the optimal set of estimates generated by the EM algorithm. The value of the first time-fixed effect is equal to zero and is set as the reference value. The shaded areas correspond to the 95\% cluster-robust confidence interval and do not account for uncertainty in group memberships.}
\end{figure}
\end{document}